\documentclass[11pt]{article}
\usepackage{amsmath, amssymb}
\usepackage[active]{srcltx}
\usepackage{amsbsy}
\usepackage{amsfonts}
\usepackage{amssymb}
\usepackage{amsmath}
\usepackage{graphicx}

\newcommand{\mathsym}[1]{{}}

\renewcommand{\title}[1]{\vbox{\center\LARGE{#1}}\vspace{5mm}}
\renewcommand{\author}[1]{\vbox{\center#1}\vspace{5mm}}

\def\IZ{\mathbb{Z}}

\def\ci{\cite}
\def \V {v}

\def \S {{\rm S}}

\def \tl {\td \l}
\def \td {\tilde}

\def \O {{\mathcal O}}

\def \N {{\mathcal N}}

\def \m {\mu}

\def \bi{\bibitem}
\def \la {\label}

\def \l {\lambda}
\def\foot{\footnote}
\def \tl  {{\tilde \l}}
\def \sql {{\sqrt{\lambda}}}
\def \adss {$AdS_5 \times S^5$\ }
\newcommand{\rf}[1]{(\ref{#1})}
\def \ov {\over}

\def\th{\theta}

\newcommand{\hw}{\hat w}
\newcommand{\hn}{\hat \nu}
\newcommand{\hk}{\hat kappa}

\def\N{{\cal N}}
\def\F{{\cal F}}

\def\cc{\circ}

\def \ha{{\textstyle{1\ov 2}}}
\def \bw {{\rm w}}

\def\r{{\rm r}}

\def \J {\mathcal{K}}

\def \S {{\cal S}}
\def \J {{\cal J}}

\def \om {\omega}

\def \bi{\bibitem}
\def \la {\label}
\def \KK {{\rm K}}

\def \l {\lambda}
\def\foot{\footnote}
\def \tl  {{\tilde \l}}
\def \sql {{\sqrt \l}}

\def \adss {$AdS_5 \times S^5$\ }

 \def \r {\rho}

\def \ov {\over}

\def \varpi {{\rm w}}
\def \OO {{\cal O}}

\def \no {\nonumber }

\def \adss {$AdS_5 \times S^5$\ }

\def \N {{\cal N}}

\def \k {\kappa}
\def \ci {\cite}

\def\tr{{\rm tr}}

\newcommand{\id}{\mathbf{1}}
\renewcommand{\imath}{i}

\renewcommand{\geq}{\,{\geqslant}\,}

\newcommand{\binner}[2]{%
  {\langle}\kern-4.15pt{\langle}#1{,}\,#2{\rangle}\kern-4.15pt{\rangle}}

\newcommand{\ffrac}[2]{\raisebox{.5pt}%
  {\footnotesize$\displaystyle\frac{#1}{#2}$}\kern1pt}

\def\id{\protect{{1 \kern-.28em {\rm l}}}}

\makeatletter
\renewcommand\section{\@startsection {section}{1}{\z@}%
                                   {-3.5ex \@plus -1ex \@minus -.2ex}%
                                   {2.3ex \@plus.2ex}%
                                   {\normalfont\large\bfseries}}
\renewcommand\subsection{\@startsection{subsection}{2}{\z@}%
                                   {-3.25ex\@plus -1ex \@minus -.2ex}%
                                   {1.5ex \@plus .2ex}%
                                   {\normalfont\normalsize\bfseries}}
\makeatother

\def \vp {\varphi}
\def \ri {{\rm i}}

\def \L {\Lambda}
\def \M {{\cal M}}
\def \F {{\cal F}}  

\numberwithin{equation}{section} \makeatletter
\@addtoreset{equation}{section}

  \def \vp {\varphi}
  \def \a {\alpha}

\def \s {\sigma}

\newcommand{\be}{\begin{eqnarray}}
\newcommand{\ee}{\end{eqnarray}}

 \def \om {\omega}

\def\kahat{{\hat\kappa}}

\def\nuhat{{\hat\nu}}

\def \ri {{\rm i}}
\begin{document}

\textwidth 170mm 
\textheight 230mm 
\topmargin -1cm
\oddsidemargin-0.8cm \evensidemargin -0.8cm 
\topskip 9mm 
\headsep9pt

\overfullrule=0pt
\parskip=2pt
\parindent=12pt
\headheight=0in \headsep=0in \topmargin=0in \oddsidemargin=0in

\vspace{ -3cm} \thispagestyle{empty} \vspace{-1cm}
\begin{flushright} 
\end{flushright}
 \vspace{-1cm}
\begin{flushright} Imperial-TP-RR-02-2010
\end{flushright}

\begin{center}

{\Large\bf
Two-loop    $AdS_5 \times S^5$ superstring:   \\
\vskip  0.2cm
testing  asymptotic Bethe ansatz and finite size corrections 
}
 
 \vspace{0.8cm} {
  S.~Giombi,$^{a,b,}$\footnote{giombi@physics.harvard.edu} 
  R.~Ricci,$^{c,}$\footnote{r.ricci@imperial.ac.uk } 
  R.~Roiban,$^{d,}$\footnote{radu@phys.psu.edu}
  and 
  A.A.~Tseytlin$^{c,}$\footnote{Also at
   Lebedev  Institute, Moscow.\ \ 
   tseytlin@imperial.ac.uk }  
 }\\
 \vskip  0.2cm

\small
{\em
$^{a}$
Center for the Fundamental Laws of Nature,
Jefferson Physical Laboratory, \\
Harvard University, Cambridge, MA 02138 USA

$^{b}$
Perimeter Institute for Theoretical Physics,
Waterloo, Ontario, N2L 2Y5, Canada

 $^{c}$  The Blackett Laboratory, Imperial College,
London SW7 2AZ, U.K. 

    $^{d}$ Department of Physics, The Pennsylvania State University,
University Park, PA 16802, USA
}

\normalsize
\end{center}

 \vskip 0.4cm

\begin{abstract}
We continue the investigation of two-loop string corrections  to
the energy of a folded string  with a spin $S$ in $AdS_5$ and an angular
 momentum $J$  in $S^5$,
in the scaling limit of large $J$ and  $S$  with 
 $\ell={\pi J \ov \sql \ln S}$=fixed. 
We compute  the generalized scaling function at
 two-loop order ${\rm f}_2(\ell)$ both for small and large values of $\ell$
 matching  the predictions based on the asymptotic Bethe ansatz.  
In particular, in the  small $\ell$ expansion, we
derive an exact integral form 
  for the $\ell$-dependent coefficient of the 
Catalan's constant term in ${\rm f}_2(\ell)$. 
Also, by resumming a certain subclass of multi-loop Feynman diagrams
 we obtain  an exact expression for the leading $\ln\ell$  part  of 
 ${\rm f}(\ell,\sql)$  which is valid to any order in the $\a' \sim {1 \ov \sql}$
  expansion. 
At large $\ell$ the string  energy has a BMN-like expansion and the first 
few leading
 coefficients 
 are expected to be protected, {\it i.e.} to be the same at weak and at 
 strong coupling.
We provide a new example of  this non-renormalization for the  
term which is  generated at two loops in string theory and at 
one-loop in gauge theory (sub-sub-leading in $1/J$). 
%
We also derive a simple algebraic formula for 
 the term of maximal transcendentality in ${\rm f}_2(\ell)$ expanded 
 at   large 
 $\ell$.
In the second part of the paper we initiate the study of 2-loop finite size
 corrections to 
the string energy by formally compactifying the spatial world-sheet
 direction 
in the string action expanded near long fast-spinning string. 
We observe that the  leading finite-size corrections  are of ``Casimir'' type  
coming from terms containing at least one  massless propagator.   
We  consider  in detail the  one-loop order
(reproducing  the leading   Landau-Lifshitz  model prediction) 
and then focus on the  two-loop  contributions to the $1\ov \ln S$ term
 (for $J=0$).
 We  find  that in a certain regularization scheme 
 used to discard power divergences 
the two-loop coefficient of  the   $1\ov \ln S$ term appears to vanish.

\def \jj {\ell}

\end{abstract}

\def \g {\gamma}
\def \F {{\cal F}} \def \V   {{ V}} 

\def \el {\ell} \def \jj {\ell} \def \ff {{\rm f}}
\def \bea{\be}
\def \eea{\ee} \def \re  {\rf} \def \K {{\rm K} }
\def \edo {\end{document}}

\def \hn {\hat \nu}
\def \hm {\hat m} 
\def \hk {\hat \k} \def \hw  {\hat m}

\def \nua  {\nu_{_e}  }
\def \hna {\hat \nu_{_e} }
\def \ge  {\gamma_{_e}}

\newpage 

\section{Introduction \label{intro}}
\setcounter{footnote}{0} 


The correspondence between     fast-spinning 
folded closed strings 
in $AdS_5\times  S^5$   and    twist  operators in  the $\N=4$ SYM theory
is a remarkable tool  for  uncovering  and checking 
the detailed structure of the  AdS/CFT correspondence.
In particular, string  perturbative computations of  quantum corrections to the spinning 
string energy,  which should correspond to strong-coupling corrections to dimensions of  gauge-theory 
operators,  provide important data for checking the integrability-based (Bethe ansatz) 
  predictions for the string spectrum.
Here we will continue the  investigation of  two-loop string corrections  to 
the energy of the folded  $(S,J)$  spinning string  \ci{gkp,ft1}
using  and extending the techniques developed in our previous papers
\ci{cusp,cuspJ}.

To put the  results of our investigation into  perspective let us  first review the general structure of 
the dependence of string  energies  or gauge-theory dimensions 
$E=\Delta( S_i,  J_m; \l) $  on  spins and string tension  $T= {\sql \ov 2 \pi}$  or `t Hooft coupling $\l$. 
In general, $E$ is a  complicated  function of  several  variables  and  even having a 
 formal set of  Asymptotic Bethe Ansatz/Thermodynamic  Bethe Ansatz   equations   describing string spectrum 
 (see, e.g., \ci{serban} for a review)     one should still understand in detail 
the various patterns of  behaviour of this  function  in various limits. 
We shall concentrate on gauge theory  states  from the 
  ``$sl(2)$  sector'' represented by 
  the operators  like     $\tr ( D_+^S \Phi^J)$  dual to strings with large 
  spin $S$ in $AdS_5$ and  large  orbital momentum $J$ in $S^5$.

In perturbative planar  gauge theory    one first  expands in  $ \l \ll 1$  for fixed spins  $(S,J)$ 
\be E\equiv \Delta =  S +  J +   \g (S,J,  \l)\  ,  \ \ \ \ \ 
\g= \l \g_1 (S,J)  + \l^2 \g_2 (S,J)  +   \l^3 \g_3 (S,J)+  ...   \ ,    \la{dima} \ee
and may then expand $\g_n$   in large  spins. 
In semiclassical  string theory limit  one first expands in $\a' $, {\it i.e.} in $ \sql  \gg  1 $,
for fixed   semiclassical  parameters  
$\S= {S \ov \sql}  , \   \J = { J \ov \sql}  $  (implying that $S$ are $J$ are assumed to be 
as large as $\sql$) 
\be  E = S + J +  e (\S,\J,  \sql)  \ , \ \ \ \ \ \ \
e= \sql   e_0 (\S,\J)  +  e_1 (\S,\J)  +   { 1 \ov \sql} e_2(\S,\J) +  ...   \ ,   \la{stri} \ee
 and  may  then expand in large $\S,\J$.
The two limits are obviously very different  and cannot be in general  compared directly. 
The gauge-string duality  relation  implies  that summing up the expansion in 
\rf{dima} (which should have a   finite radius of convergence) 
  and then  re-expanding  the result 
  at strong coupling in the semiclassical  string theory limit one should
reproduce the  string-theory expansion \rf{stri}.

There are several special sub-limits depending on the relative  values of $S$ and $J$. 
The more familiar  {\it  ``fast-string''}   limit,   generalizing the BMN limit,  
 corresponds, on the gauge-theory side,  to  taking   $J \gg 1, \ S \gg 1 $  
with ${S\ov J}$=fixed  (this is a limit of  long  but  ``locally-BPS''   operators).  
In this case $\g_n$ in \rf{dima}  happen to have the following structure   ($n=1,2,3,..$) 
\be \la{dma}
\g_n = { 1 \ov  J^{2n-1} }  \Big(a_{n1} + { a_{n2} \ov J} +   { a_{n3} \ov J^2 }  + ... \Big) \,, \qquad \qquad
  a_{nm} \equiv a_{nm} \left( { S\ov J}\right) \,, 
  \ee
where $a_{n2}, a_{n3}, ...$ are coefficients of subleading finite-size corrections 
at $n$-th loop order from the underlying spin chain point of view (see, e.g., \ci{bmsz,btz}).
The corresponding limit on the   string side is  $\J \gg 1$, \ \ $ {\S\ov \J}$=fixed 
when $e_k$ in \rf{stri}  have the following expansion  \ci{ft3, bt}  ($k=0,1,2,...$) 
\be \la{dmai}
\begin{aligned}
& e_0 = { 1 \ov \J }  \Big(b_{00} + { b_{02} \ov \J^2} +   { b_{04} \ov \J^4 }  + ... \Big)   \ ,
\   \ \ \ \  \  e_1 = { 1 \ov \J^2 }  \Big(b_{10} + { b_{12} \ov \J^2} +   { b_{13} \ov \J^3 }  + ... \Big)   \ ,\\
&    e_2 = { 1 \ov \J^3 }  \Big(b_{20}  +  { b_{22} \ov \J^2}+   ...  \Big) \ , \ \ \ \ \ 
e_3 = { 1 \ov \J^4 }  \Big(b_{30}  + { b_{31} \ov \J }+ ... \Big) \ , \ \ 
... \ ,  \ \   \ \ \ \ \ \ 
   b_{kl} = b_{kl} \left( { \S\ov \J}\right) \,.
 \end{aligned}
 \ee 
Remarkably, due to the underlying supersymmetry of the theory and the   special nature of the states the 
 two different expansions have   the same formal  dependence of the spins 
and  can be described by the following interpolating formula 
\be
  E= S + J  +  { h_1 \ov J}  + { h_2 \ov J^2 }  +  { h_3  \ov J^3 } + 
 { h_4 \ov J^4}  +    { h_5 \ov J^5} +    ...  \ , \ \ \ \ \ \    \ \ \ \ \ \  
h_n=h_n\left({S\ov J}, \l\right)\,.     \la{err} \ee
Here  in  perturbative   gauge theory ({\it i.e.}  in   \rf{dma})
\be \la{wee}
&& h_1 =\l  a_{11} \ , \ \ \ \ \  h_2 = \l  a_{12} \ , \ \ \ \ 
h_3 = \l  a_{13} + \l^2   a_{21} \ , \ \ \ \ 
h_4= \l  a_{14} + \l^2   a_{22}  \ , \ \ \ \ \\ 
&&  h_5 =  \l  a_{15} + \l^2   a_{23}  +   \l^3   a_{31} + ... \ , \ \ \ ...\ \ \
\la{pty}  \ee
while in perturbative string theory   ({\it i.e.}  in   \rf{dmai})
\be \la{see}
&& h_1 =\l  b_{00} \ , \ \ \ \ \  h_2 = \l b_{10} \ , \ \ \ \ 
h_3 = \l   b_{20}  +  \l^2  b_{02}  \ , \ \ \ \ 
h_4=  \l   b_{30} +  \l^2  b_{ 12}   \ , \ \ \ \ \\ 
&&  h_5 =  \l^3   b_{04} +  \l^{5/2}  b_{13}  +  \l^2   b_{22}    + \l^{3/2} b_{31}+ ... \ ,  \ \ \ ...\ \ 
\la{ptyy}  \ee
The functions $h_1,h_2,h_3,h_4$  are thus linear or quadratic  functions of $\l$,  
 {\it i.e.} 
the corresponding coefficients  should be the same 
in the  gauge-theory and the  string-theory expansions --  they should match as  functions  of $S \ov J$ 
\be \la{maa}
 a_{11}=  b_{00} \ , \ \ \ \   a_{12}  = b_{10} \ , \ \ \ \ \
 a_{13} =    b_{20} \ , \ \ \ \      a_{21} =  b_{02} \ , \ \ \ \ \
 a_{14}=    b_{30} \ , \ \ \ \ \ \     a_{22}  =     b_{12} \ . \ee 
At the same time,   $a_{31} \not= b_{04} $   since  $h_5$ (and also $h_6,...$)
 is a non-trivial   function of $\l$; this is 
 related to the presence of non-trivial phase in ABA,  
explaining  \ci{bt,bes}, in particular,  the well-known
 ``3-loop disagreement'' \ci{call}. 
 
The matching \rf{maa}  should be due to  the underlying supersymmetry  of 
the  large $J$ expansion 
and the structure of the  asymptotic Bethe Ansatz \ci{bes}.\foot{It may 
 be understood also 
as a consequence of  the  exactness of the coefficients of the leading 
low-derivative terms in the underlying effective Landau-Lifshitz type action  \ci{ll,mtt}.}
The equivalence 
 between the  one-loop gauge and tree-level string coefficient functions 
 $a_{11}=  b_{00}$ was explicitly demonstrated in \ci{bmsz,ft3,bfst}; the matching 
of the one-loop  gauge and the one-loop string coefficients 
$ a_{12}  = b_{10} $ was seen in \ci{btz}.  However, the 
equality of the 1-loop 
gauge and 2-loop string coefficients  $  a_{13} =    b_{20}$   was  not previously checked directly as the relevant 2-loop string computation is non-trivial to perform (the sub-sub leading finite size correction  on 
one-loop gauge theory side is also non-trivial to extract). 

One of our aims below will be to provide such a  check  in a  setting similar to the one   described above. 
We shall consider  another scaling limit -- of  {\it ``fast long strings''}  \ci{bgk,ftt,am2,frs}, 
{\it i.e.}\footnote{In the following we will use the normalization of  \cite{frs}: \  $j ={J\ov \ln S}$.}
\be \la{gaa}
  \l \ll 1, \ \ \  \  S \gg J \gg 1,  \   \ \  \  \ \ \  j\equiv  {J\ov \ln S}={\rm fixed} \ee
 on the gauge theory side and 
 \be \la{dee}
\sql \gg 1 ,\ \  \ \ \S \gg \J \gg  1 , \ \ \ \ \ \ \ 
 \ell\equiv {\pi  \J \ov \ln \S} 
= { \pi j \ov \sql}  = {\rm fixed}   \ee
   on the string theory side. 
One  can then study  also  subcases of 
 small or large $\ell$  and $ j$.
If  $ \ell \ll 1$,  {\it i.e.}    $ \ln \S \gg \J$ 
 then one finds on the string theory side 
\be && E= S  + { \sql \ov \pi}
 \ff(\ell, \sql)  \ln S  + ... \ ,\nonumber  \\ 
  && \ff(\ell, \sql) = {\rm f}_0(\ell) + { 1 \ov \sql } {\rm f}_1(\ell) + { 1 \ov (\sql)^2 } {\rm f}_2(\ell) + ... \ , \la{tre}  \\
&&
\ff_{\ell \to 0} =  \ff(\l) + 
\ell^2 \sum_{n=0}^\infty \frac{ c_n(\ln\ell)^n+d_n(\ln \ell)^{n-1}+\dots  }{(\sqrt{\lambda})^{n-1}}
 +{\cal O}(\ell^4) \  , \nonumber \ee
where 
$c_n,d_n$   are essentially  fixed by the  $O(6)$ sigma  model truncation of the string action \ci{am2}.
At tree level ${\rm f}_0= \sqrt{1 + \ell^2}$, while 
 the 1-loop coefficient  ${\rm f}_1$ is  \ci{ftt} 
  \be  \la{onlo}
&&\hskip -.6cm {\rm f}_1(\ell) = \frac{1}{ \sqrt{ 1 + \jj^2} } \Big[ 
  \sqrt{ 1 + \jj^2}  -  1
 + 2 ( 1 + \jj^2 ) \ln (1 + \jj^2) \no
  - \  \jj^2   \ln \jj^2 
  -  2( 1 + \ha  \jj^2) \ln [   \sqrt{ 2 +   \jj^2}
    (1 + \sqrt{1 + \jj^2})] 
    \Big]  \no \\
 &&~~ ={ {-{3 \ln 2}  - 2 \jj^2 \Big( \ln \jj   - { 3 \ov 4}\Big) 
 + \jj^4\Big ( \ln \jj   - { 3 \ov 8}   \ln 2   - { 1 \ov 16} \Big)
 + {\cal O}(\jj^6)}} \ . \la{jop} \ee
The  2-loop coefficient  was found in  \ci{rt1,rt2,cuspJ}
\be
&&{\rm f}_2(\ell) = -{\rm K} \  
+ \ell ^2 \Big(8\ln^2  \ell  -6 \ln  \ell-\frac{3}{2} \ln 2+\frac{11}{4}\Big) 
\no \\
&&~~~
+  \ell^4 \Big{(}\hskip-.1cm-6 \ln^2 \ell -\frac{7}{6} \ln \ell +3 \ln 2 \ln \ell
-\frac{9}{8} \ln^2 2+\frac{11}{8}\ln 2 +\frac{3}{32} {\rm K}-\frac{233}{576}\Big)
+{\cal O}(\ell^6)\la{hop}\,,
\ee
where ${\rm K}$ is the Catalan's constant.
The string   expression for $\ell^2$ term \rf{hop}
 is finally in agreement  \ci{cuspJ}  with  the ABA results
at strong coupling  \ci{gromov,basso,volin08}.  Similarly the $\ell^4$ 
term \ci{cuspJ}  also agrees with
  \ci{gromov}.
  
Extending the techniques used in \cite{cusp,cuspJ}  based
 on the AdS light-cone gauge string action 
 \cite{mt2}, we shall find additional  higher order  terms in ${\rm f}_2$  
extending the  agreement with the ABA  result of \ci{gromov}.
In particular, we shall determine  the exact form of  the  function of $\ell$ that 
multiplies the transcendental constant ${\rm K}$.

We shall  also  find that  the  
leading powers of $\ln  \ell$ terms in the $\ell\to 0$ expansion at each order 
in $1 \ov
\sql$  are
  generated by  a special class of Feynman graphs
that can be resummed to all orders in the 
worldsheet loop expansion. This then leads    to 
 the exact  form of the leading $\ln \ell$ 
 part of $\ff(\ell)$ in \rf{tre}, which is, 
  again, in agreement with the ABA  prediction  of   \ci{gromov}  
\be
{\rm f}(\ell, \sqrt{\lambda})\big{|}_{\rm lead.~log.}\  = \sqrt{1+\frac{\ell^2}{1+
\frac{4}{\sqrt{\lambda}}\ln\ell}} \la{lee}  \ . 
\ee
\def \cc {{\rm  c}}
In contrast to the fast string limit  \rf{err}  where the  spin dependence
 was the same at large and strong coupling  with the coefficients being, 
  in general,   interpolating  functions of $\l$, here the dependence 
on $\ell$  on the  string side is not the same as dependence on $j$ on 
the gauge side  where  the small $j$ 
expansion is polynomial in $j$ \ci{bgk,frs} ($q_n$ are power series in $\l$) 
\be
E-S = \big[ f(\l) + q_1 (\l) j + q_2(\l) j^3 + ...\big] \ln S + ... 
\ee  
To relate the gauge and string theory expressions it is necessary to first 
resum the gauge theory expansion   and then to  re-expand it at 
strong coupling keeping $\ell = { j \ov \sql}$ fixed \ci{rt2}.

The analog of the {\it  ``fast string''}  limit  is found by expanding   
$ \ff(\ell, \sql) $ in \rf{tre} at 
large $\ell$ on the  string side and at large $j$ on the  gauge side. 
 Namely,  assuming  $ \ell \gg 1$,  {\it i.e.}    $\J \gg   \ln \S  $  and 
 $j= { J \ov \ln S} = \frac{\sql}{\pi}\ell \gg 1 $  we should  be 
 encountering again ``locally-BPS''   states 
 for which the $\ell$ or $j$   dependence should appear to be the same 
at weak and at strong coupling.  We may write then the string result 
expressed in terms of $j= \frac{\sql}{\pi} \ell $ 
in a form similar to \rf{err} with  coefficients depending on 
${\ln \S \ov \J}$  instead of $\S \ov \J$.
To leading order in $\ln S$ we then get 
\be
\begin{aligned} 
  E&= S +  f(\l, \ell) \ln S + ... \ , \\
 f(\l, \ell)_{\ell \gg 1} &= j +   { \cc_{10}\l \ov j}  + { \cc_{11}\l  \ov j^2 }  +  { \cc_{12} \l +  \cc_{20} \l^2   \ov j^3 } + 
{ \cc_{13} \l +  \cc_{21} \l^2   \ov j^4 } +    { p_5(\l)  \ov j^5} + { p_6(\l)  \ov j^6} +    ...  \ , 
\end{aligned}
\la{rrr} \ee
with $p_5, p_6, ...$ being   non-trivial interpolating functions of  the  coupling.
Here the protected coefficients $\cc_{nm}$  are $m$-loop string theory 
contributions  which  should also match 
$n$-loop gauge theory contributions  for the same reason as discussed
 above  for \rf{err}.
  Explicit results found in  tree-level  and one-loop string theory \ci{bgk,ftt}  should  match 
the one-loop and two-loop gauge theory results \ci{bgk,beccaria} 
\be 
\cc_{10}= { 1 \ov 2 \pi^2}\ , \ \  \ \ 
 \cc_{11}= - { 4 \ov 3 \pi^3}\ , \  \   \ \ \  \cc_{20}= -{1 \ov 8 \pi^4}\ , \ \  \ \ \ \
\cc_{21}= {4 \ov 5 \pi^5}\ ,   \la{meec} \ee 
Here we shall  consider  the  $\cc_{12}$ term in  $E-S$  \rf{rrr} 
\be \la{hih}  \cc_{12} \l  { \ln^4 S \ov J^3} = \cc_{12} { 1 \ov \sql} { \ln^4 S \ov
\J^3}\ ,  \ee
 which  originates at 1 loop on the gauge theory side and at
 {\it 2 loops} on the  string theory side.
This is a sub-sub-leading finite-size term from the weak-coupling  $sl(2)$ sector 
spin-chain perspective
(with $J$ being the length of the spin chain). 
 The ABA predictions for the value of $\cc_{12}$ both at weak
\ci{volin08}  and at strong \ci{volin}  coupling   appears to be 
\be \la{pred}
\cc_{12} =  { 1 \ov 3 \pi^2}    \ ,  \ee
suggesting its non-renormalisation. 
Here  we shall compute  this coefficient directly as a  2-loop correction 
in string theory
 (still defined on $\mathbb{R}^{1,1}$),  providing  an
  intricate 2-loop  check of the ABA  at strong coupling. 
\def \OO {{\cal O}}

Another  limit that is useful to  consider is   that of 
{\it ``slow long  strings''}   which corresponds to  ``long''  far-from-BPS 
operators $\tr ( D_+^S \Phi^J)$ 
with   $ \ln S \gg J $, $J$=twist=fixed  on the gauge side and 
$\ln \S \gg \J  $, \   $\J$=fixed  on the  string side.
In this case 
\be   &&  E= S + f(\l, J) \ln S +  h(\l, J) + { u(\l, J)  \ov \ln S} + ...
 + O( {1 \ov S})  \la{egh} \ ,    \\
&&   \ff_{\l \gg 1} = c_0 \sql + c_1 + ...\ , \ \ \ \ \ \
  \ff _{\l \ll 1} = b_1  \l + b_2 \l^2 + ... \la{koe} \ . 
\ee
On the gauge theory side the 
scaling functions $f$ and $h$ are not sensitive to wrapping effects, 
{\it i.e.} they should be captured by the 
ABA  (or BES-type integral equations) 
 \ci{es,bes,frz}  and by string theory on $\mathbb{R}^{1,1}$ 
 \footnote{
In the string theory calculation of $h(\lambda, J)$ the string end-points 
turn out to be important \cite{bed}. While indeed the calculation can be carried out 
in the large $\ln S$ limit ({\it i.e.} on an decompactified worldsheet)
it requires use of the exact solution, valid on a cylindric worldsheet 
$\mathbb{R}\times S^1$. A two-loop calculation for the exact folded string solution
on $\mathbb{R}^{1,1}$
should reproduce the two-loop term in the strong coupling expansion of the virtual scaling 
function $h(\lambda, J)$ obtained from ABA in \cite{frz}. 
A linear integral equation governing $h(\lambda, J)$
was first written down in \cite{fio_large_spin}, where an alternative approach to the generalized scaling function $f(\lambda, J)$ and its subleading correction was discussed.
In the following we shall ignore $h(\lambda,J)$
and focus on the sub-sub-leading 
$1/\ln S$ term.}. 
Indeed, the wrapping contributions at weak coupling  are suppressed by  powers 
of $S$ (starting at 5 loops   with 
  $ \ln^2 S \ov S^2$  \ci{ban}).
  The $1/\ln S$ terms  are  not  present 
  in the usual perturbative expansion.
One may wonder if the 
 function $u$   may  also determined by  a linear 
 integral equation following from the ABA; it 
 appears that for fixed $J$  at weak coupling  ABA predicts $u=0$   \ci{fio}.
At the same time, $u$ is certainly non-zero at strong coupling, 
{\it i.e.}  in the semiclassical expansion in string theory as we shall review below.
Then matching the weak-coupling  and strong-coupling forms of \rf{egh}
would require resummation.

For $J=0$, considering the 1-loop correction to the  folded spinning string 
energy  on $\mathbb{R} \times S^1$, {\it i.e.}  by 
replacing the integral over the 
continuous  spatial momentum  by a
 discrete summation 
\be
\int dp \to  \frac{2 \pi}{ L}  \sum_n\,\,\,,\qquad \qquad L\sim  \ln S\,, 
\ee
 one finds a finite-size ``Casimir effect''-type  correction 
to the string 1-loop energy coming from five massless  modes:
\be \la{fik}
\J=0: \ \ \   \ \ \    \ u = k_1 + { k_2 \ov \sql } + ...\ , \ \ \  \ \ \ \ \ 
   k_1= - { 5\pi  \ov 12}    \ . \ee    
This correction was first  found in \ci{zama}  (by formally extending to    
$\mathbb{R} \times S^1$
world sheet the sum over the characteristic frequencies  found in \ci{ft2} in the 
infinite spin limit)  and then 
confirmed rigorously in \ci{bed}
(by  starting with the exact form of the folded string solution).

As we shall discuss below,  if one starts with the 1-loop expression in the case of 
 $\J\not=0$ \ci{ftt} (that leads to   \rf{onlo} in the case of the string on 
 $\mathbb{R}^{1,1}$)
and extends  it to  $\mathbb{R} \times S^1$ world sheet    one finds instead only one 
massless mode
 contribution; taking the $\J\to 0$ limit would lead  to the conclusion 
 that $ k_1= - { \pi \ov 12} $, in an apparent contradiction with \rf{fik}.
More precisely, 
the massless mode contribution to the 1-loop string energy  producing the 
``Casimir'' ${1 \ov L}  \sim { 1 \ov \ln S} $ term 
can be written as 
\be  \la{jjj} 
 \Delta E_1=   (E_1)_{\rm massless}=- { \pi  \ov 12} {1 \ov  \ell^2 + 1   } { 1 \ov  \ln \S}=
- { 1  \ov 12} {\ell \ov  \ell^2 + 1   } { 1 \ov  \J }
 =    - { 1 \ov 12\pi } { \l \ov J^2 + {\l \ov \pi^2} \ln^2 \S } \ln \S \ . 
\ee
The relation to the $\J=0$ case \rf{fik}   can be understood  by also 
taking into account the contribution of  four massive 
(mass $\sim \J$) modes producing {\it exponential}  $\sim {\rm exp } ( - c \J)$
    corrections that should be  added to \rf{jjj} and 
 resummed  before taking the $\J\to 0$ limit. 
From the point of view of comparison to the  $sl(2)$ sector Bethe ansatz  result, 
the massless  mode contribution should be  a finite size effect captured by ABA   while the exponential 
 contributions of 4 light massive modes  should be  ``L\"uscher'' corrections corresponding to 
wrapping contributions  at weak coupling. The  full 1-loop string semiclassical result  must match 
the TBA results as was demonstrated in \ci{grom}. 

The  last  form of the expression in \rf{jjj}  which is formally  analytic in $\l$ suggests that 
  at least first two terms in its large $J$ expansion 
\be  \la{kkk} 
  \Delta E_1
 =    - { 1 \ov 12\pi } { \l \ov J^2}  \ln  S  +   { 1 \ov 12\pi^3 } { \l^2 \ov J^4}  \ln^3  S   + ...  
\ee
  may  not  be renormalised, {\it i.e.} should 
  appear also with the same coefficients at weak   coupling, 
  originating from finite-size corrections in the  $sl(2)$ sector Bethe Ansatz.
Like  the leading order $\l$  term in the classical string energy 
({\it i.e.} the  $\cc_{10}$ term in \rf{rrr}  or the analog of  
$b_{00}$ term in \rf{see})\foot{As in \ci{bgk} here 
 $\ln S$ may be replaced by $\ln { S \ov J}$ but we will ignore this  detail.
 The  $\ln^2 S$ term was found in \ci{ft1} on the  string side and reproduced 
 from the  one-loop  gauge-theory BA in \ci{bfst}.}
\be  
E_0 -S = 
\la{cla} E_0 - S = \sqrt{ J^2 + {\l \ov \pi^2}  \ln^2 S}  + ...
=   J  +   {\l \ov 2\pi^2 J }  \ln^2 S  + ... \ ,  
\ee
which  matches   the 1-loop  gauge theory result, 
the leading terms in \rf{kkk} 
should be   similar to  the  protected $b_{10}$ and 
$b_{12}$ terms in \rf{see},\rf{ptyy} and
 $\cc_{11}$ and $\cc_{21}$ in \rf{rrr},\rf{meec}. Note, however, that the terms in \rf{rrr} 
are found in a different limit than \rf{jjj}: by first 
fixing $\ell$ and extracting the coefficient of the $\ln S$ term 
and then expanding this coefficient in  large $\ell$. 
 Here instead we are discussing the 
 coefficient of the $1 \ov \ln S$ term at 
 fixed $\ell$ and then  expand in large  $\ell$. 

As in other similar cases of ``fast string'' states   \ci{btz} the non-renormalization of the
 leading  term in \rf{kkk}, {\it i.e.} that it 
 can    be obtained  also  directly  at weak coupling  from the one-loop  $sl(2)$ sector spin chain  Hamiltonian,  
 suggests that it  should be reproduced  as  finite-size correction from  the corresponding Landau-Lifshitz (LL) model. 
  The LL model \ci{krt}  describes, on the one hand,  fast strings moving in $AdS_3 \times S^1$ 
and,  on the other hand,     the corresponding coherent  states  of the 1-loop $sl(2)$ sector 
  Hamiltonian \ci{ste,ptt}.  Indeed, from the string theory point of view, the LL model keeps  the contribution 
of the ``massless'' mode  in the $AdS_3$ part  that is responsible  for the  ``Casimir'' term in 
 \rf{jjj}.  We shall discuss the details of the LL model relation in Appendix~A. 
  In contrast,    the order $\l$  term in \rf{hih} appears to   receive contributions from 
 several massive  string modes  in the 2-loop string  computation that  we shall describe below 
and is {\it  not}  captured by the 1-loop LL model.

While  the  matching of  the 1-loop  string and the  strong-coupling TBA results for semiclassical
($\J = { J \ov \sql}$,etc.  kept fixed)   string states 
appears to be guaranteed  \ci{grom}, this still  remains to be verified at the 2-loop string level. 
This should provide further non-trivial tests of {quantum integrability}  
of  \adss  superstring  and of consistency of TBA. 
Below we shall initiate the study of 2-loop string finite size corrections by formally extending 
the computation of the string partition function in the  folded string background 
done in \ci{cusp}  on $\mathbb{R}^{1,1}$  to the $\mathbb{R} \times S^1$ world sheet. 
 We  shall   discuss in detail  the computation  of the 2-loop
string    correction to the $1 \ov \ln S$ term  in   \rf{egh} at $J=0$
and show that  it appears to be zero in a natural regularization scheme, 
in contrast to the  non-vanishing 1-loop term 
\rf{jjj}.

The detailed plan for the rest of the paper is as follows. In section \ref{LCreview} 
 we discuss the 
AdS light-cone gauge for the $AdS_5\times S^5$ superstring. This is the gauge we will use 
for all our computations. We also introduce the generalized cusp background with 
vanishing winding
(which is related to the spinning folded string by a conformal transformation \ci{krtt}).  
In section 
\ref{gensc} we review the relation between the partition function for the long 
$(S,J)$ spinning
 string and the corresponding quantum corrected AdS energy.  
The various contributions to the 2-loop string partition function in the generalized cusp 
background are discussed in section \ref{2Loop}. Compared to our previous work
\cite{cuspJ} we  try to compute the various Feynman integrals exactly 
 rather than perturbatively at small $\ell$. With these results at hand,  in section
  \ref{Gensection} we discuss  the calculation of the generalized scaling function
   ${\rm f}(\ell)$ at 2-loop order both for small and large values of $\ell$. 
    In particular, in the small $\ell$ expansion, we show that the 
    coefficients of $\ln 2$ and $\ln^2 2$ match all the values reported in \cite{gromov}.
      We also present an exact formula for the
$\ell$ dependent coefficient of the Catalan  constant ${\rm K}$  
in ${\rm f}_2(\ell)$ in terms of an integral 
generating function, see eq.~\rf{CatalanFunction}. Still in the context of the small $\ell$
 expansion, we obtain, by resumming all ``maximally non 1-PI" diagrams, an exact expression 
  for the leading logarithm coefficient in  ${\rm f}(\ell)$  which is valid at any loop
   order, see eq.~\rf{LeadingLog}.  In the case of the large $\ell$ expansion, we derive
    a simple algebraic formula for the maximal transcendentality piece which turns out 
    to be proportional to $\pi^2$, see eq.~\rf{pi2} which we match against the ABA prediction.
This result is the large $\ell$ analog of the coefficient of the ${\rm K}$  constant at 
small $\ell$.
%
%
The non-renormalization of the coefficient $c_{12}$ in
      \rf{rrr} mentioned above  is  discussed in section \ref{nonrenorm}. 

In the second part of the paper we study the finite size $1/\ln S$  corrections
 to the string energy 
by placing the string sigma model on a cylinder. In section \ref{pf1loop}, after replacing
 the integral over continuous spatial momentum  with a discrete summation, we extract the 
 finite size correction to the 1-loop energy for both the folded string  with $J=0$
 (matching the result of \cite{zama,bed})
   and for $J \not=0$. 
     In section \ref{pf2loop}
    we apply the same strategy to the 2-loop computation of the folded string energy. We
     observe that the only relevant finite size contributions come from terms containing
      at least a massless propagator, the purely massive ones producing 
       exponentially suppressed  contributions.
       The final result is presented in section \ref{FStotal} where an ambiguity in the 
       regularization prescription and its possible resolution  are discussed.  Finally, 
       in Appendix~A  we study the  LL model and show that it captures the leading finite 
       size correction to the 1-loop energy of the long folded $(S,J)$ string reproducing
        the result of section  \ref{pf1loop}. 
 Other appendices 
	contain  technical details on various aspects of our  calculations.


\section{AdS light-cone action and the generalized cusp solution}
\label{LCreview}

We begin by reviewing the superstring action in the AdS 
light-cone gauge \cite{mt2}.  A great advantage of this gauge is its simplicity;
for example the $AdS_5\times S^5$ gauge-fixed action is at most quartic in the fermions.
Expanding the action around the folded spinning string in the scaling limit of infinite spin, one finds that the bosonic propagator is almost diagonal \cite{cusp,cuspJ}. This renders this gauge more efficient than the conformal gauge for higher-loop computations. 

The AdS light-cone gauge is defined  in the Poincar\'e parametrization in $AdS_5\times S^5$  in which the 
 10d metric  may be written as ($m=0,1,2,3; \ M=1,...,6$)
\be \la{mei}
\begin{aligned}
& ~~~~~~~~~~ds^2 = z^{-2} (dx^m dx_m + d z^M d z^M) = z^{-2} (d x^m d x_m +
d z^2) + du^M du^M\ ,\\
& x^m x_m =
x^+ x^- + x^* x  ,\ \ \ x^\pm = x^3 \pm x^0 ,\ \ x,x^* = x^1 \pm  \mathrm{i} x^2 , \ 
~~z^M=z ,\ \ u^M , ~~ u^Mu^M=1\ .
\end{aligned}
\ee 
The parametrization  of $S^5$ we will use is the following\
 ($a=1,2,3,4$): 
\be
u^{a} = \frac{y^{a}}{1+{ { { \frac{1}{4}}}}y^2}\ , \qquad \
u^{5} = \frac{1-{ { { \frac{1}{4}}}}y^2}{1+{ { { \frac{1}{4}}}}y^2} \cos\varphi\ , \qquad\
u^{6} = \frac{1-{ { { \frac{1}{4}}}}y^2}{1+{ { { \frac{1}{4}}}}y^2} \sin\varphi\ . 
\la{nee}
\ee
 The angle $\varphi$
parameterizes a large circle $S^1 \subset  S^5$ at $y^a=0$.

The  AdS light-cone gauge is defined by imposing $ \Gamma^+ \theta^I=0$ 
on the two  Majorana-Weyl  fermions  in the superstring  action 
 as well as 
\begin{equation}
  \label{ga}
  \sqrt{-g} g^{\alpha\beta} ={\rm diag}(-z^2, z^{-2})\ , \qquad \qquad x^+ = p^+ \tau \ .
\end{equation}
Our calculations will be conveniently performed in a Euclidean formulation.  
After  the rotation $\tau \to  -\mathrm{i} \tau, \ p^+ \to  \mathrm{i} p^+$,  and
 after setting $p^+=1$, the $AdS_5\times S^5$    superstring  action  can 
 be written as \cite{mt2} 
\be
\label{s}
I &=&  \frac{1}{2} T \int d \tau \int
 d \sigma \; \mathcal{L}_E\ , \quad  \quad  \quad T =
\frac{R^2}{2 \pi \alpha'} = \frac {\sqrt{\lambda}}{2 \pi} \ , \\
\mathcal{L}_E& =& \dot{x}^* \dot{x} + (\dot z^M  + \mathrm{i}  z^{-2} z_N 
\eta_i {\rho^{MN}}^i{}_j \eta^j)^2  + \mathrm{i}  (\theta^i \dot{\theta}_i +
       \eta^i\dot{\eta}_i - h.c.) -  z^{-2} (\eta^2)^2 \nonumber \\
   &&+  z^{-4} ( x'^*x'  + {z'}^M {z'}^M) + 2 \mathrm{i} \Big[\ 
       z^{-3}\eta^i \rho_{ij}^M z^M (\theta'^j - \mathrm{i}
       z^{-1} \eta^j  x') + h.c.\Big]\;.  \label{euc}
       \ee
This  action has manifest $SO(6)\simeq SU(4)$ symmetry.  The fermions are
complex $\th^i = (\th_i)^\dagger,$ $\eta^i = (\eta_i)^\dagger\ $
$(i=1,2,3,4)$ transforming in the fundamental representation of $SU(4)$. 
 $\rho^{M}_{ij} $ are  off-diagonal blocks of
six-dimensional gamma matrices in chiral representation and 
$(\rho^{MN})_i^{\hphantom{i} j} = (\rho^{[M}
  \rho^{\dagger N]})_i^{\hphantom{i} j}$ and
$(\rho^{MN})^i_{\hphantom{i} j} = ( \rho^{\dagger [M}
  \rho^{N]})^i_{\hphantom{i} j}$ are  the $SO(6)$ generators.

Since the AdS light-cone gauge is adapted to the  Poincar\'e patch, we apply a 
conformal transformation
to the spinning folded string in global $AdS_5\times S^5$ to work with 
coordinates as in \rf{mei}. In the scaling limit
\rf{dee} this conformal transformation gives us the so called generalized 
null cusp background,
 see  \cite{krtt,cuspJ}. This is a  bosonic solution for which  only the radial 
 coordinate $z$ and one isometric angle $\varphi$ of 
$S^5$   are nontrivial ({\it i.e.} $x = x^* = 0$, $y^a = 0$)
\begin{equation}
 \label{SJ}
   z = \sqrt{\frac{\kappa}{\mu}} \sqrt{ \frac{\tau}{\sigma}}\ , \ \ \ \ \ \ \ 
   x^+ = \tau\ , \qquad x^- = -\frac{\kappa }{2 \mu}\ \frac 1
  \sigma \ , 
   \qquad\ \ \ \ 
  \varphi = \frac{\nu_e}{2\kappa} \ln\tau
 \ ,
\end{equation}
where the parameters entering in the solution satisfy the following constraint 
\be
\kappa^2+\nu_e^2=\mu^2\,.
\ee
Since $ x^+ x^-  = - \ha z^2$ this bosonic solution ends on a null cusp at the 
boundary $z=0$ of $AdS_5$.
A generalization of this background which includes a winding parameter 
$w$, $ \varphi \sim w \ln\sigma$, is also possible  \cite{cuspJ}, but for simplicity
it will not be considered here.

It is useful to define the 
 fluctuations around the $z$  solution   with extra rescalings 
\be \label{fluct}
\begin{aligned}
& z=\sqrt{\kahat}\, \sqrt{\frac{\tau}{\sigma}}\ {\tilde z}
~,~~~z^M=
\sqrt{\kahat}\,\sqrt{\frac{\tau}{\sigma}}\ {\tilde z}^M~
\ , \ \ \ \ 
x = \sqrt{\kahat}\,\sqrt{\frac{\tau}{\sigma}} {\tilde x}
~,~~~~
\theta=\frac{1}{\sqrt{\sigma}}{\tilde\theta}
~,~~~~
\eta=\frac{1}{\sqrt{\sigma}}{\tilde\eta}\ , \\
& 
{\tilde x}={\tilde x}_1+{\rm i} {\tilde x}_2
\ , \ \ \ \ ~~~~
{\tilde z}\equiv e^{\tilde\phi}=1 +\tilde\phi+ ...
\ , \ \ \ \  ~~~~
{\tilde z}^M={\tilde z} u^M 
\ , \ \ \ \  ~~~~
u^M u^M=1~~,
\end{aligned}
\ee
where we introduced the rescaled parameters
\begin{equation}
\hn_e \equiv  \frac {{\nu_e}}{ \mu }\  , \ \ \ \ \ \ \qquad
  \hat{\kappa} \equiv  \frac \kappa \mu = \sqrt{ 1 - \hn_e^2  } \ 
  . \la{resa} 
\end{equation} 
Defining the fields as in (\ref{fluct}) renders constant 
the coefficients in the Lagrangian for the 
bosonic fluctuations, {\it i.e.} independent of the worldsheet variables. To obtain 
the constant coefficients also in front of the fermionic fluctuations one has  to additionally shift the $S^5$ angle $\varphi$ by a quantity ${\check\varphi}$
\begin{equation}
\check\varphi=\frac{\hat\nu_e}{2\hat\kappa} \,\ln\tau  + \, \tilde\varphi\ ,
\label{rot-par}
\end{equation}
where  $\tilde\varphi$ represents the fluctuation around the classical background.

It is also useful to replace the worldsheet coordinates $(\sigma,\tau)$ with
\be
t=\frac{\ln \tau}{\hat\kappa}\,,~~~~~~~~~~~~s=\ln \sigma\,,
\ee
which puts the induced worldsheet metric in the conformal gauge
\be
ds^2=\frac{1}{4}(dt^2+ds^2)\,.
\ee 
The fluctuation spectrum can be derived by expanding the Lagrangian 
to quadratic order in the fluctuations.   It consists of eight bosonic 
 and eight fermionic massive fields. The precise form of the spectrum can be
  read off the pole structure of the bosonic and fermionic propagators 
  given in Appendix~\ref{propagators}.

\section{Generalized scaling function from the partition function}
\label{gensc}

Before moving to the discussion of the 2-loop calculation we briefly review how the knowledge of the string partition function can be used to compute the generalized scaling function. Note that to compare to the gauge theory predictions we need to compute the partition function for a worldsheet with Lorentzian signature. We will therefore present all results  as function of the Lorentzian parameter $\hn=-{\rm i}\hn_e $. 

The effective action
$W$ has the following expansion
\be
W=-\ln Z_{\rm string}=\frac{\sqrt{\lambda}}{2\pi}V  {\cal F}(\hn)\ , 
~~~~~~~   \ \ \ \ \  \   {\cal F}= 1 + { 1 \over \sqrt{\lambda} }  {\cal F}_1
 +  { 1 \over (\sqrt{\lambda})^2 }  {\cal F}_2+ ...\,,  
\label{defW}
\ee
where $Z_{\rm string}$ is the string partition function.
For a generic 2d sigma model the expectation values of 2d conserved quantities in semiclassical approximation can be found using a thermodynamical approach,  for more details see \cite{rt2,cuspJ}. In the present case the relevant conserved quantities are $E-S$ and $J$ for which it is possible to show that the following relations hold
\be
  \label{pqr}  
E-S&=& \  {\cal M} \sqrt{1+{\hat\nu}{}^2 }\ 
\big[ {\cal F}({\hat\nu})-{\hat\nu}\frac{d{\cal F}({\hat\nu})}{d{\hat\nu}}\big] \ , 
\label{q} \\
J 
&=&  \  \M \  \big[ {\hat\nu}{\cal F}({\hat\nu})-
(1+{\hat\nu}{}^2) \frac{d{\cal F}({\hat\nu})}{d{\hat\nu}}\big] \ , \la{o}
\ee
where ${\cal M} $ is the ``string mass'' (tension 
$\times$  length):
\be
{\cal M}  = {\sql \ov 2 \pi} L= {\sql \ov  \pi}\ln S= \sql  \mu \,.
\ee

Defining 
\be 
\la{defrmf}
&&{\rm f} (\ell) 
\equiv  { E-S \ov {\cal M}}  \ , \  \ \ \ \ \ \ \ \ \ 
\ell\equiv \frac{ J}{ {\cal M}  } = \hn + { 1 \over \sql } \ell_1 (\hn) 
 +  { 1 \over (\sql)^2 } \ell_2(\hn) + ...    \ , \la{ln} 
\ee
we find from \rf{q} and \rf{o}
\be
 {\rm f} (\ell) &=& \sqrt{1+{\hat\nu}{}^2 }\ 
\big[ {\cal F}({\hat\nu})-{\hat\nu}\frac{d{\cal F}({\hat\nu})}{d{\hat\nu}}\big] \ , 
\la{quu} \\
\ell 
&=&  \  {\hat\nu}{\cal F}({\hat\nu})-
(1+{\hat\nu}{}^2) \frac{d{\cal F}({\hat\nu})}{d{\hat\nu}} \ , \la{ouuu}
\ee
allowing one to  compute  $ {\rm f} (\ell)$,  given 
${\cal F}({\hat\nu})$,  by  solving for $\hn$. 

Expanding  \rf{quu}, \rf{ouuu} perturbatively in $1/\sql$ leads to  
the following expressions for the  
quantum   corrections to  the  generalized scaling function $ {\rm f}(\ell)$ 
in terms of ${\cal F}$ 
\be
{\rm f}(\ell)&=&{\rm f} _0(\ell)+{ 1 \over \sql } {\rm f} _1(\ell)
 +  { 1 \over (\sql)^2 } {\rm f} _2(\ell) + ... \\
{\rm f}_0 &=& \sqrt{1+\ell^2}\vphantom{\Big|} \ ,  \ \ \ \ \ \ \ \ \ 
{\rm f}_1 = \frac{{\cal F}_1(\ell)}{\sqrt{1+\ell^2}} \ ,  \la{gg} \\ 
{\rm f}_2&=&\frac{{\cal F}_2(\ell)}{\sqrt{1+\ell^2}}  +\frac{1}{2}(1+\ell^2)^{3/2}
\Big(\frac{d{\rm f}_1}{d\ell}\Big)^2 \ . \la{pp}
\ee
Higher-loop corrections can be obtained analogously.

\section{Two-loop partition function}
\label{2Loop}

\begin{figure}
\begin{center}
\includegraphics[width=90mm]{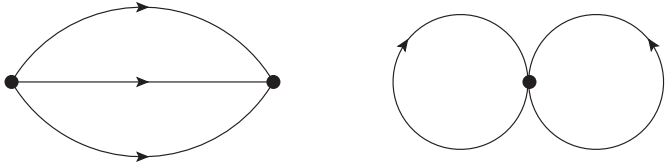}
\parbox{13cm}{\caption{The 2-loop 1PI topologies: ``sunset'' and ``double-bubble''. The propagators 
here are  either bosonic or fermionic.}
\label{1PI}}
\end{center}
\end{figure}

\begin{figure}
\begin{center}
\includegraphics[width=55mm]{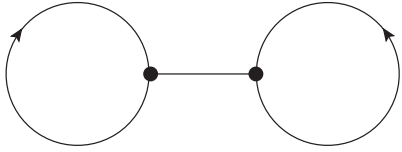}
\parbox{13cm}{\caption{The 2-loop tadpole topology. 
The non-vanishing graphs   have the internal line corresponding  to a $\tilde\phi$-propagator
 while the propagators in the loops   can be either bosonic or fermionic.}
\label{non-1PI}}
\end{center}
\end{figure}

In this section we evaluate the diagrams which contribute 
 to the effective action $W =-\ln Z_{\rm string}$ at 2-loop order 
\begin{equation}
\begin{aligned}
W_2 &= W_{2{\rm B\ sunset}}+W_{2{\rm B\ double-bubble}}+W_{2{\rm F\  sunset
}}+W_{2{\rm F\ double-bubble}}+W_{2{\rm\ tadpoles}}\\  
        &\equiv \frac{V }{2\pi\sqrt{\lambda}} {\cal F}_2(\hat\nu)\,.
\end{aligned}
\end{equation}
This will be later used to extract the 2-loop term ${\rm f}_2(\ell)$ in the 
generalised scaling function.  The relevant connected vacuum diagrams are
 shown in figs.\ref{1PI} and \ref{non-1PI}.   Note that both one-particle 
 irreducible (1PI) topologies and the non-1PI ÒtadpoleÓ topology contribute. 
 As already discussed in \cite{cusp,cuspJ}, the presence of non-1PI tadpole graphs 
 is crucial in this gauge for the cancellation of UV divergencies and to  extract
  the correct expression of ${\rm f}_2(\ell)$.


For the computation of the various topologies we need to expand the action up to fourth order in the fluctuations. Details on this expansion can be found in \cite{cuspJ}.
All manipulations of tensor structures appearing in the evaluation of the Feynman diagrams are performed in $d=2$, and the
 resulting scalar integrals are computed using an analytic regularization scheme in which power divergent
contribution are set to zero, 
\be
\int \frac{d^2p}{(2\pi)^2} (p^2)^n=0\,,\  \qquad n \geq 0\,.
\ee
We also use the following notation
\be
{\rm I}[m^2]&=& \int \frac{d^2p}{(2\pi)^2}\frac{1}{p^2+m^2}\,.
\label{int-basis}
\ee
Note that the integral
 ${\rm I}[m^2]$  is UV divergent when $m^2>0$ and both UV and IR divergent when the mass vanishes. 
  
  We now begin with the analysis of the 1PI  diagrams and end this section with the calculation of the tadpole.
   
\subsection{Bosonic and fermionic double-bubble}
The double-bubble diagrams arise from the quartic terms in the Lagrangian. If we introduce the following one-loop integrals
\be
\begin{aligned}
  J_B(i,j) &= \int \frac{d^2 p}{(2 \pi)^2} \frac {p_0^i p_1^j}{p^4 +
    p^2 + \hat{\nu}^2  p_0^2 },\\
  J_F(i,j) &= \int \frac{d^2 p}{(2 \pi)^2} \frac {p_0^i p_1^j}{p^2 + \frac {4  +3 
  \hat{\nu}^2 }{16} - \frac {\mathrm{i}} 2 \hat{\nu} p_0}
\end{aligned}
\label{JBJF}
\ee
we can express the bosonic and fermionic contributions as follows
\be
\begin{aligned}
W_{2{\rm B~double-bubble}}=&-2\big(J_B(0,2)+J_B(2,0)\big)
\left(J_B(0,2)+(1+\hat\nu^2)J_B(2,0)\right)\\
&+\frac{1}{2}\Big(3 J_B(0,2)^2+2(3+2\hat\nu^2)J_B(0,2)J_B(2,0)+3J_B(0,2)^2\Big)\\
&-\hat\nu^2\big(J_B(0,2)+J_B(2,0)\big) \,{\mbox{I}}[m_{\tilde x}^2]-2\hat\nu^2 J_B(2,0) 
 {\rm I}[m_y^2]-\frac{1}{2}\hat\nu^2 {\rm I}[m_y^2]^2\,,\\
W_{2{\rm F~double-bubble}}=&-\hat\nu^2 J_F(0, 0)^2 + 16 J_F(1, 0)^2 
+ 8\, i\hat\nu  J_F(0, 0) J_F(1, 0) \\
&- 4 \big(J_B(0, 2) + J_B(2, 0)\big)\big [J_F(0, 0) + 
    2(J_F(0, 2) + J_F^*(0, 2))\big]\\ 
 &   + \Big{[}2 (2 + \hat\nu^2) J_F(0, 0)  
 + 8 \big(J_F(0, 2) - i\hat\nu J_F(1, 0) + J_F^*(0, 2)\big)\Big] {\rm I}[m_y^2]\,.
\end{aligned}
\ee
Here we have introduced the notation 
\be m^2_{\tilde x}\equiv\frac{1}{4}(2+\hat\nu^2)\ , \ \ \ \ \ \  \ \ \
m_y^2\equiv\frac{1}{4}\hat\nu^2  
\ee 
for the masses of the 
$\tilde x=x_1+\mathrm{i} x_2$ AdS fluctuation and of the four $y^a$ 
fluctuations on $S^5$, respectively.
%
 
Using the expressions for the one-loop integrals (\ref{JBJF}) given in 
Appendix~\ref{sec:1loop_int}  we can find exact expressions for the $\hat\nu$
dependence of all double-bubble diagrams.

\subsection{Bosonic Sunset}
The contributions to the sunset topology arise from the cubic interactions in 
the Lagrangian. We begin with the bosonic case. We can arrange the various terms
 depending on their denominator structure. From the form of the interactions
we have two possibilities 
\be
&&\int \frac{d^2p\, d^2q\, d^2r}{(2\pi)^4} \delta^{(2)}(p+q+r) \frac{{\cal 
N}(p,q,r)}{{\cal D}_B(p)(q^2+m^2)(r^2+m^2)} \,,~~~~m^2=m_{\tilde x}^2 ~{\rm or}~ m_y^2\nonumber\\
 &&\int \frac{d^2p\, d^2q\, d^2r}{(2\pi)^4}  
  \delta^{(2)}(p+q+r) \frac{{\cal N}(p,q,r)}{{\cal D}_B(p){\cal D}_B(q){\cal D}_B(r)} \,,
\label{sunset-structures}
\ee
where 
\be
{\cal D}_B(p)\doteq p^2(p^2+1)+\hat\nu^2 p_0^2
\ee
is the denominator appearing in the propagator of the mixed fields
$\tilde\phi$ and $\tilde\varphi$.

There are two contributions with a single ${\cal D}_B(p)$ in the
 denominator, with $m^2=m_{\tilde x}^2\,, m_y^2$. These may be evaluated 
 exactly. A convenient strategy is, for example, to solve the momentum 
 conservation constraint by expressing the momenta as $p=P,\ q=Q,\ r=-P-Q$, 
 and  to perform the integration over $Q$ first. The remaining expression, which 
 contains the non-Lorentz invariant factor  ${\cal D}_B(P)$, can be evaluated at 
 the end. The final results for this structure can be written as follows
\be
\begin{aligned}
&W_{2{\rm B~sunset}; m_x}=
-\frac{1}{8\pi}{\rm I}\big[\frac{2+\hat\nu^2}{4}\big]\left(2+\hat\nu^2-2\sqrt{1+\hat\nu^2}
+8\pi\hat\nu^2\,{\rm I}\big[\frac{1}{4}(1+\sqrt{1+\hat\nu^2})^2\big]\right)\\
&~~~~~~~~~~~~~~~~  +\int_0^1 du \,\frac{(1+\hat\nu^2)\,{\rm arctanh}\, 
u}{2\pi^2\big[\sqrt{1+\hat\nu^2+u^2}+\sqrt{1+(1+\hat\nu^2)u^2}\big]^2}\ ,
\end{aligned}
\label{sunsetmx}
\ee
\be
W_{2{\rm B~sunset}; m_y}=\frac{\hat\nu^2}{2}{\rm I}\big[\frac{\hat\nu^2}{4}\big]\left({\rm I}\big[
\frac{\hat\nu^2}{4}\big]+2~{\rm I}\big[\frac{1}{4}(1+\sqrt{1+\hat\nu^2})^2\big]\right)
\ee
While the integral in the second line of (\ref{sunsetmx}) cannot be evaluated 
in closed form in terms of elementary functions, it can be computed to any 
desired order in the small or large $\hat\nu$ expansion. In particular, this
 integral produces Catalan's constant term  at small $\hat\nu$ and $\pi^2$ term 
 at large 
 $\hat\nu$, see Section 5 below.

The remaining bosonic sunset corresponds to the more complicated 
structure with three ${\cal D}_B$ denominator factors, as in the
 second line of (\ref{sunset-structures}), which arises from the three-vertices
  involving only $\tilde\phi$ and $\tilde\varphi$ fluctuations. The presence
   of three Lorentz non-invariant denominators makes an exact evaluation of 
   this contribution more involved. We have computed it in the small $\hat\nu$ 
   expansion up to sixth order:
\be\label{three-DB-nu6}
\begin{aligned}
W_{2{\rm B~sunset};({\cal D}_B)^3}&=\frac{1}{2}{\rm I}[1]^2
+\left(\frac{1}{2}{\rm I}[1]^2-\frac{1}{8\pi}{\rm I}[1]\right) \hat\nu^2+\left(-\frac{7}{64\pi} {\rm I}[1]
+\frac{5}{512\pi^2}\right) \hat\nu^4\\
&+\left(\frac{7}{192\pi} {\rm I}[1]
+\frac{3}{512\pi^2}\right) \hat\nu^6+{\cal O}(\hat\nu^8)\,.
\end{aligned}
\ee
Notice that the result only contains the UV divergent integral ${\rm I}[1]$ and a 
rational finite part. We expect this to remain true to all orders in the small 
$\hat\nu$ expansion. In particular, we do not expect this term to contain 
irreducible 2-loop sunset-type integrals such as ${\rm I}[1,1,1]$
and its variations with higher powers of the 
denominators\footnote{Here we use the notation
\begin{equation}
{\rm I}[m_1^2,m_2^2,m_3^2]=\int \frac{d^2pd^2q}{(2\pi)^4}\frac{1}{(p^2+m_1^2)
(q^2+m_2^2)((p+q)^2+m_3^2)}.
\nonumber
\end{equation}
In particular one finds that ${\rm I}[1,1,1]=\frac{1}{288\pi^2}
\left(\psi^{'}(1/3)-\psi^{'}(2/3)+\psi^{'}(1/6)-\psi^{'}(5/6)\right)$,
 where $\psi(z)=\frac{\Gamma^{'}(z)}{\Gamma(z)}$ is the digamma function.}: 
  these would  introduce new transcendental numbers
in the small $\hat\nu$
 expansion besides $\ln 2$ and Catalan's constant
  (which do not appear for $\nu=0$ and   
    in the Bethe ansatz solution of \cite{gromov}). Likewise, 
    at large $\ell$ we do not expect this term to contribute 
    to the highest transcendentality part
     of the free energy, on which we wish to concentrate in this paper. Therefore,
      we will not attempt here an exact evaluation of this specific term.
      Its contribution is,    of course, 
       crucial for complete cancellation of UV divergences. 
       We will assume that this cancellation occurs, as it was 
       rigorously checked up to order $\hat\nu^4$ in \cite{cuspJ}.

\subsection{Fermionic Sunset}

The possible structures of the two-loops integrals in this case are
\begin{equation}
\begin{aligned}
&\int \frac{d^2p\, d^2q\, d^2r}{(2\pi)^4} \delta^{(2)}(p+q+r) \frac{{\cal N}(p,q,r)}{{\cal D}_F(p)
{\cal D}_F(q)(r^2+m^2)}+ {\rm c.c}\,,~~~~m^2=m_{\tilde x}^2 ~{\rm or}~  m_y^2 \\
&\int \frac{d^2p\, d^2q\, d^2r}{(2\pi)^4} \delta^{(2)}(p+q+r) \frac{{\cal
 N}(p,q,r)}{{\cal D}_F(p){\cal D}^*_F(q){\cal D}_B(r)}+ {\rm c.c}\,,
\label{fsun-struc}
\end{aligned}
\end{equation}
where ${\cal N}(p,q,r)$ is a sum of tensors of rank up to four, and
 ${\cal D}_F(p)$ is the characteristic denominator which appears 
 (together with its complex conjugate) in the fermion propagator
\begin{equation}\label{def}
{\cal D}_F(p) = (p_0- \frac{i}{4}\hat\nu)^2+p_1^2+\frac{1}{4}(1+\hat\nu^2)\,.
\end{equation} 

The presence of more than one non-Lorentz invariant term in the 
denominators makes the exact computation of the fermionic sunset difficult. A possible 
approach is to shift the momenta $p$ and $q$ associated to the non-Lorentz 
invariant factor  ${\cal D}_F$. Indeed,
 performing the shift $p_0\rightarrow p_0+\frac{i}{4}\,\hat\nu$, or its 
 conjugate version if we have ${\cal D}^*_F(p)$, reduces ${\cal D}_F(p)$ to a
  Lorentz invariant expression:
\be
p_0\rightarrow p_0+\frac{i}{4}\,\hat\nu\,\,, \qquad\qquad {\cal D}_F\rightarrow p^2+\frac{1+\nu^2}{4}\,.
\ee
Of course, the remaining momentum $r$ in \rf{fsun-struc} should be accordingly shifted
 to preserve overall momentum conservation. After such shift, one is left with two out
  of three Lorentz invariant propagators, and one can proceed with an exact evaluation 
  as described in the previous subsection for the bosonic terms. We should point out,
   however, that such a shift of momenta is potentially dangerous due to the divergent 
   nature of the integrals involved. It effectively amounts to a ``change of scheme" 
   in regularizing the integrals, which is not necessarily compatible with the way we
    have computed all other 2-loop contributions. Nonetheless, we have observed that, in practice, 
     such issue 
      only affects the ``rational part" of the free energy, both at large and small 
      $\hat\nu$ (see eq.~(\ref{F2-small}) and (\ref{F2-large}) below). Therefore,
       for the purpose of this paper, we will compute these fermionic 
       integrals by shifting momenta, keeping in mind that in the following
        results only terms containing {\it transcendental} numbers should be 
	trusted when summing up all diagrams.

As in the bosonic case, the result can be divided into three contributions, the first 
two corresponding to the first structure in (\ref{fsun-struc}) with 
 $m^2=m_{\tilde x}^2$ or $m^2=m_y^2$ and the third to the second structure in (\ref{fsun-struc}). 
 After performing the momentum shift and evaluating the integrals, we obtain
\be
W_{2{\rm F~sunset}; m_x}=-\frac{\hat\nu^2}{8\pi}\, {\mbox{I}}[\frac{1}{4}(1+\hat\nu^2)]+\hat\nu^2\, {\mbox{I}}[\frac{1}{4}(1+\hat\nu^2)]\,{\mbox{I}}[\frac{1}{4}(2+\hat\nu^2)]
-(\hat\nu^2-\frac{1}{2})\,{\mbox{I}}[\frac{1}{4}(1+\hat\nu^2)]^2+{\cal W}_{1}
\label{Fsunsetmx}
\ee
\be
\begin{aligned}
W_{2{\rm F~sunset}; m_y}=-\frac{5\hat\nu^2}{4\pi}\, {\mbox{I}}[\frac{1}{4}(1+\hat\nu^2)]-2\,{\mbox{I}}[\frac{\hat\nu^2}{4}]\,{\mbox{I}}[\frac{1}{4}(1+\hat\nu^2)]
 +(2\hat\nu^2+1){\mbox{I}}[\frac{1}{4}(1+\hat\nu^2)]^2+{\cal W}_{2}
\end{aligned}
\label{Fsunsetmy}
\ee
\be
&&W_{2{\rm F~sunset};{\cal D}_B}=-\frac{1}{8\pi}
{\mbox{I}}[\frac{1}{4}(1+\hat\nu^2)]\left(12+5\hat\nu^2
-12\sqrt{1+\hat\nu^2}+8\pi(\hat\nu^2-2){\mbox{I}}[\frac{1}{4}
\left(1+\sqrt{1+\hat\nu^2}\right)^2]  \right)\no  \\
&& ~~~~~~~~~~~~~ ~~~~~~~~~~-\big(\hat\nu^2-\frac{1}{2}\big)\ {\mbox I}
[\frac{1}{4}(1+\hat\nu^2)]^2+{\cal W}_3\,,
\label{FsunsetDB}
\ee
where  ${\cal W}_1$, ${\cal W}_2$ and ${\cal W}_3$ are finite and have  integral 
representations which we include in  Appendix~D.

\subsection{Tadpole contribution}

  Let us finally take into account the contribution of the non-1PI diagrams. 
The only fluctuation  that can acquire a non-trivial expectation value  is $\tilde\phi$. 
  Therefore,  the relevant non-1PI 2-loop diagrams are obtained by sewing 
  together two 1-loop tadpoles with a $\tilde\phi$ propagator at zero momentum. 
Exact expressions for these tadpoles were already found in \cite{cuspJ}. 
Here we summarise the final result:
\be
&&A_{\mbox{\tiny tadpole}} = -\frac{1 }{4\pi}\Big{[}\left(1-\sqrt{1 + \hat\nu^2} \right)
+\frac{ \hat\nu^2}{2} \left( 
\ln(2+ \hat\nu^2)-4 \ln (1+\hat\nu^2)+2\ln  \hat\nu^2 
+2 \ln \left(\sqrt{1 +  \hat\nu^2}+1 \right)
\right)\Big{]}\nonumber\\
&&~~~~~~~~~~~~~+2 {\mbox{I}}[   \frac{1}{4}(1+\hat\nu^2)]\,.
\ee
Then  the total contribution of the non-1PI graphs
 is 
\begin{equation}
W_{2~{\rm tadpoles}}= -\frac{1}{2} A_{\mbox{\tiny tadpole}}^2\,. \label{W2tadpoles}
\end{equation} 

\section{Generalized scaling function}
\label{Gensection}

Collecting together all the partial results for $W_2$ we can extract
 the generalized scaling function as outlined in  sec.\ref{gensc}.
In particular,  we will extract the piece of maximal 
transcendentality, which is proportional to ${\rm K}$  in 
the small $\hat\nu$ expansion,  and  proportional to  $\pi^2$ 
in the large $\hat\nu$ expansion.  Furthermore,
 we will also be able to verify that the coefficient of the $1\ov \ell^3$ term \rf{hih}
  in ${\rm f}_2$ is not renormalized, {\it i.e.} is the same as at weak coupling.

At small $\hat\nu$, one may 
represent  the structure of the 2-loop free energy (and analogously the structure 
of the generalized scaling function) in the form
\begin{equation}
\begin{aligned}
{\cal F}_2 (\hat\nu) = \alpha_2(\hat\nu)\ln^2\hat\nu
+\alpha_1(\hat\nu)\ln \hat\nu
+\beta_2(\hat\nu)\ln^2 2+\beta_1(\hat\nu)\ln 2
+\gamma(\hat\nu){\rm K} +\delta(\hat\nu)\,,
\label{F2-small}
\end{aligned}
\end{equation}
where $\alpha_2(\hat\nu),\ldots,\delta(\hat\nu)$ can
 be expressed as analytic Taylor series around $\hat\nu=0$.
  We recall that the presence of $\ln^2\hat\nu$ and $\ln\hat\nu$
   is due to the massive $S^5$ fluctuations which become very 
   light as $\hat\nu\rightarrow 0$. The  functions 
   $\beta_i(\hat\nu),\gamma(\hat\nu) ,\delta(\hat\nu)$ were
    computed to order $\hat\nu^4$ in \cite{cuspJ}, while the 
    functions $\alpha_2(\hat\nu), \alpha_1(\hat\nu)$ multiplying the logarithmic
     terms could actually be extracted in closed form. The results of 
     the present paper allow us to also obtain $\beta_i(\hat\nu)$ and $
     \gamma(\hat\nu)$ to arbitrary order, as discussed below.
      However, we will not be able to fix 
      the purely rational term $\delta(\hat\nu)$ due to the fact 
      that we do not exactly compute the complicated bosonic contribution,
       second line of (\ref{sunset-structures}), and also due to limitations of our approach
        to computation of the  fermionic sunset integrals, see previous subsection.  

At large $\hat\nu$, on the other hand, the results of this paper (which match  the Bethe
 ansatz calculations of \ci{gromov,volin08}) 
 indicate that the 2-loop free energy
  may be written in the form 
\begin{equation}
\begin{aligned}
{\cal F}_2 (\hat\nu) = p(\hat\nu)~ \pi^2 +r(\hat\nu)\ , 
\label{F2-large}
\end{aligned}
\end{equation}
where $p(\hat\nu)$ and $r(\hat\nu)$ are analytic 
around $\hat\nu=\infty$ and contain only rational numbers as 
coefficients in their expansion 
at large $\hat\nu$. 
As in the small $\hat\nu$ expansion, our
  analysis allows us to extract 
  exactly only the coefficient $p(\hat\nu)$ of the highest transcendentality
   constant $\pi^2 = 6\zeta(2)$, but not of the 
    purely rational part $r(\hat\nu)$.

\subsection{ $\ln^2 2$ and $\ln 2$ terms at small $\hat\nu$}
A curious ``experimental" observation, which we have explicitly verified up
 to order $\hat\nu^{10}$, is that after summing up all 2-loop diagrams the $\ln 2$ terms 
  appearing in the small $\hat\nu$ expansion can
   be eliminated from the free energy by a simple rescaling 
\be 
\hat\nu \rightarrow 2^{-3/4} ~\hat\nu\,.
\ee
In other words, the pattern in 
which $\ln^2 2$ and $\ln 2$ can appear
 is completely determined by the coefficients of the 
  $\ln^2\hat\nu$ and $\ln \hat\nu$ terms.  
This observation applies only to the complete free energy and does not hold 
integral by integral or even diagram by diagram. The small $\hat\nu$ 
form (\ref{F2-small}) of the 2-loop free energy can then be written more compactly 
as \footnote{Note that the function $\tilde\alpha_1(\hat\nu)$ differs from 
the function $\alpha_1(\hat\nu) $ in (\ref{F2-small}); they are related as 
$\alpha_1(\hat\nu)={\tilde\alpha}_1(\hat\nu)+\frac{3}{2}\alpha_2({\hat\nu})\ln 2$.
}
%
\be
{\cal F}_2(\hat\nu) = \alpha_2(\hat\nu)\left(\ln\hat\nu+\frac{3}{4}\ln 2\right)^2+
\tilde\alpha_1(\hat\nu)\left(\ln \hat\nu+\frac{3}{4}\ln 2\right)+\gamma(\hat\nu)\KK+\delta(\hat\nu)\,.
\label{F2-small-new}
\ee   
Explicitly, the coefficients of the logarithmic terms are
\begin{equation}
\begin{aligned}
&\alpha_2(\hat\nu)=-2\hat\nu^4 \ , \\
&\tilde\alpha_1(\hat\nu)=-\frac{1}{\hat\nu^2}\left(12+14\hat\nu^2+2\hat\nu^4
-2\hat\nu^2(2+\hat\nu^2)\sqrt{1+\hat\nu^2}\right)+\frac{4}{\hat\nu^4}\left(3+4
 \hat\nu^2+\hat\nu^8\right)\ln(1+\hat\nu^2)\\
&~~~~~~~~~~-2\hat\nu^4 \ln\Big[\frac{1}{2}\sqrt{1+\frac{\hat\nu^2}{2}}
(1+\sqrt{1+\hat\nu^2})\Big]=-2\hat\nu^2+\frac{17}{6}\hat\nu^4
+\frac{7}{5}\hat\nu^6-\frac{83}{160}\hat\nu^8+\ldots
\label{log-coeff}
\end{aligned}
\end{equation}
Computing the generalized scaling function 
${\rm f}_2(\ell)$ by plugging (\ref{F2-small-new}) with 
the values (\ref{log-coeff}) in eq. (\ref{pp}) we find that 
%
%
the coefficients of $\ln^2 2$ 
and $\ln 2$ indeed match the Bethe ansatz result of \cite{gromov} 
through the order $\ell^6$ explicitly given there (the coefficients of 
$\ln^2\ell$ and $\ln \ell$ were already shown in \cite{cuspJ} to match the results of 
\cite{gromov} to all orders in $\ell$).

\subsection{The Catalan constant term  in the small $\ell$ expansion of $\ff_2$}

Here we present a closed-form 
 expression for the coefficient   of the term  in $\ff_2$ 
proportional
 to the Catalan constant appearing 
 in the small $\ell$ expansion of
  the generalized scaling function $\ff_2(\ell)$
  (we will denote this term  $\ff_{2; {\rm K}}(\ell) $).
  It can be expanded to any order in $\ell$ extending the ${\cal O}(\ell^4)$ result 
  obtained in \cite{cuspJ}. 

From the discussion in section~\ref{2Loop} it follows that 
the only bosonic contribution to the coefficient of the Catalan constant 
${\rm K}$ comes from the integral term in (\ref{sunsetmx}) whose 
small $\hat\nu$ expansion  reads\footnote{Here we restored the 
overall factor $\frac{2\pi}{\sqrt{\lambda}}V_2
= \frac{8\pi}{\sqrt{\lambda}}V$ in $W_2$.}
\be
\begin{aligned}
& \int_0^1 du \,\frac{8(1+\hat\nu^2)\,{\rm arctanh}\, u}{\big[
\sqrt{1+\hat\nu^2+u^2}+\sqrt{1+(1+\hat\nu^2)u^2}\ \big]^2}\\
&=\Big(1+\frac{1}{2} \hat\nu^2-\frac{7}{32}\hat\nu^4+\frac{7}{64}
\hat\nu^6-\frac{61}{1024} \hat\nu^8+...\Big) {\rm K}\\
&~~~~~-\frac{1}{64}\hat\nu^4+\frac{1}{128
}\hat\nu^6-\frac{11}{6144}\hat\nu^8+... \  . 
\end{aligned}
\label{Bcatalan}
\ee
Similarly, the fermions contribute only through
 the integral  ${\cal W}_1$ in (\ref{Fsunsetmx}) and 
 we observe that their net effect is to simply change
  the sign of the coefficient of the bosonic contribution to ${\rm K}$ term in (\ref{Bcatalan}), 
   see the Appendix~\ref{W1W2W3}. 
   Using (\ref{pp}), to obtain $\ff_{2; {\rm K}}$ we need to
    divide (\ref{Bcatalan}) by $\sqrt{1+\ell^2}$, while replacing $\hat\nu \rightarrow \ell$, 
    and change the overall sign to account for the fermion 
    contribution.\footnote{Here we do not 
    need to include the piece proportional to the 1-loop partition 
    function in (\ref{pp}), as it does not contain terms proportional to K.}
        Therefore, we find  the following integral representation 
	which can be expanded to any order in $\ell$
\be
\label{CatalanFunction}
\begin{aligned}
&\ff_{2; {\rm K}}(\ell)=- 
 \int_0^1 du \,\frac{8 \sqrt{1+\ell^2}\,{\rm arctanh}
\, u}{\big[\sqrt{1+\ell^2+u^2}+\sqrt{1+(1+\ell^2)u^2}\ \big]^2} \ 
\Bigg{|}_{_{\rm K}}\\
&~~~~~~~~~~
=\Big(-1+\frac{3}{32}\ell^4-\frac{3}{32}\ell^6+\frac{81}{1024} \ell^8+...\Big){\rm K}\,.
\end{aligned}
\ee
Up to order $\ell^6$ this precisely matches the  ABA result of
\cite{gromov}, while higher order terms were not explicitly given
 there.\foot{Higher order terms can be found from 
 the general expressions given in Appendix C  of \ci{gromov}.} 

The fact that the fermionic contribution   simply changes the sign of the
 bosonic contribution to the coefficient of the Catalan
  constant was first observed for the ordinary cusp
   anomaly ($J=0$) in \cite{rt1}. It is remarkable that the 
   same  applies  to all orders in $\hat \nu$.

\subsection{Leading logarithms at small $\ell$: all-loop resummation}
The $n$-loop term in the strong coupling expansion of the generalized scaling function ${\rm f}(\ell,\sqrt{\lambda})$ at small $\ell$ is expected to contain the leading logarithmic term 
\be 
{\rm f}_n (\ell)=\hat c_n(\ell) \ln^n\ell+\ldots\ ,  \qquad \qquad
\hat c_n(\ell)=\ell^2 c_n+{\cal O}(\ell^4) \ .
\ee
The leading coefficient $c_n=(-1)^n 2^{2n-1}$ is completely captured by the $O(6)$ sigma 
model \cite{am2}, while the exact function $\hat c_n(\ell)$ was derived in \cite{gromov} from 
the asymptotic Bethe ansatz. In this subsection, we show that the AdS light-cone gauge 
action can be used to obtain an all-loop string theory prediction for this coefficient function 
which exactly matches the Bethe ansatz result. 

At one and two loops, the leading logarithmic terms in the string free energy are
\be 
{\cal F}_1 = -2\hat\nu^2 \ln \hat \nu+\ldots\ , 
 \qquad \qquad {\cal F}_2 = -2\hat\nu^4 \ln^2\hat\nu+\ldots\ .
\ee 
As recalled earlier, the string theory calculation makes it clear that the appearance of
logarithms in the small $\hat\nu$ expansion is due to the presence of the light $S^5$ 
fluctuations $y^a$ with mass $m \sim \hat\nu$. Moreover, it was observed in \cite{cuspJ} 
that the leading logarithmic term at 2-loops comes solely from the one-particle reducible 
diagram in Fig.~\ref{non-1PI} with the $S^5$ fields running in the two loops and 
the $\tilde{\phi}$ propagator at zero momentum. 

\begin{figure}
\begin{center}
\includegraphics[width=65mm]{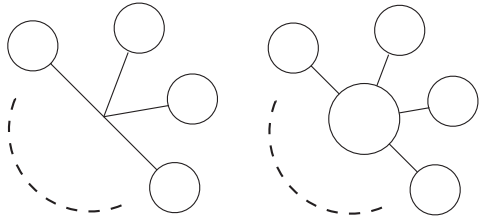}
\parbox{13cm}{\caption{Multi-loop maximally reducible diagrams contributing to the leading 
logarithmic terms in the string free energy. The loops are made of the light $S^5$ 
fluctuations and the propagators correspond to the constant mode of the AdS fluctuation $\tilde\phi$.}
\label{resum}}
\end{center}
\end{figure}

It is natural to expect (and not difficult to show) that the same will hold at all loops, 
namely that the leading logarithms at the $n$-loop order come from the ``maximally
non-1PI" diagrams containing $n$ loops of the $S^5$ fields connected by trees of 
$\tilde\phi$ propagators, as depicted in Fig.~\ref{resum}. Notice that due to 
momentum conservation, the $\tilde\phi$ propagators are always at zero momentum. 

It is,  in fact,  not difficult to exactly resum such diagrams by directly computing the path integral for an appropriate truncation of the AdS light-cone action (\ref{euc}). Since the relevant diagrams only contain 1-loop subdiagrams of the $S^5$ fields, it is sufficient to truncate the action to quadratic order in $y^a$, while keeping the exact dependence on the {\it constant} mode of $\tilde\phi$ (since its propagator should be   at zero momentum). In the following we will denote this constant mode as $\tilde\phi_0$. The truncated action of interest is then
\be 
&&S_{{\rm lead.~log.}} = \frac{\sqrt{\lambda}}{4\pi}\int dtds ~{\cal L}_{{\rm lead.~log.}} \nonumber\\
&&{\cal L}_{{\rm lead.~log.}} = \frac{1}{2}\cosh 2\tilde\phi_0+e^{2\tilde\phi_0}(\partial_t y^a)^2+
e^{-2\tilde\phi_0}(\partial_s y^a)^2+\frac{1}{4}\hat\nu^2 e^{2\tilde\phi_0}y^a y^a\,.
\ee 
We can now integrate out  the $y^a$ fields exactly. The corresponding one-loop determinant is 
\be 
W_{1}^{(y^a)} = 2 V \int \frac{d^2p}{(2\pi)^2}~ 4 \ln\left(e^{2\tilde\phi_0}p_0^2+e^{-2\tilde\phi_0}p_1^2+\frac{1}{4}\hat\nu^2 e^{2\tilde\phi_0}\right)
\ee 
where $V=\frac{1}{4}\int dt ds$. Performing the momentum integral one finds (discarding quadratic divergences) 
\be 
W_{1}^{(y^a)}= 2V\Big( {\rm I}[1]-\frac{1}{4\pi}\ln \big(\frac{\hat\nu^2}{4} e^{2\tilde\phi_0}\big) \Big) \hat\nu^2 e^{2\tilde\phi_0} 
\ee 
Notice that this is of course UV divergent, due to the presence of ${\rm I}[1]$. This 
divergence is supposed to cancel once we include all 
(bosonic and fermionic) modes in the theory. For the present purpose, we can just retain the 
finite piece proportional to $\ln\hat\nu$, so that after integrating out $y^a$ we end up with 
the following effective action for the constant mode $\tilde\phi_0$
\be 
\label{SeffTruncation}
S_{{\rm eff}}(\tilde\phi_0) = V \left(\frac{\sqrt{\lambda}}{2\pi} \cosh 2\tilde\phi_0 
-\frac{\hat\nu^2}{2\pi} e^{2\tilde\phi_0} \ln \hat\nu^2\right)\,.
\ee 
Now the exact path-integral for this reduced model can be obtained by performing the 
integration over the constant mode. In fact, since the relevant diagrams of Fig.~\ref{resum} 
only contain $\tilde\phi_0$ at tree level, all we have to do is solve the classical equation of 
motion for $\tilde\phi_0$: 
\be 
\frac{\delta S_{{\rm eff}}}{\delta \tilde\phi_0} = 0 \quad \rightarrow \quad e^{2\tilde
\phi_0} =\frac{1}{\sqrt{1-2\frac{\hat\nu^2}{\sqrt{\lambda}}\ln \hat\nu^2}}\,.
\ee 
Plugging back into the effective action (\ref{SeffTruncation}), and recalling that in 
our normalizations we define $S_{\rm eff}=V \frac{\sqrt{\lambda}}{2\pi} {\cal F}$, 
we arrive at the following all-loop free energy for leading logarithms
\be 
{\cal F}_{\rm lead.~log.} = \sqrt{1-2\frac{\hat\nu^2}{\sqrt{\lambda}}\ln \hat\nu^2}= \sqrt{1+\frac{2}{\sqrt{\lambda}}{\cal F}_{1~{\rm lead.~log.}}}\ .
\ee 
In the second equality we have stressed that this 
expression can be written entirely in terms of the leading 
logarithmic part of the 1-loop free energy. 

To obtain the coefficient of the leading logarithms in the generalized scaling function, we need just  to plug this result for the free energy into eq.~(\ref{quu})-(\ref{ouuu}). Eliminating $\hat\nu$ in favor of $\ell$ from (\ref{ouuu}) and restricting to leading logarithms yields the following answer
\be
\label{LeadingLog}
{\rm f}(\ell, \sqrt{\lambda})\big{|}_{\rm lead.~log.} = \sqrt{1+\frac{\ell^2}{1+\frac{2}{\sqrt{\lambda}}\ln\ell^2}}\,.
\ee 
This precisely agrees with the Bethe ansatz result of \cite{gromov}. In particular, the first few terms are
\be 
{\rm f}_1= -\frac{2\ell^2}{\sqrt{1+\ell^2}}\ln \ell +\ldots , \ \ \ 
{\rm f}_2 = \frac{8 \ell^2+6 \ell^4}{(1+\ell^2)^{3/2}}\ln^2\ell+\ldots  , \ \ \
{\rm f}_3 =-\frac{32\ell^2+48\ell^4+20\ell^6}{(1+\ell^2)^{5/2}}\ln^3\ell+\ldots  
\ee    

\subsection{The $\pi^2$   term  in the large $\ell$ expansion of $\ff_2$}

Let us now  consider the large $\hat\nu$ expansion and 
again focus on the part of $\ff_2$  of maximal transcendentality which in this case 
turns out to be proportional to $\zeta(2)$ or $\pi^2$
 (we will denote this  term   ${\rm f}_{2; \pi^2}$). 
The $\pi^2$ term in  the partition function turns out to be rather simple
\be
{\cal F}_{2;\,\pi^2}=\pi^2\left(\frac{1}{3\hat\nu^2}+\frac{1}{4\hat\nu^4}\right)
\label{F2-pi}
\ee     
Notice that the expansion stops at next to leading order.
 It is interesting to discuss  various partial
  contributions to this coefficient. Being of maximal
   transcendentality at 2 loops in large $\ell$ expansion, 
   $\pi^2$ 
   can arise only from the sunset diagrams. 
   Specifically, it originates from the integral (\ref{Bcatalan}) in 
   $W_{2{\rm B~sunset}; m_x}$ and from ${\cal W}_1, {\cal W}_2,
    {\cal W}_3$ in the fermionic sunsets, see  Appendix~\ref{W1W2W3}. 
    As an example, let us explain how to obtain the large 
    $\hat\nu$ expansion of the integral (\ref{Bcatalan}) in the bosonic sunset
    (a similar approach also works for the integrals ${\cal W}_1,{\cal W}_2,
     {\cal W}_3$ in the fermionic sunsets).
In contrast with the small $\hat\nu$ case, a direct expansion of the integrand in 
equation (\ref{Bcatalan}) leads to divergent integrals at sufficiently high orders in 
${\hat\nu}^{-1}$.
This signals that the large $\hat\nu$ expansion is non-analytic as a function 
of $\hat\nu^2$. A consistent expansion can be constructed by first using the identity
\be 
\frac{{\rm arctanh} u}{u}=\int_0^1 dy \frac{1}{1-u^2y^2}
\ee
to evaluate in closed form the $u$ integral in equation (\ref{Bcatalan}). The 
integrand of the resulting $y$ integral can be expanded at large $\hat\nu$,
the integral of each term being finite. 
%
%
  The absence of divergences indicates the consistency of this procedure. In 
  this way we obtain
\be
\begin{aligned}
&\int_0^1 du \,\frac{8(1+\hat\nu^2)\,{\rm arctanh}\, u}{(\sqrt{1+\hat\nu^2+u^2}+\sqrt{1+(1+\hat\nu^2)u^2})^2}\\
&=2+(6-\pi^2)\frac{1}{\hat\nu^2}+\frac{16}{3}\frac{1}{\hat\nu^3}+(4-\frac{\pi^2}{2})\frac{1}{\hat\nu^4}
-\frac{104}{45}\frac{1}{\hat\nu^5}+\ldots
\label{PiBose}
\end{aligned}
\ee
The presence of odd powers of $1/\hat\nu$
 exposes the expected non-analyticity in ${\hat\nu}^2$ of the large 
 $\hat\nu$ expansion. Notice
  also that the result contains $\pi^2$ as well as rational numbers;
   as was already mentioned above,
    this 
   is a general feature of the large $\ell$ expansion of the generalized scaling function.

Combining this with the fermionic terms listed in Appendix~\ref{W1W2W3}, 
we then have the following partial contributions to the coefficient of $\pi^2$ 
in ${\cal F}_2$
\be 
\label{details}
\begin{aligned}
&W_{2{\rm B~sunset}; m_x}\rightarrow \pi^2 \Big(-\frac{1}{\hat\nu^2}-\frac{1}
{2\hat\nu^4}-\frac{1}{4\hat\nu^6}+\frac{7}{16\hat\nu^8}-\frac{49}{64\hat\nu^{10}}+\ldots\Big)\\
&W_{2{\rm F~sunset}; m_x}\rightarrow \pi^2 \Big(\frac{4}{3\hat\nu^2}+
\frac{3}{4\hat\nu^4}+\frac{1}{4\hat\nu^6}-\frac{7}{16\hat\nu^8}+\frac{49}{64\hat\nu^{10}}+\ldots\Big)\\
&W_{2{\rm F~sunset}; m_y}\rightarrow \pi^2
\Big(-\frac{4}{3\hat\nu^2}-\frac{1}{\hat\nu^4}\Big)\ , \ \ \ \ \ \ \ \ 
W_{2{\rm F~sunset}; {\cal D}_B}\rightarrow \pi^2 \Big(\frac{4}{3\hat\nu^2}+
\frac{1}{\hat\nu^4}\Big)\,, \\
\end{aligned}
\ee
while $W_{2{\rm B~sunset}; m_y}$ and $W_{2{\rm B~sunset}; ({\cal D}_B)^3}$ do not yield terms proportional to $\pi^2$. Note that the contributions in the last two lines precisely cancel each other, while the contributions of $W_{2{\rm B~sunset}; m_x}$ and $W_{2{\rm F~sunset}; m_x}$ cancel each other beyond order $\frac{1}{\hat\nu^4}$ and leave (\ref{F2-pi}) as a net result. Notice in particular that even the leading term $\frac{\pi^2}{3}\frac{1}{\hat\nu^2}$ receives both bosonic and fermionic contributions.

To extract the $\pi^2$ coefficient in the generalized scaling function ${\rm f}_2(\ell)$, we then simply need to compute
\be
\label{pi2}
{\rm f}_{2;\,\pi^2}(\ell)=\frac{{\cal F}_{2;\,\pi^2}(\ell)}{\sqrt{1+\ell^2}}=\frac{\pi^2(3+4\ell^2)}{12\ell^4\sqrt{1+\ell^2}}=\pi^2\left(\frac{1}{3\ell^3}
+\frac{1}{12\ell^5}-\frac{1}{96\ell^9}+\ldots\right)
\ee
Again,  we did not include the terms induced by the 1-loop partition function as 
they cannot contain contributions proportional to $\pi^2$. 

This answer can be compared directly to the asymptotic Bethe Ansatz 
expression  for ${\rm f}_2$ derived in \cite{gromov} which can be written as
($g= { \sql \ov 4 \pi}$) 
\be\label{gromov}
{\rm f}_2^{\rm{ABA}}=\frac{16\pi^2}{\sqrt{\ell^2+1}}\left( 
\frac{2 g^2\partial_a {\cal {\tilde F}}^2(a_0)}{\sqrt{\ell^2+1}}
-\frac{2g^2 {\cal \tilde F}^2(a_0)}{\ell^2+1}+2 g^2\delta{\cal F}
-\left(\frac{5}{256 \ell^6}+\frac{3}{64 \ell^4}+\frac{1}{32 \ell^2}\right)\right)
\ee
Here $a_0=\sqrt{1+\ell^2}$ and we refer the reader to \cite{gromov} for more 
details on the definition of the functions ${\cal \tilde F}$ and $\delta{\cal F}$.
All the pieces in this formula can be analytically computed at large $\ell$. The 
first terms in this expansion are
\be
{\rm f}_2^{\rm{ABA}}=\frac{\pi^2}{3} \frac{1}{\ell^3}+\left(-\frac{32}{9}
+\frac{\pi^2}{12}\right) \frac{1}{\ell^5}-\frac{232}{45} \frac{1}{\ell^6}
+\frac{16}{5} \frac{1}{\ell^7}+\frac{20416}{1575} \frac{1}{\ell^8}
-\left(\frac{3614}{1575}+\frac{\pi ^2}{96}\right)\frac{1}{\ell^9}+...
\ee
 It turns out that the only relevant contributions to the $\pi^2$ coefficient 
 arise from the last term in parenthesis in (\ref{gromov}), {\it i.e.} 
$
-(\frac{5}{256 \ell^6}+\frac{3}{64 \ell^6}+\frac{1}{32 \ell^2})$
and 
a term  in $\delta{\cal F}$ (see  Appendix  C in  \cite{gromov})
\be
\delta{\cal F}=\dots+
\frac{1}{g^2}\Big(\frac{5}{512 \ell^6}+\frac{1}{32 \ell^4}+\frac{5}{192 \ell^2}\Big) 
+\dots   \ . 
\ee
Plugging these two expressions in eq.~(\ref{gromov}) we
 reproduce  our string theory result (\ref{pi2}).
%

\subsection{Non-renormalization of the leading terms in the large $\ell$ expansion}
\label{nonrenorm}

While the  small $\ell$ expansion  of  the  string theory   result for 
${\rm f}(\ell,\lambda)$ should be compared with results
 of the all-loop Bethe Ansatz expanded at strong coupling, the  large $\ell$
 expansion 
  (or large $J$ ```BMN-type'' expansion) 
makes contact with  perturbative 
 gauge theory results: as discussed  in Introduction, 
 coefficients of  the leading terms in this expansion may 
 be protected, {\it i.e.} the same  at strong and weak coupling.
 

On general grounds, the string energy is expected 
to have the expansion  given in \rf{rrr}  with $j=\frac{\sql }{\pi}\ell$. 
   Rewritten in terms of $\ell$, the generalized scaling function in \rf{rrr} 
   takes the form
\be\label{strExp}
{\rm f}(\ell,\l)= { \pi \ov \sql}  f(\l, \ell)
= \big(\ell+\frac{\pi^2\cc_{10}}{\ell}+\frac{\pi^4 \cc_{20}}{\ell^3}+...\big)
+\frac{1}{\sql}\big(\frac{\pi^3 \cc_{11}}{\ell^2}+\frac{\pi^5 \cc_{21}}{\ell^4}+...\big)+
\frac{1}{\lambda}\big(\frac{\pi^4 \cc_{12}}{\ell^3}+...\big)  
\ee
 The  protected coefficients  
  appear   at one $(\cc_{10}\,, \cc_{11}\,, \cc_{12})$   and two $(\cc_{20}\,, \cc_{21})$ 
   loops in gauge theory, while in string theory they appear at 
   tree level $(\cc_{10}\,,\cc_{20})$, one loop $(\cc_{11}\,, \cc_{21})$
   and two loops ($\cc_{12}$). 

From  tree-level and one-loop string results \ci{ftt} 
we find
\be
\begin{aligned}
& {\rm f}_0=\sqrt{1+\ell^2}=\ell+\frac{1}{2\ell}-\frac{1}{8\ell^3}+...\ \ \rightarrow\ \ \ 
 \cc_{10}=\frac{1}{2\pi^2}~,~~ \ \  \ \ \cc_{20}=-\frac{1}{8\pi^4}\\
& {\rm f}_1= \frac{{\cal
F}_1(\ell)}{\sqrt{1+\ell^2}}=-\frac{4}{3\ell^2}+\frac{4}{5\ell^4}+...\ \ 
 ~\rightarrow \ \ \    \cc_{11}=-\frac{4}{3\pi^3}~,~~\ \  \cc_{21}=\frac{4}{5\pi^5} \ .
\end{aligned}
\ee
On the  gauge-theory side, 
 the coefficients $\cc_{10}$ and $\cc_{11}$  were 
   obtained  from  
    finite size corrections to the one-loop
    $sl(2)$ spin chain in \cite{bgk};  
    the coefficients  $\cc_{20}$ and $\cc_{21}$ were  found from 
     the analysis of the integral equation \ci{frs} for the generalized scaling function 
       in \cite{beccaria}.
 
Our results  allow us to extract the  expression for 
the term  $\cc_{12}\ov \ell^3$ or \rf{hih}, 
 which is the leading two-loop contribution 
 in the string sigma model.
 As it turns out to be proportional to $\pi^2$, its computation is unambiguous 
 (as discussed above,   the shift of momenta performed in the fermionic sunset diagram
 does not affect $\pi^2$ terms).\footnote{To be precise, since we do not 
 have a complete handle on rational terms, in this paper we have 
 not proven that at order $1\ov \ell^3$ there are no rational contributions 
 coming from 2-loop worldsheet diagrams. However, the full agreement with 
 the ABA seen in \cite{cuspJ} at small $\ell$ up to order $\ell^4$ strongly 
 suggests that 
 no such terms should be present.} 
 This coefficient can then be 
  read off equation~$(\ref{pi2})$:
\be
{\rm f}_2=\frac{\pi^2}{3\ell^3}+...\ \ \ \rightarrow\ \ \  \cc_{12}=\frac{1}{3 \pi^2}\,.
\ee
The same result was obtained on the weakly coupled  gauge-theory 
side (as a finite-size  $sl(2)$ spin chain correction) 
 in  \cite{volin08}. This provides  the first
     direct check that the non-renormalization theorem for the leading
      terms in (\ref{strExp}) at two-loop level in string theory.\footnote{It is 
      interesting to compare 
the terms which contribute to the leading coefficients $\cc_{11}$ and 
$\cc_{12}$ in the string expansion (\ref{strExp}).  The only contributions 
to $\cc_{11}$ turn out to be coming from the  AdS fluctuations $\tilde\phi$ 
and $\tilde\varphi$ while as observed before (cf.~discussion after
 eq.~(\ref{details})) this is not the case for $\cc_{12}$. 
 This is not
  necessarily in disagreement with the expectation that an effective  Landau-Lifshitz 
  model 
  based on the AdS fluctuations should capture the leading protected terms in
   the expansion.
    Indeed,  a calculation of $\cc_{12}$  would be a  
   two-loop one  in the ``one-loop'' (in gauge-theory sense) 
    LL model  and thus  would   require
    counterterms which would effectively take into account 
    the contributions of  other fluctuation fields.
    }


\section{String finite size corrections:  computations  on $\mathbb{R} \times S^1$
 worldsheet}

In the  previous sections we discussed properties of the generalized scaling function 
in various limits of its argument $\ell={\pi J\ov \sql\ln S}$. An 
 interesting
 question 
is about  finite size corrections in the case of small $J$  which are 
proportional to $1 \ov \ln S$. 
 As we will see in the following at the  one loop order, such 
finite size corrections
 provide a sharp distinction between the $\ell=0$
and $\ell\ne 0$ cases. 
The limit $\ell\rightarrow 0$ of the latter should involve a 
resummation of infinitely many exponential corrections which may 
yield  polynomial terms  in $1\ov \ln S$. 

As already discussed in the Introduction, a calculation of finite size corrections 
should potentially require use of the exact finite spin solution on a finite size 
worldsheet. 
This is indeed the case for the virtual scaling function $h(\lambda, J)$ 
whose string theory evaluation, while possible on an $\mathbb{R}^{1,1}$ worldsheet,
requires use of the exact folded string solution.
It was  noticed  in \cite{bed}, at 1-loop order and for the leading 
$1\ov \ln S$ correction, that use of the exact finite spin solution is not 
actually necessary 
and the correct result may be obtained by
  considering 
 the folded string solution in its simplified  scaling-limit form 
 and using it  in  the $\mathbb{R} \times S^1$ world sheet  computation.  
We will assume that this short-cut  applies  also 
at  higher-loop  orders  as well. 

Following \cite{cusp,cuspJ}, in the previous sections we used the AdS light-cone gauge and the 
equivalence between a minimal surface describing a null cusp Wilson loop and 
the fast-spinning  folded  string in \adss.
A simple inspection of 
the spectra of fluctuations around the folded spinning string \cite{ftt} and around 
the generalized cusp surface reveals that they are the same only up to a rescaling of 
worldsheet coordinates by a numerical factor. While this  rescaling is not 
relevant on $\mathbb{R}^{1,1}$ worldsheet, it 
should be  accounted for on $\mathbb{R} \times S^1$. 
Since it is the folded spinning string  (dual to 
gauge theory twist  operator)    we are interested in,
 we will normalize
 the calculation to the closed string spectrum, 
even though we will formally use the same fluctuation action as 
in the ``open string'' (cusp Wilson loop)  case. 

In sections 7 and 8 we will 
 evaluate the leading finite size corrections to the energy of the folded string. 
We will comment along the way on the differences with the folded spinning string in the
scaling limit  with 
{\it i.e.} $\ell\ne 0$. 
It was  mentioned in \cite{cuspJ}  that  for finite size systems 
differences may appear between the thermodynamic reasoning that led to 
the expressions (\ref{gg}) and $(\ref{pp})$  for the quantum corrections 
to the target space 
energy and the calculation of the expectation values of the energy and spin 
operators. Below in section 7 we  will  show that 
 no differences appear at the one-loop order.


Let us first  comment on 
 the map between the  open and closed string  normalizations 
  and introduce
the two-dimensional momenta on a cylindric worldsheet with spatial length $L$.
By inspecting the open and closed string worldsheet volumes it is easy to see
that the relation between them is given by ($\beta$ is time interval) 
\be
V\equiv 
\frac{1}{4}\int_{-\beta_\text{open}/2}^{+\beta_\text{open}/2} dt_o \int_0^{L_\text{open}}
ds_o
=
2\beta\ln S
=\int_{-\beta_\text{closed}/2}^{+\beta_\text{closed}/2} d t_c \int_0^{L_\text{closed}}
d s_c
%
~~.
\ee
From here it follows that the relation between coordinates is just
\be
(t,s)_\text{open}=2 (t,s)_\text{closed}~,~~~~~~~~~p_\text{open}=\frac{1}{2}p_\text{closed}~~.
\label{remap}
\ee
In particular, the length of the open string worldsheet is twice that of the closed 
string worldsheet:
\bea
L\equiv L_{\text{closed}}=\frac{1}{2}L_{\text{open}}=2\pi\mu=2\ln S~~.
\label{length_def}
\eea
The transformation (\ref{remap}) simply rescales the open string spectrum 
by a factor 4 which then becomes the spectrum of the fluctuations around the closed 
string background \cite{ft1}. 
For $J=0$ this consists of  one field ($\tilde\phi$) with $m^2=4$, 
two fields ($\tilde x,\tilde{x}^*$) with $m^2=2$, five massless fields ($y^a$) and 
 eight fermionic degrees of freedom  with $m^2=1$.

In the calculation of the leading finite size corrections at one and two loops we will label
momenta as $p, q, r$, subject to momentum conservation $p+q+r=0$. 
On a cylindrical world sheet  the two components 
of momenta  ($p_0$, $p_1$)
%
%
should be treated independently: the first  is continuous while the second 
 is discrete, being labeled by an integer
\be
\label{pdiscrete}
p_1=\frac{2\pi}{L}n\,,~~~~~~~~~~~~n\in \IZ
\ee
with $L=2\ln S$ 
being the length  of the 
worldsheet cylinder.  The two-dimensional loop momentum integration is therefore 
replaced by a one-dimensional integral over $p_0$ and a summation over the 
discrete values of $p_1$:
\be
\label{intvssum}
\int d^2 p\rightarrow \int dp_0 \sum_{p_1}= \frac{2\pi}{L} \int dp_0\sum_{n}~~.
\ee

\section{Leading finite size correction to the folded string energy 
 at one-loop order}

One-loop finite size corrections may be  computed either  in terms of the partition 
function (by directly applying the discussion in section \ref{gensc}) 
or  by evaluating directly the expectation values of the energy and 
spin  operators.  We discuss both approaches in some detail and identify the precise 
origin of  the leading $1\ov \ln S$ terms. The resulting 
 observations will simplify the two-loop 
calculation in the next section by allowing us 
to focus  only on  a small set of terms.

\subsection{Partition function approach \label{pf1loop}}

As discussed in section~\ref{gensc}, the one-loop correction  to the energy of the 
folded string is simply given by 
\be
(E-S)_1=\frac{1}{\beta} W_1~~,~~~~W_1 = \frac{1}{2\pi}V{\cal F}_1=\frac{1}{2\pi}V
\left({\cal F}_1^{L=\infty}+\Delta {\cal F}_1\right)~~,
\ee
where $W_1=-(\ln Z)_1$ is the one-loop effective action, ${\cal F}_1$ is the one-loop 
free energy, $V$ is the worldsheet volume and $\beta$ is the length of the 
 non-compact worldsheet 
direction.
Generalizing the expression in \ci{ft1} in the long string limit 
to $\mathbb{R} \times S^1$
the  one-loop free energy  is given by 
\bea
{\cal F}_1&=&{ 1 \ov 2} \times  \;
\frac{2\pi}{L} \int_{-\infty}^{\infty} \frac{dp_0}{ 2\pi }\sum_{n=-\infty}^\infty\Big[
\ln\Big(p_0^2+\big(\frac{2\pi n}{L}\big)^2+4\Big)
+
2\ln\Big(p_0^2+\big(\frac{2\pi n}{L}\big)^2+2\Big) \cr
&&
~~~~~~~~~~~~~~~~~~~~~
+
5\ln\Big(p_0^2+\big(\frac{2\pi n}{L}\big)^2\Big)
-
8\ln\Big(p_0^2+\big(\frac{2\pi n}{L}\big)^2+1\Big)
\Big]~~.
\eea
Integrating by parts and noticing that the integrand vanishes as $p_0^{-4}$ at large values of $p_0$  leads to
\bea
&&{\cal F}_1=
-\frac{1}{L}  \int_{-\infty}^{\infty} {dp_0}\sum_{n}\Big[
\frac{p_0^2}{p_0^2+\left(\frac{2\pi n}{L}\right)^2+4}
+
2\frac{p_0^2}{p_0^2+\left(\frac{2\pi n}{L}\right)^2+2}\cr
&& \ \ \ \ \ \ \  \  ~~~~~~~~~~~~~~~~ +\  5\frac{p_0^2}{p_0^2+\left(\frac{2\pi n}{L}\right)^2} 
-
8\frac{p_0^2}{p_0^2+\left(\frac{2\pi n}{L}\right)^2+1}
\Big]~~.\label{IBP1}
\eea
%
Sums of this type have been discussed previously in \cite{zama} and are reviewed in 
Appendix~\ref{appInt}. They may be computed exactly; if the denominator of the 
summand is not singular as $n,p_0\rightarrow 0$, then the $L$ dependence is 
exponentially suppressed (cf.~(\ref{1loopsum})). Thus, only the third term in (\ref{IBP1})
can contribute to 
$L^{-1}\propto(\ln S)^{-1}$ dependence. The other 
terms are, of course, crucial to ensure the finiteness of the free energy. Discarding the 
exponential dependence on $L$ we are therefore led to:
\bea
{\cal F}_1&=&-\ha 
 \; \int_{-\infty}^{\infty}{dp_0} \Big[
\frac{p_0^2}{\sqrt{p_0^2+4}}
+
2\frac{p_0^2}{\sqrt{p_0^2+2}}
+
5|p_0| \coth\big(\ha L|p_0|\big)
-
8\frac{p_0^2}{\sqrt{p_0^2+1}}
\Big]~~.
\eea
Isolating and subtracting  the leading large $L$ contribution we are left with
\be
\Delta {\cal F}_1 = \frac{20}{L^2}
\int_0^\infty {dx}\ x\ (1-\coth x) = - \frac{4}{L^2} \;\frac{5\pi^2}{12}
=-\frac{1}{\ln^2 S} \;\frac{5\pi^2}{12}~~,
\label{final_1loop_F}
\ee
where  we used eq.(\ref{length_def}).
Some comments are in order. First, as was  mentioned above, this finite size correction
is completely controlled by the massless worldsheet fields. Restricting to a subset of 
the contributions is bound however to yield a divergent result. 
 The subtraction done in 
(\ref{final_1loop_F}) is effectively the same 
as  the evaluation of $\zeta(-1)=-{1\ov 12}$. One may expose 
the relevant sum in the massless term of eq.~(\ref{IBP1}) by first 
carrying out the $p_0$ integral. 
Indeed,  partial fractioning so  that 
the resulting integral over $p_0$ is convergent, the massless contribution to ${\cal F}_1$
can be written as
\be
({\cal F}_1)_\text{massless} = \frac{5}{L}\sum_{n=-\infty}^{\infty}\int {dp_0}
\frac{\left(\frac{2\pi n}{L}\right)^2}{p_0^2+\left(\frac{2\pi n}{L}\right)^2}
=\frac{20\pi^2}{L^2}\sum_{n=1}^\infty|n|\ \ \mapsto \ \ \frac{4}{L^2} \;{5\pi^2}\zeta(-1)~~.
\ee
The other terms in eq.~(\ref{IBP1}) provide the necessary regularization 
of this sum. In the following we will use this observation to simplify the evaluation of the 
finite size corrections by first evaluating the integral over the continuous parameters
and regularizing the resulting sums  over $n$ 
 using  zeta-function technique.

As a result, we find that the above expression for the 
partition function implies that the leading finite size correction
to the energy of the  long folded string is
\footnote{As mentioned above, we ignore the 
terms that are independent of $\ln S$, whose evaluation requires use of the exact 
folded string solution, valid on a cylindrical worldsheet.}
\be
(E-S)_1=(E-S)_1{}_{_{L\rightarrow\infty}}
+\Delta (E-S)_1=\frac{1}{\pi}\left[-3\ln 2 \;\ln S -\frac{5\pi^2}{12\,\ln S}+
{\cal O}\big({ \ln S \ov S}\big ) \right]~~.
\label{1loop_cor}
\ee

\subsection{Expectation value approach \label{Evev1}}

As a test of the validity of thermodynamic arguments on finite-size worldsheets 
it is instructive to compute  the one-loop finite size correction to the energy of the 
folded string by directly evaluating the expectation value of the energy operator 
\cite{cuspJ}
\be
\label{EmSo}
E-S=\frac{\sqrt{\lambda}}{2\pi}\int_0^L ds\Bigl[1+2{\tilde\phi}
                             +(2{\tilde\phi}^2+|x|^2) \Bigr]~~.
\ee
At  the tree  and one-loop level  the expectation value of $E-S$ is\foot{In what follows 
we will use the notation $(...)_1 = \langle ... \rangle_1$.}
\be
\label{Epieces}
\frac{2}{L} (E-S)_0=\frac{\sqrt{\lambda}}{\pi}\ , 
~~~~~~~~~~~~~~~~~
\frac{2}{L} (E-S)_1 = {\cal E}_1+{\cal E}_2
\ee
where
\be
\label{E1c}
{\cal E}_1&=&2\langle{\tilde\phi}\rangle=-\int 
\frac{d p_0}{(2\pi)^2}\sum_{p_1}
\Bigl[4\frac{p_0^2-p_1^2}{P(p,0)}
-\frac{4+4p_1^2}{P(p,2)}
+2\frac{(p_0^2-p_1^2)(p^2+{2})}{P(p,0)P(p,4)}
+4\frac{4+4p_1^2}{P(p,1)}\Bigr]
\\
\label{E2c}
{\cal E}_2&=&\int\frac{d p_0}{(2\pi)^2}\sum_{p_1}~
\Bigl[\frac{4}{P(p,2)}
          +\frac{4}{P(p,4)}\Bigr]\ , \ \ \ \ \ \ \ \ \ 
	 P(p,m^2)\equiv p_0^2+p_1^2+m^2  \ , 
\ee
where ${\cal E}_1$ and ${\cal E}_2$ are the contribution of the tadpole and the 
quadratic term in eq.~(\ref{EmSo}), respectively.

These sums and integrals are of the same type as those appearing in the evaluation of 
the one-loop partition function (see Appendix~\ref{appInt}). Choosing to first 
carry out the summation over the discrete component of the momentum we find that the
leading $1\ov \ln S$ corrections are given by an expression 
analogous to eq.~(\ref{final_1loop_F}). The complete 
contribution arises from ${\cal E}_1$, 
in particular,  from the first and the 
third terms in eq.~(\ref{E1c}) as these
 are the only ones containing massless propagators. The final result reproduces 
eq.~(\ref{1loop_cor}), 
%
%
confirming the validity of the thermodynamic arguments at the one-loop level.

\subsection{Finite size corrections to the energy of 
 folded spinning string  with $J\not=0$}

The calculation in the two previous subsections may be extended without difficulty to the folded 
spinning string with an angular momentum $J$ on the $S^5$. 
We will again find that thermodynamic arguments still hold on a finite size worldsheet. We will also 
find that the limit $J\rightarrow 0$ is  subtle: 
 if taken in the final answer, it leads 
to a correction different from the one found above.
 We will discuss the origin of this difference.

Let us begin with the partition function approach. As was argued above, 
 only the massless 
modes contribute to the $1\ov \ln S$ terms. Here the light mode
 arises from the mixed $\tilde\phi$ and $\tilde\varphi$ fields. 
The relevant part of the partition function is then
\be
({\cal F}_1)_\text{massless}
= 
\ha 
\int \frac{dp_0}{ 2\pi}\ \frac{2\pi}{L}\sum_n 
\ln \det K_{\tilde\phi\, \tilde\varphi}
= {1 \ov2 L} \int {dp_0}{ } \sum_n\ln \left[p^2(p^2+4)
+4{\hat\nu}^2 p_0^2\right]~~,
\ee
where $p^2=p_0^2+\left(\frac{2\pi n}{L}\right)^2$.
Performing the integration over $p_0$, 
expanding to leading order in $L$ and replacing $\hat\nu\rightarrow\ell$
we obtain for the leading finite size term 
\be
\Delta{\cal F}_{1}= 
\frac{4}{L^2}\; \frac{\pi^2\,\zeta(-1)}{\sqrt{1+\ell^2}}~~.
\ee
From equations (\ref{defrmf}) and (\ref{gg}) it follows then that
\be
\Delta(E-S)_1=\frac{1}{\pi}\;\frac{\Delta{\cal F}_{1}}{\sqrt{1+\ell^2}}\,\ln S
= - \frac{\pi}{12}\, {1 \ov 1+\ell^2}\;\frac{1}{\ln S}~~.
\label{SJpart}
\ee
Note that the limit $\ell\rightarrow 0$ of this expression  is different
from the corresponding term in eq.~(\ref{1loop_cor}). The difference may be traced
 to the fact that, as 
$\ell\rightarrow 0$, four more massless modes emerge. 
They have been included in 
eq.~(\ref{1loop_cor}) but they produce  only exponential  corrections to 
eq.~(\ref{SJpart}) at $\ell\not=0$.
 A resummation of these corrections should yield, in the limit 
$\ell\rightarrow 0$,  the missing $- { 4 \ov 12}$  contributions.

Let us now turn to the calculation of the expectation values of the energy (\ref{EmSo})
and the angular momentum operators. We will first compute the expectation value of $E-S$.
Since $x$ is a massive field, we only need to compute the contributions  proportional  to the 
tadpole $\langle \tilde\phi \rangle$ and to $\langle \tilde\phi^2 \rangle$  which we 
called ${\cal E}_1$ and ${\cal E}_2$, respectively.
As in the computation of the partition function, the only relevant contributions arise from 
the mixed fields $\tilde\phi$ and $\tilde\varphi$. They are:
\be
&&{\cal E}^\text{massless}_1=-{\hat\kappa}\int \frac{dp_0}{(2 \pi)^2}
\sum_{p_1}  \frac{2 (p_0^4-p_1^4)
+4(1+2 {\hat \nu}^2)p_0^2-4p_1^2}{p^2(p^2+4)+4{\hat \nu}^2 p_0^2}
\ , \\
&&{\cal E}^\text{massless}_2= {\hat\kappa}\int \frac{dp_0}{(2 \pi)^2}
\sum_{p_1} 
\frac{ 4p^2}{p^2(p^2+4)+4{\hat\nu}^2 p_0^2}
\ee
Note that the latter contribution was not important in the $J=0$ case but becomes 
relevant here due to the mixing of $\tilde\phi$ and $\tilde\varphi$ which introduces 
a small-momentum singularity whose nature is slightly different from that of a regular
massless field.
Summing ${\cal E}^\text{massless}_1$ and ${\cal E}^\text{massless}_2$ 
and proceeding as above  leads to the following finite size correction
\be
\Delta (E-S)_{1}=\frac{2\pi(1+2{\hat\nu}^2)\zeta(-1)}{(1+{\hat\nu}^2)\;L^2}\,\ln S\,.
\ee
To express $(E-S)_{1}$ in terms of $J$ it is necessary 
to compute the expectation value 
of  the angular momentum  $J$. Focusing again only on the 
terms which are sensitive to $1/L$ 
corrections, we have 
\be
\begin{aligned}
J=\frac{\sqrt{\lambda}}{2\pi}\int ds\, {\cal J}~\ , ~~~~~~~~~~~~~~~~~
{\cal J}={\hat\nu}+2{\hat\nu}\tilde\phi+2{\hat\nu}\tilde\phi^2~~,
\end{aligned}
\ee
leading to 
\be
\ell=\frac{2\pi}{\sqrt{\lambda}}\frac{\langle {\cal J}\rangle}{\ln S}=
{\hat\nu}+\frac{2\pi}{\sqrt{\lambda}} \frac{4 \pi{\hat \nu} \zeta(-1)}
{\sqrt{1+{\hat \nu}^2}}\frac{1}{L^2} \ . 
\ee
This  can be inverted to express $\hat \nu$ as a function of $\ell$
\be\label{ellcorr}
{\hat\nu}\simeq
\ell-\frac{2\pi}{\sqrt{\lambda}} \frac{4 \pi\ell \zeta(-1)}{\sqrt{1+\ell^2}}\frac{1}{L^2}
+{\cal O}(L^{-4})\,.
\ee
Then 
\be
E-S=(E-S)_{0}({\hat\nu}(\ell))
+
(E-S)_{1}({\hat\nu}(\ell))+...\ , \ \ \ \ \ \qquad 
(E-S)_{0}(\hat\nu)=\sqrt{1+{\hat\nu}^2}\, . 
\ee
Using (\ref{ellcorr}) and expanding to leading order we obtain for 
the finite size correction\foot{Note that the classical contribution $(E-J)_{0}$ has 
an effect on the one-loop result
because of the one-loop expansion (\ref{ellcorr}).} 
\be
\Delta(E-S)_{1}(\ell)=
\frac{4 \pi\,\zeta(-1)}{(1+\ell^2)\;L^2}\,\ln S
=- \frac{\pi}{12} \, \frac{1}{(1+\ell^2)}\frac{1}{\ln S}\,.
\ee
 We have thus  reproduced the result (\ref{SJpart})
obtained in the partition function approach, supporting the expectation that 
the thermodynamic arguments are still valid  for the leading 
 finite size  correction. We shall therefore use the
 free energy based approach also in the 2-loop computation below.\foot{
While we have not discussed explicitly 
the renormalization of the spin $S$, it is possible to argue \cite{rt2} that correction to it
 are 
suppressed by $S^{-n}$ factors and thus are subleading to the 
${1 \ov \ln S}$ corrections we are interested in. 
}

\section{Leading finite size correction to the folded string  energy at two-loop order}
\label{pf2loop}

As we have seen above,   only massless 
fields can yield ${1 \ov \ln^{n}S}$  finite size 
 contributions and  to compute
them it suffices to first evaluate the integral(s) over the continuous 
component of the 
loop momentum and then evaluate the sum over the discrete momentum  using  zeta
function regularization.
 Here we  will follow this
strategy  in the two-loop calculation.
 


The relation between the two-loop partition function and 
 the energy of the folded 
string follows from the discussion in section \ref{gensc}; the contributions to the former 
follow from Feynman diagrams with topologies shown in figures~\ref{1PI} and 
\ref{non-1PI}:
\be
W_2&=&\beta E_2= -(\ln Z)_2=\frac{V}{2\pi\sql}\, {\cal F}_2
= \beta\,\frac{\ln S}{\pi\sql}\, {\cal F}_2\ , 
\\
{\cal F}_2 &=&
-4\pi^2 \left(
A_{\text{sunset}}^{\text{BBB}}
+A_{\text{double-bubble}}^{\text{BB}}
+A_{\text{sunset}}^{\text{BFF}}
+A_{\text{double-bubble}}^{\text{BF}}
+A_{\text{double-bubble}}^{\text{FF}}
+A_{\text{non-1PI}}
\right)
\label{sixTerms}
\ee
As discussed in \cite{cusp,cuspJ}, for a non-compact worldsheet the partition 
function receives nontrivial contributions both from 1PI (fig.~\ref{1PI}) 
and non-1PI (fig.~\ref{non-1PI}), the role of the latter being to render the result 
finite. We shall see that the non-1PI graphs contribute nontrivially to finite size 
corrections as well.

The explicit expressions of the six terms in equation~(\ref{sixTerms}) are rather 
lengthy and are collected in Appendix~\ref{ExplicitList}. Their structure is determined 
by the topology of the Feynman graphs:
\be
\label{contribs_structure}
\begin{aligned}
&A_{\text{sunset}}=\int\frac{dp_0 dq_0 dr_0}{(2\pi)^4}\sum_{p_1,q_1,r_1}
\delta^{(2)}(p+q+r) \frac{f(p,q)}{P(p,m_1^2) P(q,m_2^2) P(r,m_3^2)}+\dots\\
&A_{\text{double-bubble}}=\int\frac{dp_0 dq_0 dr_0}{(2\pi)^4}\sum_{p_1,q_1,r_1}
\delta^{(2)}(p+q+r)  \frac{g(p,q)}{P(p,m_1^2) P(q,m_2^2)}\\
& A_{\text{non-1PI}}=\int\frac{dp_0dq_0}{(2\pi)^4}\sum_{p_1,q_1}\frac{1}{2} T(p) T(q)\ , 
\end{aligned}
\ee
where $f(p_0,q_0)$ and $g(p_0)$ have a polynomial dependence of degree 4 in 
$p_0$ and $q_0$,
\be\label{tad1}
T(p)=+\frac{1+\frac{1}{2}p_0^2}{P(p, 4)} 
         +\frac{5}{4}\,\frac{p_0^2-p_1^2}{P(p,0)}-4\,\frac{p_0^2}{P(p,1)} +\frac{1+p_0^2}{P(p,2)} 
\ee
is the integrand of the one-loop tadpole for the field ${\tilde \phi}$ and the ellipsis 
in the sunset contribution stand for terms with one cancelled propagator. Such 
terms combine naturally with those arising from the double-bubble topology.

In obtaining the  contributions listed in Appendix~\ref{ExplicitList} we have 
discarded power-like divergences in the continuous momentum integral. The various 
terms have been organized such that the summation over the space-like momenta is 
manifestly finite. The integral over the  continuous (Euclidian time-like)
momenta produces  all the divergences which should cancel out 
when all integrals are added 
up. 

The one-loop calculation described above suggests that only diagrams with at least one
massless field ({\it i.e.} at least one factor $P(p, 0)=p^2 $) can yield a polynomial dependence in 
$1 \ov \ln S$. To demonstrate  that 
this is indeed the case let us briefly discuss 
the  $L$-dependence of the integrals that can appear
 in the two-loop partition function.

\subsection{On the $L$-dependence of two-loop integrals}

From eqs.~(\ref{contribs_structure}) and  the explicit expressions in 
Appendix~\ref {ExplicitList} it is clear that the integrals that enter the calculation 
of the two-loop partition function fall into  two classes: $(a)$ iterated one-loop integrals, 
and $(b)$ sunset-type integrals involving three propagators. 

Integrals of the first type are, up to numerator factors, similar to the integrals 
that enter the one-loop partition function. As in that case, a polynomial dependence 
in the inverse length of the string can arise only if the integrand is generated by a 
massless field. The precise $L^{-1}$ dependence of the result depends strongly on the 
 numerator factors; these factors determine whether only one or both integral 
factors yield such contributions. For example, using the summation formulae in 
Appendix~\ref{appInt} it is easy to see that for $m \geq 1$
\be
\int_{-\infty}^{+\infty} dp_0\sum_{p_1}\,\frac{p_0^m}{p_0^2+p_1^2} =
\pi\int_{-\infty}^{+\infty}dp_0 \,p_0^{m-1}\coth\left(\ha Lp_0\right)
=(\text{divergent})+ \frac{1}{L^{m}}\times (\text{finite})~~.
\ee
Thus, the leading finite size contribution of the product of two such integrals 
contains one of the integrals evaluated in the $L\rightarrow\infty$ limit. 

In the sunset-type two-loop integrals, the sums over the discrete components of momenta
are generically of the type
\be
S(a,b,c)=\sum_{m,n}\frac{1}{\left[(\frac{2\pi n}{L})^2+a^2\right]
                                             \left[(\frac{2\pi m}{L})^2+b^2\right]
                                             \left[(\frac{2\pi (n+m)}{L})^2+c^2\right]}
\ee
for some typically different $a,b,c$ with $a^2=p_0^2+m_a^2$, $b^2=q_0^2+m_b^2$ 
and $a^2=(p_0+q_0)^2+m_c^2$. If all masses are nonvanishing, $m_{a,b,c}\ne 0$, 
such sums may be
evaluated by a repeated application of the contour integral trick of \cite{zama}.
We choose two copies of a contour that runs parallel to the real axis above and below 
it and write $S(a,b,c)$ as
\be
S(a,b,c)=\frac{1}{(2\pi i)^2}\int_{{\cal C}_z} dz \int_{{\cal C}_y} dy
\frac{\cot \pi z\;\cot \pi y}
             {\left[\left(\frac{2\pi z}{L}\right)^2+a^2\right]
              \left[\left(\frac{2\pi y}{L}\right)^2+b^2\right]
              \left[\left(\frac{2\pi (y+z)}{L}\right)^2+c^2\right]}
\ee
For an integrand with suitable properties a contour deformation argument implies that 
the sum is given by the residues of the purely imaginary poles given by the rational part 
of the integrand. In these residues, $a, b, c$ or some combination thereof will appear in 
the argument of the $\cot$ function;
moreover, since these poles occur at purely imaginary values of $p_0$ and $q_0$, 
the $\cot$ function will in fact become $\coth$.\footnote{The $L\rightarrow\infty$ 
limit should be taken with care if the numerator polynomial has a high degree, as this 
makes the integral divergent. Formally, this limit amounts to the formal replacement
$
\coth(\cdot)\mapsto {\rm sgn}(\cdot)~.$
}
We conclude that, if all propagators are massive ({\it i.e.} if $m_{a,b,c}\ne 0$), the $L$ 
dependence is exponentially suppressed. A slightly more involved analysis is necessary 
if some masses are equal (but nonvanishing); the conclusion, however, is unchanged.
In Appendix~\ref{appInt} we illustrate this conclusion by numerically evaluating the 
integral $I[1, \ha, \ha]$ which, in the $L\rightarrow\infty$ limit, yields the complete 
two-loop energy, see Fig.~\ref{decay}.

It therefore follows that, among all sunset-type integrals, only those with at least one
massless field can yield a polynomial dependence on $1\ov \ln S$. For the purpose of 
finding the leading finite size corrections to the energy of the folded string it suffices to 
focus our attention only on these contributions. The other terms are, of course, crucial to 
guarantee the cancellation of UV divergences.

Let us note also that the  arguments above require first to  evaluate  the sum over the discrete 
momentum. The $L$-dependence,  however, is not expected to change if we first carry 
out the integral over the continuous momentum components in the presence of a 
suitable regulator ({\it i.e.} a regulator which does not depend on $L$). 
For example, it is possible to verify that
\be
\int_{-\Lambda_1}^{\Lambda_1} dp_0 \sum_{p_1}  
\frac{1}{p_0^2+p_1^2+m^2}=\int_{-
\infty}^{\infty} dp_0 \sum_{p_1=-\Lambda_2} ^{p_1=\Lambda_2}  
\frac{1}{p_0^2+p_1^2+m^2}
\ee
for suitable cutoffs $\Lambda_1$ and $\Lambda_2$. 
Thus the formal consideration about the exponential suppression 
of all-massive sunset-type integrals should also hold
if the integration over the continuous variables is performed first.
We adopt this technically simpler strategy in our two-loop calculation; as 
seen at the one-loop level, the zeta-function regularization of the resulting sums should
then  
yield the correct result.

\subsection{Contribution of massless integrals}

The discussion in the previous subsection implies that the only terms from 
Appendix~\ref{ExplicitList} which potentially contribute to the finite size correction 
in the two-loop effective action involve at least one massless field. 
Here and in Appendix~\ref{ExplicitList} we denote by $B_{m_1^2,m_2^2}$ and 
$B_{m_1^2,m_2^2,m_3^2}$ the terms in the integrand of the bosonic sunset 
and double-bubble diagrams with masses as indicated; 
similarly $C_{m_1^2,m_2^2}$,  $F_{m_1^2,m_2^2,m_3^2}$ and 
${\cal A}^{\text{non-1PI}}_{m^2}$ denote, respectively, the terms from 
the mixed bosonic-fermionic double-bubble, fermionic sunset and non-1PI 
diagrams.\footnote{Purely fermionic double-bubble diagrams are identically 
vanishing; even if they  were not, they  could 
 contribute only exponentially suppressed terms, 
as all fermions are massive.}

With this notation and after accounting for the various cancelations discussed 
in Appendix~\ref{ExplicitList}, the only terms in the two-loop integrand which 
may yield ${\cal O}(1/L^2)$ contributions upon integration over the continuous 
 momenta and summation over the discrete ones are:
%
%
\be
B_{0,0,4}&=&
-\frac{5}{4}\, \frac{1}{P(p,0)P(q,0)}
+\frac{5}{2}\,  \frac{1-2p_0^2}{P(p,0)P(r,4)}
+5\, \frac{(1+p_0 q_0)^2}{P(p,0)P(q,0)P(r,4)}
\\[3pt]
F_{0,1,1}+C_{0,1}&=&
\,\frac{10 p_0 q_0 r_0^2}{P(p,1)P(q,1)P(r,0)}
\\[3pt]
{\cal A}^{\text{non-1PI}}_{0}&=&\,\frac{5}{4}\frac{p_0^2-p_1^2}{P(p,0)}T(q)
\ee
To identify the relevant massless contribution coming from the non-1PI term, which is 
proportional to $T(p)T(q)$, we used the observation following from the calculation in 
section~\ref{Evev1} that the leading finite size correction to the tadpole term 
is already of the desired  ${\cal O}(L^{-2})$ order. Thus, in the product $T(p)T(q)$ we 
need to keep the massless contribution from only one of the two tadpole graph factors, 
while the other one can be treated in the $L\rightarrow\infty$ limit. 

Except for the third term in $B_{0,0,4}$ and for $F_{0,1,1}+C_{0,1}$ all other terms
factorize into a product of one-loop integrals which may be easily evaluated using 
($p^2=p^2_0 + p^2_1$) 
\be
&&\int_{-\infty}^{+\infty}  dp_0 \frac{1}{p^2+m^2} =\frac{\pi}{\sqrt{p_1^2+m^2}}
\\
&&\int_{-\infty}^{+\infty}  dp_0 \frac{p_0^2}{p^2+m^2} = \int dp_0\left(1-\frac{p_1^2+m^2}{p^2+m^2}\right)
\mapsto -\pi \sqrt{p_1^2+m^2}\,.
\label{intEG}
\ee
In the second integral above we discarded a linearly-divergent term; such terms are 
analogous to quadratically divergent terms which are discarded in the 
$L\rightarrow\infty$ calculation.\footnote{Indeed, in the $L\rightarrow\infty$ limit  the 
integral (\ref{intEG}) is just
$$
\int d^2p \frac{p_0^2}{p^2+m^2} = \frac{1}{2}\int d^2p \frac{p^2}{p^2+m^2}\mapsto
-\frac{m^2}{2}\int \frac{d^2p}{p^2+m^2}~~.
$$} They should be cancelled by contributions of the path integral measure
(or discarded using analytic regularization). 

To carry out the integrals of products of three propagators it is useful to first 
Fourier-transform the integrals over the 0-th component of momenta to 
position space
\be
\int_{-\infty}^{+\infty}  
dp_0\frac{e^{ip_0 x}}{p_0^2+p_1^2+m^2}=\pi \frac{e^{-|x|\sqrt{p_1^2+m^2}}}
{\sqrt{p_1^2+m^2}}~.
\ee
In this form, the numerator factors depending on the 0-th component of momenta
are realized as derivatives with respect to the position variable. 
The three relevant two-loop integrals, with constant numerator and with a numerator
bilinear in the integration variables, are evaluated in Appendix~\ref{2Lints}.
Now we will discuss the evaluation of the integrals of $B_{0,0,4}$, $F_{0,1,1}+C_{0,1}$ 
and  the non-1PI contribution.

\subsubsection{$B_{0,0,4}$}

Applying the strategy described above to the integral of $B_{0,0,4}$ leads to
\be\label{BB}
\int_{-\infty}^{+\infty}  dp_0 dq_0 B_{0,0,4}=\frac{1}{4}\left[
-\frac{15}{2}\pi^2+\frac{5}{2} \pi^2 |p_1| |q_1|
+\frac{5 \pi^2  \, p_1{\rm sgn}(q_1)}{\sqrt{4+(p_1+q_1)^2}}
-5 \pi^2 |q_1| \,\frac{p_1^2+p_1 q_1+ 1}{\sqrt{4+(p_1+q_1)^2}}\right]\,.~
\ee
To obtain this expression we performed some convenient relabelling of the discrete 
momenta $p_1$ and $q_1$. We also dropped terms
which are odd in the discrete momenta  and therefore vanish after summation over 
a symmetric domain.  

In the last two terms, the remaining sums over the discrete momenta $p_1$ and 
$q_1$ are coupled due to the presence of nontrivial denominators. They may be 
decoupled by shifting one of the summation variables, e.g. $p_1\rightarrow p_1-q_1$.
Since these sums are clearly divergent, such manipulations should be treated with care.
It is not hard to check that, with some regulator $R(p)$,  
\be
\sum_{p_1}R(p_1)\left[\frac{(p_1+q_1)^n}{\sqrt{(p_1+q_1)^2+4}}-
\frac{p_1^n}{\sqrt{p_1^2+4}}\right]=q_1^n \,{\cal O}(L^{-1})={\cal O}(L^{-(n+1)})
\label{difference}
\ee
for all exponents $n\ne 0$. The result  depends strongly on the regulator $R$; 
however, the $L$ dependence is such that this difference is of 
too high an order in $L^{-1}$
to contribute to the leading $L^{-2}$ correction. 
Analyzing separately a constant numerator factor of the last term in 
eq.~(\ref{BB}) (which corresponds to the $n=0$ terms in eq.~(\ref{difference})) 
shows that the shift $p_1\rightarrow p_1-q_1$ does not affect the value of the sum either.


Decoupling the sums in the last two terms in \rf{BB} by appropriately 
shifting the summation variables leads to
\be
\sum_{p_1,q_1} B_{0,0,4}= \frac{1}{4}\sum_{p_1,q_1} \left(
-\frac{15}{2}\pi^2+
\frac{5}{2} \pi^2 |p_1| |q_1|-5 \pi^2 |q_1| \,\frac{p_1^2+2}{\sqrt{4+p_1^2}}\right)~~.
\label{B001}
\ee
In deriving this expression we further discarded terms which are odd in the discrete 
momenta and thus vanish when summed over a symmetric domain.
An example illustrating the discarded terms is the following:
\bea
\sum_{p_1,q_1}\frac{p_1}{\sqrt{4+(p_1+q_1)^2}}
=\frac{1}{2}\sum_{p_1,q_1}\frac{p_1+q_1}{\sqrt{4+(p_1+q_1)^2}}
=\frac{1}{2}\sum_{p_1,r_1}\frac{r_1}{\sqrt{4+r_1^2}}~,
\eea
where we discarded a term odd under the interchange of $p_1$ and $q_1$. The remaining
sum over $r_1$ also vanishes since the summand is odd. Alternatively, we can use the
the zeta-function regularization, which we assume, to show that the sum over $p_1$ 
can be set to zero. Indeed, 
\bea
\sum_{p_1}1=\frac{2\pi}{L}\sum_{n=-\infty}^{+\infty}1=\frac{2\pi}{L}\Big(
1+2 \sum_{n=1}^{+\infty}1\Big)=\frac{2\pi}{L}(1+2\zeta(0))=0~~.
\label{zeta0}
\eea
As in  the integral in eq.~(\ref{intEG}), 
such manipulations are similar to discarding quadratic divergences in two-loop 
integrals in the $L\rightarrow\infty$ limit.

With this prescription the constant term in (\ref{B001}) vanishes. It is easy to see that, 
if in the second term we take both sums to contribute $L^{-1}$ terms, then the result is 
of order $L^{-4}$ and thus too high an order. It follows therefore that one of the sums 
should be evaluated in the $L\rightarrow\infty$ limit:
\be
\sum_{p_1,q_1}|p_1||q_1|=2\left(\frac{2\pi}{L}\right)^2(2\zeta(-1))
\int_{-\infty}^{+\infty}|q_1|\, dq_1~~.
\label{p1q1contrib}
\ee
This integral is a pure quadratic divergence, similar to other quadratically divergent
{\it integrals} which have been discarded; thus, it may be discarded as well.

For the last term in \rf{B001} we observe that carrying out the sum over $q_1$ already 
yields a term of order  $1/L^2$
\be
\begin{aligned}
&-5\pi^2 \sum_{p_1,q_1} |q_1| \,\frac{p_1^2+2}{\sqrt{4+p_1^2}}=-5 \pi^2 \, \frac
 {2\pi}{L}\, \sum_{m} \frac{2\pi |m|}{L}\,\sum_{p_1}\frac{p_1^2+2}{\sqrt{4+p_1^2}}\\
&~~~~~~~~~~~~~~~= -{10\pi^2}\left(\frac{2\pi}{L}\right)^2\;\zeta(-1)\;\sum_{p_1}\frac{p_1^2+2}{\sqrt{4+p_1^2}}
 \end{aligned}
\ee
It is therefore appropriate to approximate the remaining sum over $p_1$ with 
the corresponding integral.\foot{Alternatively, one may argue that
 the difference between the
sum and the integral is  exponentially suppressed due to the mass-like constant term in the denominator.} 
We can therefore write the contribution of $B_{0,0,4}$ as
\be
\int_{-\infty}^{+\infty} dp_0dq_0\sum_{p_1,q_1} B_{0,0,4}=
-\frac{5\pi^2}{2}\left(\frac{2\pi}{L}\right)^2\;\zeta(-1)\;{\cal I}_B\,,
\qquad \ \ \ {\cal I}_B\equiv\int_{-\infty}^{+\infty}  dp_1 \frac{p_1^2+2}{\sqrt{4+p_1^2}}\,.
\label{finB}
\ee
By power-counting ${\cal I}_B$ is quadratically divergent; a closer inspection reveals that 
it does not contain logarithmic divergences. We will postpone 
its discussion   until we  analyze
other terms contributing to the  leading finite size correction.

\subsubsection{$F_{0,1,1}+C_{0,1}$}

The integrals over the continuous momentum components 
$p_0$ and $q_0$ for the term $F_{0,1,1}+C_{0,1}$ are very similar to those 
appearing in $B_{0,04}$; the result is
\be
\begin{aligned}
\int_{-\infty}^{+\infty} dp_0dq_0\,(F_{0,1,1}+C_{0,1})=
5 \pi^2\left(1 - p_1 q_1+ 
2 q_1 \sqrt{1 +  p_1^2} \,{\rm sgn}(p_1 + q_1) -  \sqrt{1 +  p_1^2} 
\sqrt{1 +  q_1^2}\right)
\end{aligned}
\ee
The sum over $p_1$ and $q_1$ of the 
first two terms vanishes due to  zeta-function regularization and 
summation over a symmetric domain while the third term can be argued to contain only 
exponential dependence on $L$ and may therefore be ignored. 
The remaining contribution, after 
shifting $q_1$ and 
dropping a term odd under $p_1\rightarrow -p_1$,  becomes
\be\label{FC2}
\begin{aligned}
&\int_{-\infty}^{+\infty} dp_0dq_0\sum_{p_1,q_1}(F_{0,1,1}+C_{0,1})=
10 \pi^2 \left(\frac{2\pi}{L}\right)^2\sum_m  |m| \sum_{p_1}  \sqrt{1+ p_1^2}
\\
&~~~~~=20\pi^2\left(\frac{2\pi}{L}\right)^2\;\zeta(-1)\,{\cal I}_F~~, 
\qquad\ \ \ \ \ \ \qquad{\cal I}_F\equiv\int_{-\infty}^{+\infty} dp_1 \sqrt{1+ p_1^2}
\end{aligned}
\ee
Similarly to ${\cal I}_B$ in $B_{0,0,4}$, the integral ${\cal I}_F$ is quadratically 
divergent; unlike ${\cal I}_B$, however, ${\cal I}_F$ also exhibits subleading 
logarithmic divergences which cannot be removed by, {\it e.g.},  an analytic regularization
scheme. As we shall see, these divergences cancel out once all finite-size 
contributions are combined.

\subsubsection{Non-1PI}

The finite size contributions of this type arise entirely from the  factor
\be
T_0=\int_{-\infty}^{+\infty}  
dp_0 \sum_{p_1}\frac{5}{4}\,\frac{p_0^2-p_1^2}{p_0^2+p_1^2}\,.
\ee
In the continuum limit this term can be neglected since it clearly vanishes due to the
$p_0\leftrightarrow p_1$ antisymmetry of the integrand. As was noted 
 in the calculation of the 
expectation value of the energy operator at one loop, this is no longer so once $p_1$ 
is discrete. Following the same steps as for the evaluation of the contributions of 
$B_{0,0,4}$ and $F_{0,1,1}+C_{0,1}$ and carrying out first the integral 
over $p_0$ we find
\be
\begin{aligned}
T_0&=\frac{5}{4}\sum_{p_1}  \int_{-\infty}^{+\infty}  
dp_0\left(1-2\frac{p_1^2}{p_0^2+p_1^2} \right)=
-\frac{5}{2} \sum_{p_1} \int_{-\infty}^{+\infty}  
dp_0 \frac{p_1^2}{p_0^2+p_1^2}\\
&=-\frac{5}{2}\pi \sum_{p_1} |p_1|=
-5\pi \left(\frac{2\pi}{L}\right)^2  \;\zeta(-1)~~,
\end{aligned}
\ee
where one factor $\frac{2\pi}{L}$ arises from the definition of the summation 
over $p_1$  in  eq.(\ref{intvssum})
while the second one from the definition $p_1 = { 2 \pi n \ov L}$  in 
 eq.~(\ref{pdiscrete})).
The constant term on the first line was discarded  due to   zeta-function 
regularization (see eq.~(\ref{zeta0}))  and also 
 because its $p_0$ integral is linearly divergent.

To complete the calculation we should evaluate the continuum analog of this 
tadpole  contribution 
following the same steps as in the  discrete version of the 
calculation. Carrying first the $q_0$ integral we 
find that
\be
\int_{-\infty}^{+\infty}  dq_1 \int_{-\infty}^{+\infty} dq_0\, T(q)&=&\pi 
\int_{-\infty}^{+\infty}  dq_1 \Big(
-\frac{5}{2}|q_1|+4{\sqrt{q_1^2+1}}
-\frac{1}{2}\frac{q_1^2+2}{\sqrt{q_1^2+4}}-\frac{q_1^2+1}{\sqrt{q_1^2+2}}\Big)~.~
\label{TPcont}
\ee
The first term is a pure quadratic divergence similar to (\ref{p1q1contrib}) and other 
quadratic divergent integrals which have been discarded; we will discard it as well.
The contribution of the non-1PI graphs is therefore
\be
\int_{-\infty}^{+\infty}  dp_0 dq_0 \sum_{p_1, q_1}{\cal A}^{\text{non-1PI}}
=-5\pi^2 \big(\frac{2\pi}{L}\big)^2 \zeta(-1)
\Big[ 4{\cal I}_F-\frac{1}{2}{\cal I}_B-\int_{-\infty}^{+\infty}  dq_1\,
\frac{q_1^2+1}{\sqrt{q_1^2+2}}
\Big]
\label{finNON1pi}
\ee

\subsection{Summing up}
\label{FStotal}

We are now in position to assemble the leading $1 \ov \ln S$ term in the 
two-loop correction to 
the energy of the folded string.
Combining eqs.~(\ref{finB}), (\ref{FC2}) and (\ref{finNON1pi}) and reconstructing 
the finite size correction $\Delta {\cal F}_2$ to the free energy as defined in 
eq.~(\ref{sixTerms}) we find that the divergent integrals ${\cal I}_F$ and ${\cal I}_B$ 
cancel out and  the leading finite size correction to the free energy is\footnote{
Here we have restored a factor of $\frac{1}{(2\pi)^4}$ coming from the loop 
momentum integration. An additional multiplicative factor of $(-16\pi^2)=
(-4\pi^2)\times(4)$ arises from 
the definition of ${\cal F}_2$ in eq.~(\ref{sixTerms}) and from the definition of 
$A_{\text{sunset}}^{\text{BBB}}$,
$A_{\text{double-bubble}}^{\text{BB}}$,
$A_{\text{sunset}}^{\text{BFF}}$,
$A_{\text{double-bubble}}^{\text{BF}}$,
$A_{\text{double-bubble}}^{\text{FF}}$ and
$A_{\text{non-1PI}}$ in Appendix~\ref{ExplicitList}.}
\be
\label{F2corr}
\Delta{\cal F}_2 = -5
\left(\frac{2\pi}{L}\right)^2  \;\zeta(-1)
\int_{-\infty}^{+\infty} dq_1\,
\frac{q_1^2+1}{\sqrt{q_1^2+2}}
~.
\ee
The remaining integral in $\Delta{\cal F}_2$ is clearly divergent.
It is, however,  free of logarithmic divergences as these 
cancelled out in a nontrivial way between  
various contributions to $\Delta {\cal F}_2$. 
The result then depends on how we deal with the remaining quadratic divergences.

It is easy to see that the quadratic divergence in eq.~(\ref{F2corr}) 
is of the type $\int dq_1|q_1|$, {\it i.e.}  
it is of the same nature as quadratic divergences that 
have been discarded in the calculation in the $L\rightarrow\infty$ limit; they are
also similar to divergences that have been discarded in the process of reorganizing 
the integrands  of the 2-loop Feynman integrals.  
One option then is  to discard them here as well by simply replacing
\be
\int_{-\infty}^{+\infty}  dq_1\,
\frac{q_1^2+1}{\sqrt{q_1^2+2}}
\ \  \rightarrow\ \ 
\int_{-\infty}^{+\infty}  dq_1\,\Big(\frac{q_1^2+1}{\sqrt{q_1^2+2}}-|q_1|\Big)=1~~.
\label{intF}
\ee
If we adopt this prescription\footnote{Notice 
that subtracting the quadratic divergence as
\be
\int_{-\infty}^{+\infty}  dq_1\,
\frac{q_1^2+1}{\sqrt{q_1^2+2}}
=\int_{-\infty}^{+\infty}  dq_1\,
\Big[
\sqrt{q_1^2+2}-\frac{1}{\sqrt{q_1^2+2}}
\Big]
\ \  \rightarrow\ \ 
-\int_{-\infty}^{+\infty}  \frac{dq_1}{\sqrt{q_1^2+2}}
\nonumber
\ee
is not valid, as it artificially introduces a logarithmic divergence.
} 
 we  end up with\foot{As in \rf{defW} here we define 
$(E-S)_n$ without the explicit loop-counting factor $1 \ov (\sql)^n$. 
Also, as at 1 loop (\ref{1loop_cor}), we ignore the $\ln S$-independent term.} 
\be
\Delta{\cal F}_2 = \frac{5\pi^2}{12\ln^2 S} \ , 
~~~~~~~~~~~~~~~~~~
(E-S)_2= \frac{1}{\pi} \Big( - \KK \ln S   +  \frac{5 \pi^2}{12 \ln S} \Big)   \ . \la{tw}
\ee
This result, however, may seem strange: 
 such an evaluation of the integral in  \rf{F2corr}
leads to a departure from the pattern of  transcendentality 
of coefficients 
 noticed at one loop order: while there  the coefficient of the 
  finite size correction had one 
additional unit of transcendentality compared to the leading term
({\it i.e.} $\pi^2$  vs. $\ln 2$ in \rf{1loop_cor}), 
the corresponding  coefficients in the
candidate  2-loop expression \rf{tw}  have the same  transcendentality
($\pi^2 = 6 \zeta(2)= 6 \sum^\infty_{n=1} {1 \ov n^2}  $  vs.
$\KK= \sum^\infty_{n=0} {(-1)^n \ov (2 n +1)^2 } $). 
This observation  may be considered as a  hint that a different definition 
of the integral in  \rf{F2corr} may be more appropriate.

Notice that the term
surviving in (\ref{F2corr}) is
the last terms in eq.~(\ref{TPcont})
which is nothing but the $q_0$ integrals of 
the 
fourth terms in $T(q)$, see
eq.~(\ref{tad1}). 
Interpreting it this way and evaluating the integral using the two-dimensional 
Lorentz invariance of the denominator
it is easy to see that, up to quadratic divergences,
this integral vanishes when evaluated  in the 
``decompactified'' (continuum spatial momentum) case: 
\be 
\int d^2q\,\frac{q_0^2+1}{q^2+2}=\int d^2q\,\frac{\frac{1}{2}q^2+1}{q^2+2}=
\frac{1}{2}\int d^2q\ \rightarrow\ 0
\ee
This then suggests that the integral  in \rf{F2corr} should 
not have a finite part after the quadratic divergences  are subtracted out  
\be
\Delta{\cal F}_2 \propto\int dq_1\,
\frac{q_1^2+1}{\sqrt{q_1^2+2}}
\ \  \rightarrow\   \ 0 \ . 
\label{LOR}
\ee
This prescription then implies  the vanishing of the leading 
 finite size  two-loop correction to the  energy 
of  the folded string.

The values (\ref{intF}) and (\ref{LOR}) may be interpreted as 
corresponding to different regularization schemes, 
each preserving different  amount of symmetries. 
Carrying out the momentum integrals
iteratively obscures the fact that in the $\ln S\rightarrow\infty$ limit the quadratic part 
of the action is invariant under the 2d 
Lorentz transformations. The second prescription corresponds to insisting on that symmetry 
in the limit $\ln S \to \infty$. A bonus is that, as a result, one 
avoids  violation of the pattern of 
transcendentality of coefficients  observed at one loop.

Let us finally comment on the case of $J \not=0$. 
At 1-loop order we saw in detail that turning on a non-zero value of  
angular momentum on $S^5$ exposes  the fact that part of the leading 
finite size corrections at $J=0$ arises from the resummation 
of infinitely many exponentially small corrections at $J\ne 0$ (with $\ell$ held fixed).
From the Bethe ansatz  standpoint such exponential terms may be interpreted as
``L\"uscher'' corrections (or wrapping corrections in gauge theory).
A similar  picture is expected  at 2 loops: it 
 would be interesting 
to identify in the two-loop calculation the terms that become exponentially suppressed
as $J$ is  switched on.
 The structure of the $J=0$ result \rf{F2corr}  and an analogy with the 1-loop case 
 (the overall coefficient 5, which is related to the number of massless fields, replaced by 1) 
 suggest that 
the leading term in the small $\ell$ expansion of $\Delta{\cal F}_2$  for $J \not=0$ 
should be  
\be
\label{F2corrJ}
\Delta{\cal F}_2 = -
\left(\frac{2\pi}{L}\right)^2  \;\zeta(-1)
\int_{-\infty}^{+\infty} dq_1\,
\frac{q_1^2+1}{\sqrt{q_1^2+2}}
~~.
\ee
The final numerical value  depends again on a regularization prescription
used to subtract quadratic divergences and  is thus   zero if we 
adopt the ``2d Lorentz-invariant'' prescription in \rf{LOR}.


\

{\bf Acknowledgments}\\
We would like to thank N. Gromov and D. Volin for many useful discussions. 
This work was supported in part by the US National Science Foundation under 
DMS-0244464 (S.G.), PHY-0608114 and PHY-0855356 (R.Ro.), the US Department of 
Energy under contracts DE-FG02-201390ER40577 (OJI) (R.Ro.), the Fundamental Laws 
Initiative Fund at Harvard University (S.G.) and the A.P. Sloan Foundation (R.Ro.). It was 
also supported by EPSRC (R.Ri.). The work of S.G. is also supported by Perimeter Institute 
for Theoretical Physics. Research at Perimeter Institute is supported by the 
Government of Canada through Industry Canada and by the Province of Ontario through
the Ministry of Research $\&$ Innovation. R.Ri. would like to thank the Simons Center for 
Geometry and Physics for hospitality during the 8th Simons workshop on Physics and 
Mathematics.

\newpage

\appendix

\section{Long folded spinning string: 1-loop  finite size corrections\\
and Landau-Lifshitz model}

Let us  start with a review of the form of the 1-loop correction to the energy of the long 
folded $(S,J)$ string \ci{ftt}.
In this case 
 $\mu= {1 \ov \pi} \ln S  \to \infty$ with 
$\el \equiv { \J \ov \mu}$=fixed and 
\be \la{jo} 
&& E_1 = { 1 \ov \kappa}  E_{2d} = { 1 \ov \mu \sqrt{ 1 + \el^2}}  E_{2d}\  , \la{syy}\\
&& 
  E_{2d}= { 1 \ov 2} \sum_{n=-\infty}^{\infty} \Big[ \O_{n-} + \O_{n+}
  +  2 \sqrt{ n^2 + (\el^2 + 2) \mu^2 } + 4  \sqrt{ n^2 + \el^2 \mu^2 }
  -8 \sqrt{ n^2 + (\el^2 + 1) \mu^2 } \Big]  \no
  \ee
where 
\be  \la{fre} 
 \O_{n\pm } = \sqrt{ n^2  + 2 \mu^2(1  + \el^2) \pm   2\mu  \sqrt{ n^2 \el^2 + \mu^2(1+  \el^2)^2} }  \   \ee
are the contributions of the two ``mixed''   $AdS_3$ modes. 
  We would like to determine the  leading 
contribution to $\mu^{-n}$  corrections coming from this  sum over characteristic  frequencies. 
It is easy to see that   for non-zero $\el$  the  massive modes give  sums of exponential corrections but
there is one special mode that becomes  light in the  $\mu \to \infty$  limit: this is 
the lighter of two $AdS_3$  modes in \rf{fre}, {\it i.e.} 
\be \la{ex}
\O_{n-} = { n \ov \sqrt{ 1 + \el^2}} \Big[
 1  + { n^2 \ell^4 \ov 8 \mu^2 (1 + \el^2)^2 }  + \OO  \Big( { 1 \ov \mu^4} \Big)  \Big] \ . 
\ee
As a result, the leading $ 1 \ov \mu$  contribution to   the  1-loop correction to the energy 
comes from the first term in \rf{ex}:  
\be 
(E_1)_{1\ov \mu}  =  { 1 \ov 2 \mu ( 1 + \el^2) }   \sum_{n=-\infty}^{\infty} n 
=  - { 1 \ov 12}  { 1 \ov  \mu ( 1 + \el^2) } 
 =  - { 1 \ov 12 \pi } {\l  \ln S  \ov   J^2  +  {\l \ov \pi^2}  \ln^2 S  } \ , \la{mui}
\ee
where we used that $ \ha \sum_{n=-\infty}^{\infty} n =  \zeta(-1) =  - { 1 \ov 12} $. 
Since the sum in $ E_{2d}$ in \rf{syy} is UV  finite, one may  interchange  summation over $n$ with 
taking the large $\mu$ limit   and the  use of the 
$\zeta$-function  regularization is just a short-cut  to extract the relevant term  in that finite sum. 

If we take  $\el \ll 1$ or   $J^2 \ll  {\l \ov \pi^2}  \ln^2 S$ we get 
$(E_1)_{1\ov \mu}=   - { 1 \ov 12} { \pi \ov  \ln S }$ which is  the 
``non-wrapping''  (from the BA  point of view  \ci{gro}) 
  part of the total string coefficient $- { 5\ov 12} $  \ci{zama,bed} 
found for the $1\ov \ln S $ coefficient in the limit when $J$  can be ignored. 
The  distinction between the ``non-wrapping'' and ``wrapping'' 
contributions becomes  clear  for nonzero $J$:     to recover the extra 
$- { 4\ov 12} $  contribution from  four   $S^5$ modes that  become   massless 
 in the strict $J=0$ limit we   need to resum  the exponential (``L\"uscher'') contributions
  corresponding
 to them before taking the large $\mu$ limit.\foot{It is  only  in the massless 
 or conformal limit
 that the contribution of a 2d mode  is given by the  Casimir effect
   on a cylinder, {\it i.e.} is proportional to 
 $ - { 1 \ov 12} { T \ov L} $ where $T$ is the time interval   and $L$ is the 
 length of the spatial circle.} 

As discussed in the Introduction, the analytic dependence of  \rf{mui} on $\l$ 
suggests that the order  $\l$  term  there is not renormalized, {\it i.e.} its value is the
same also at weak coupling.  Then it can be reproduced  as a 1-loop  correction in the
 corresponding 
Landau-Lifshitz  model.  This  is  the aim of this Appendix.

\def \L {{\cal L}}


Here we shall follow \ci{ptt} and \ci{mtt} (Appendix ~D there). 
 The semiclassical  states  from $sl(2)$ sector 
correspond to  strings rotating in $AdS_3$ part of $AdS_5$ and whose
center of mass is moving along big  circle of $S^5$, {\it i.e.} their
energy depends on the  two spins $(S,J)$. The fast string
limit is when  $J$  is  large with  $\tl = {\l \ov
J^2}$  being fixed.
On the  gauge theory side we assume  $J$ is large 
and consider only the 1-loop (order $\l$) term   in the dilatation operator. 
In the previous  discussions  it was assumed that $S/J$ is fixed in this limit 
but as we shall see below   the LL description captures also 
the case when  $\ell$  or $ \tl \ln^2 S$  is fixed.\foot{We may assume that 
 $\ln S$ should be
 replaced by $\ln ({S\ov J})$  with $S\ov J$   fixed, see below. 
 }  
The corresponding  LL action  \ci{ste} derived from  $sl(2)$  spin chain Hamiltonian 
(or from the  bosonic 
string action  in $AdS_3 \times S^1 $ by fixing an analog of the  static gauge \ci{krt,ptt})
is  
\begin{equation}
I= J \int dt \int^{2\pi}_0   { d \s \ov 2 \pi}  \ \L \ , \ \ \ \ \ \ \ \ \ 
L=-  2 \sinh^ 2\rho\ \dot{\eta}-\frac{\tilde{\lambda}}{2}
(\rho'^2+\sinh^2 2\rho\ \eta'^2 )\ , \ \ \ \ \  \tl \equiv  { \l \ov J^2}= {1 \ov \J^2}  \  .     \label{LLs}
\end{equation}
Here $\rho$  and $\eta=\ha
(t-\phi)$   are combinations of  the    $AdS_3$ coordinates:\  
$ds^2 = - \cosh^2 \rho  \  dt^2 + d \rho^2 + \sinh^2 \rho \ d \phi^2$. 

The
folded string solution  
 is given by $t=\kappa
\tau$, $\phi=w \tau$,
  $\rho=\rho(\sigma)$, $\vp= \nu  \tau$,
 To leading order in the $1/\J$ expansion, the corresponding
 solution of the LL equations  is
\be
&&\eta=\om \tau\ ,\ \  \ \ \ \ \ \om =\ha ({\kappa-  w })  \ , \ \ \ \  \ \ \ 
\rho''  + 2\bw \sinh 2\rho=0\ , \ \ \ \ \ \ \ \ \ \ \
   \bw\equiv  {\om \ov \tl} \ ,   \label{SG2}\\
&& \rho'^2=2\bw  ({\cosh}\ 2\rho_0 -{\cosh}\ 2\rho )\ , \ \ \ \ \ \ \ \   0 < \r < \r_0   \ .  \la{pi}
\ee
As discussed in \ci{mtt}, one may follow the same steps as in the $SU(2)$ sector 
and derive  the  Lagrangian for small  fluctuations of 
$\rho$ and $\eta$ near  the given solution 
\be 
\td \L = 2 g \dot  f  - \ha \tl  \Big[ g'^2 + f'^2   + 4\bw (3\cosh2\rho-2\cosh 2\rho_{0})\ g^2 
+ 4\bw  \cosh 2\rho\ f^2 \Big] \ . \la{fll} \ee 
Here  $f$ and $g$ are properly redefined
fluctuation fields, {\it i.e.} linear combinations of $\td \r$ and $\td \eta$, 
   and  $\rho(\s)$  is a solution of \rf{pi}. 
The short string limit when  $\rho_{0}\to  0$  was discussed  in \ci{mtt};
here we consider instead the long string limit when $\rho_{0}\to  \infty$.
In this case $w=\kappa$ so that $\omega=0$, {\it i.e.} 
\be \la{ds}
\eta =0 \  ,\ \ \ \ \ \ \ \ \ \ \   \ \r= \mu \s  \ . \ee 
To describe  this  case as a limit  of eqs. \rf{pi} and \rf{fll}
we may take the limit
\be \la{liu}
\omega\to 0, \ \ \   \rho_{0}\to  \infty \ , \ \ \ \ \ \ 
 \mu^2 = 2\bw  \ {\cosh}\ 2\rho_0={\rm fixed} \  , \  \ \ \ 
  \bw\ {\cosh}\ 2\rho \to 0 \ . \ee
$\mu$ may be related to spin  since to 
 leading  $\tl$-order  the expression for  the $AdS_3$  spin $S$
is \cite{ptt} (this  follows  directly from the action in  \rf{LLs})
\begin{equation}
S= 4 J\int_{0}^{\pi/2 }\frac{d\sigma}{2\pi}\  \sinh^2 \rho  \ , 
\end{equation}
where we integrate over one stretch  of the string  and the factor 4  accounts for the whole 
 $(0, 2 \pi)$ interval. Using that $\rho = \mu \s$   this gives 
 \be 
 \mu = {1 \ov \pi} \ln { S \ov J} + {\rm const}   \ . \ee 
 Even if $J$ is  large, we are still allowed   to assume 
  ${ S \ov J} \gg 1$ and even $\mu \gg 1$. 
  The classical energy   of this asymptotic LL solution 
  is then \foot{Since on the string theory side the LL action is derived in the gauge $t=\tau$, 
 the 2d energy 
  corresponding to the action in  \rf{LLs} is the same as the  target space energy.}
  \be 
   E^{(LL)}_0 = \ha  J \tl \mu^2 = { \l \ov 2\pi^2  J } \ln^2 { S \ov J}    \ . \ee
  That agrees   with the expansion  of the  original classical  string energy in \rf{cla}. 
 
In the  limit \rf{liu}   the fluctuation Lagrangian  \rf{fll}  becomes 
\be 
\td L = 2 g \dot  f  - \ha \tl  \big( g'^2 + f'^2   - 4 \mu^2  g^2  \big) \ , \la{oll} \ee 
and so that  the  characteristic  frequencies on $\mathbb{R} \times S^1$ are 
found to be $\pm \tilde \O_n $ where 
\be 
\la{ghu}
\tilde \O_n =  { \l  \ov 2 J^2 } n \sqrt{ n^2 + 4 \mu^2}  \ ,  \ee
and the correction to the energy is given by their sum over $n$, 
\be \td E_1 = \ha \sum_{n=-\infty}^\infty  \tilde \O_n     \ . \la{lala} \ee
Not too surprisingly,  
this is the  same expression that one 
 finds by expanding the contribution  of the mode \rf{ex} to $E_1$ 
 ({\it i.e.}   $ { 1 \ov \kappa}   \O_{n-} =  { 1 \ov \mu \sqrt{ 1 + \el^2} } \O_{n-}$) 
in \rf{jo} 
first in  large $\el$ or large $J$ to isolate the leading term corresponding to 
the LL model\foot{To match the LL model we need to take the large $J$ limit first.
The large $\el$ limit 
of the second  $AdS_3$  mode $\O_{n+}$  with sign $+$ in \rf{fre}
is  $ { 1 \ov \mu \sqrt{ 1 + \el^2} } \O_{n+} \to  2 + { \l n^2  \ov 2 J^2} +   O({\l^2  \ov J^4})$
and is 
 not seen in LL model (cf. the discussion in \ci{ptt}).}   
\be \la{gh} 
\Big[{ 1 \ov \mu \sqrt{1 + \el^2}} \O_{n-}\Big]_{\el \to \infty} = 
 { \l  \ov 2 J^2 } n \sqrt{ n^2 + 4 \mu^2}   - { \l^2 \ov 8 J^4}  n (n^2 + 4 \mu^2)^{3/2} + ...
   \ . \ee 
This implies that  the  LL model should capture the leading finite  size correction  \rf{mui}  discussed  above:
indeed, taking now $\mu$ large   gives  the leading term as 
${ \l \ov J^2}  n\, \mu$  which is  the same as  \rf{mui}  after summing over $n$.

To compare this  to  the discussion  in \ci{btz}
let us look at the $n=0$   contribution to the full string result in 
\rf{jo}:
\be 
&& E^{(0)}_1 = { 1 \ov 2 \mu \sqrt{ \el^2 + 1 }}  \Big( 0 + 2 \mu \sqrt{  \el^2 + 1 }
  +  2\mu  \sqrt{ \el^2 + 2  } + 4  \mu  \el
  -8 \mu  \sqrt{ \el^2 + 1 } \Big) \no \\ 
  &&
   =  - 3  +  \sqrt{  \el^2 + 2 \ov \el^2 + 1  }
  +   2  \sqrt {   \el^2 \ov \el^2 + 1  } \ .  \la{opt} 
  \ee
  Note that there is no zero mode contribution from the lightest $AdS_3$ 
mode.
  The zero-mode contribution  is thus not contributing to the $1\ov \mu$ 
  expansion at  fixed $\el$. 
Expanding  \rf{opt}  at  large $\el$   gives 
\be \la{equ} [E^{(0)}_1]_{\el \to \infty}  =   - { 1 \ov  2\el^2 }  + { 1 \ov  8 \el^4 } + ... 
=  - {  \l \m^2  \ov  2J^2 }  + O \Big({  \l^2  \ov  J^4 } \Big) \ . \ee
 Expanding the  non-zero mode  part of \rf{jo} we get 
 \be \la{jj}
 [E_1]_{\el \to \infty} 
 =  { \l  \ov 2 J^2 }  \sum_{n=1}^\infty   \Big( n \sqrt{ n^2 + 4 \m^2 } - n^2 - 2\m^2  \Big) 
 + O \Big( { \l^2  \ov  J^4 } \Big) \ . \ee The sum in \rf{jj}
is UV finite, with  the ``regulator''   $-n^2 - 2 \mu^2$ terms  coming from other 
modes  not seen in the  $AdS_3$   LL model.\foot{This expression  is  essentially equivalent to  the expressions in \ci{btz} 
found for a circular string solution.} If  we expand  \rf{jj} in large  $\mu$ first we would 
get order $\mu$ term   with coefficient $\zeta(-1)= - { 1 \ov 12} 
$  and order $\mu^2$ term with coefficient $\zeta(0)
= - \ha$.  The latter  cancels against 
the 0-mode contribution in \rf{equ} so we reproduce again the result \rf{mui}. 
This confirms that this contribution is  correctly captured by the lightest $AdS_3$ mode 
accounted for in the LL model. 

By analogy with the circular string in $AdS_3 \times S^1$ case discussed in sect 3.1
 in \ci{btz} 
we expect the  full expression, {\it i.e.}  the sum of \rf{equ}  and \rf{jj}, 
\be \la{sd}
\td E_1 = - {  \l \m^2  \ov  2J^2 }  + 
 { \l  \ov 2 J^2 }  \sum_{n=1}^\infty   \Big( n \sqrt{ n^2 + 4 \m^2 } - n^2 - 2\m^2 \Big ) 
 + O \Big( { \l^2  \ov  J^4 } \Big) \ , \ee
 can be  reproduced from the BA equations  for $sl(2)$ sector  spin chain  model. 
Note that the  expression in  \rf{sd}  starts,  in fact,  with a $\mu^3$ term. 
This  leading  $\m^3$ term comes from replacing sum by integral. 
Expanding in large $\el$ the 
1-loop string  result  in \rf{onlo}   (found \ci{ftt} by replacing the  summation over $n$ by integration)   we get 
\be 
(E_1)_{asympt.}= - { 4 \m \ov 3\el^2 } + ... =  - { 4 \l \m^3 \ov 3  J^2 } + ...   \  . 
\ee 
This term comes from 
usual 1-loop ``non-anomaly'' part  of  BA (see,  e.g.,  \ci{krc}) 
  while here  we are interested  in true
 finite size corrections.

The  conclusion is  that  the  ``non-wrapping'' string result \rf{mui}  should be captured by the 
ABA   since the  LL model  follows from  the spin chain description.  

\section{Propagators}
\label{propagators}

We present here the expressions for the bosonic
and fermionic propagators:
\be
\label{Bo}
 K_B^{-1}(p) &=&
\begin{pmatrix}
0 & \frac{2}{p^2+\frac{1}{4}(1+ \kahat^2)} & 0  & 0 & {\bf 0}_{1\times 4} \cr
\frac{2}{p^2+\frac{1}{4}(1+ \kahat^2)} & 0& 0  & 0 & {\bf 0}_{1\times 4} \cr
0 & 0 & \frac{p^2}{{\cal D}_B(p)} & \frac{\nuhat p_0}{{\cal D}_B(p)} 
                      & {\bf 0}_{1\times 4} \cr
0 & 0 & \frac{-{\nuhat} p_0}{{\cal D}_B(p)} & \frac{1+p^2}{{\cal D}_B(p)} & {\bf 0}_{1\times 4} \cr
 {\bf 0}_{4\times 1} & {\bf 0}_{4\times 1} & {\bf 0}_{4\times 1} & {\bf 0}_{4\times 1} & 
 \frac{{\bf 1}_{4\times 4}}{p^2+\frac{1}{4}\nuhat^2}
\end{pmatrix}
\\[5pt] \la{Bob}
{\cal D}_B(p)&\equiv & p^2(p^2+1)+\hat\nu^2 p_0^2\ , 
\ee
\be
K_F^{-1}(p)=\frac{N_+(p)}{{\cal D}_F(p)} + \frac{N_-(p)}{{\cal D}_F^*(p)} 
~,  \ ~~~~~\ \ \ 
{\cal D}_F(p)=\left(p_0 - \frac{\ri \nuhat}{4}\right)^2+p_1^2 
+ \frac{1+{\hat\nu}^2 }{4} \ .
\la{fep}\ee
The precise form of $N_+(p)$ and $N_-(p)$ appearing in the fermionic propagator is given in \cite{cuspJ}.

\section{Useful 1-loop integrals
\label{sec:1loop_int}}

We have used the notation
\be
{\rm I}[m^2]= \int \frac{d^2p}{(2\pi)^2}\frac{1}{p^2+m^2}\,.
\ee
This integral is logarithmically UV divergent. For zero mass it is also IR divergent. Integrals with different masses can be related thanks to the following identity 
\be 
{\rm I}[m_1^2]-{\rm I}[m_2^2] = \int \frac{d^2p}{(2\pi)^2}\frac{m_2^2-m_1^2}{(p^2+m_1^2)(p^2+m_2^2)}=\frac{1}{4\pi}\left(\ln m_2^2-\ln m_1^2\right)\,.
\ee 
Other convenient integrals, appearing in the evaluation of the double bubble topology, are
\be
  J_B(k,r) = \int \frac{d^2 p}{(2 \pi)^2} \frac {p_0^k p_1^r}{p^4 +
    p^2 + \hat{\nu}^2  p_0^2 },\ \ \ \ \ \ \ \ \ \ \ 
  J_F(k,r) = \int \frac{d^2 p}{(2 \pi)^2} \frac {p_0^k p_1^r}{p^2 + \frac {4  +3 
  \hat{\nu}^2 }{16} - \frac {\mathrm{i}} 2 \hat{\nu} p_0}\ .
\ee
Their evaluation yields
\be 
&&  J_B(0,0) = -\frac 1 {\sqrt{1 + \hat{\nu}^2}}
    \Big(\frac 1 {4 \pi} \ln \frac {(\sqrt{1 + \hat{\nu}^2} +
 1  )^2} 4  - {\rm I}[0] + {\rm I}[1 +
      \hat{\nu}^2]\Big),\\
 && J_B(1,0) = J_B(0,1) = 0,\ \ \ \ \ 
  J_B(2,0) = - \frac {1}{8 \pi \hat{\nu}^2} (\sqrt {1 + \hat{\nu}^2} - 1)^2 +
  \frac 1 2 {\rm I}\Big[\frac {(\sqrt{1 + \hat{\nu}^2} +
          1)^2} 4\Big],\\
 &&  J_B(0,2) = \frac {1}{8 \pi \hat{\nu}^2} (\sqrt {1 + \hat{\nu}^2} -1)^2 +
  \frac 1 2 {\rm I}\Big[\frac {(\sqrt{1 + \hat{\nu}^2} +
       1)^2} 4\Big],\\
&&  J_B(1,1) = 0,\ \ \ \ \
  J_F(0,0) = {\rm I}\big[\frac{1+\hat{\nu}^2}4\big],\ \ \ \ 
  J_F(1,0) = \mathrm{i} \frac {\hat{\nu}}4 \left({\rm I}\big[\frac{1+\hat{\nu}^2}4\big] - 
  \frac 1 {4 \pi}\right),\\
&&  J_F(0,1) =0,\ \ \ \ 
  J_F(2,0) = - \frac{2 + 3 \nu^2}{16} {\rm I}\big[\frac{1+\hat{\nu}^2}4\big] + \frac{7 
  \hat{\nu}^2 }{256 \pi},\\
&&   J_F(0,2) = - \frac{2 + 2 \hat{\nu}^2 }{16} {\rm I}\big[\frac{1+\hat{\nu}^2}4\big] +
   \frac{\hat{\nu}^2 }{256 \pi},\ \ \ \ \ 
  J_F(1,1) =0.
\ee

\section{${\cal W}_1$, ${\cal W}_2$, ${\cal W}_3$ in the fermionic 
sunset \label{W1W2W3}}

We list below the explicit expression for the integral quantities appearing in the various expressions (\ref{Fsunsetmx}), (\ref{Fsunsetmy})  and (\ref{FsunsetDB})
 contributing to the fermionic sunset 
\be
{\cal W}_i = \int_0^1du~~{\rm arctanh}\,u\,\    U_i\ , \ \ \ \ \ \ \ \ \ \ i=1,2,3 \ee
$$ 
 U_1=\frac{ 
 A \big{[}(1 + u^2)^3 + \hat\nu^4 u^2 (3 + 6 u^2 - u^4) + 
    \hat\nu^2 (1 + 7 (u^2 +  u^4) + u^6)\big{]}-(1 + \hat\nu^2) 
    (1 + (2 + 4 \hat\nu^2) u^2 + u^4)^2}{4 \pi^2 \hat\nu^4  u^2 ( u^2-1)^3 A}$$
\be && A\equiv \sqrt{1+u^2(2+8\hat\nu^2+4\hat\nu^4)+u^4} \ , \nonumber
\\
&& U_2=\frac{\hat\nu^4 (-1 + 6 u^2 + 3 u^4) + 8 u^{3} \left(u - \sqrt{1 + \hat\nu^2} \sqrt{\hat\nu^2 + u^2}\right) 
   + 4\hat\nu^2 u^2 \left(1 + 3 u^2 - 2 u \, \sqrt{1 + \hat\nu^2}\sqrt{\hat\nu^2 + u^2}  \right) }{
  2 \hat\nu^4 \pi^2 ( u^2-1)^3 u}\nonumber\\
&&U_3=  \frac{8 - 8 \sqrt{(1 + \hat\nu^2\, u^2)} (1 + \hat\nu^2)^{3/2} +
 4 \hat\nu^2 (3 + u^2) +\hat\nu^4 (3 + 6 u^2 - u^4)}{4 \hat\nu^4 \pi^2 (-1 + u^2)^3}\no
\ee
The large $\hat\nu$ expansion of ${\cal W}_1$, ${\cal W}_2$ and ${\cal W}_3$ is
\be
&&{\cal W}_1=-\frac{9}{64 \pi ^2}+\frac{\left(-51+8 \pi ^2+24 \ln 2-48 
\ln\hat\nu\right)}{96 \pi ^2}\frac{1}{\hat\nu^2}+\frac{\left(-77+3
 \pi ^2+16 \ln 2-32 \ln\hat\nu\right)}{64 \pi ^2}\frac{1}{\hat\nu^4}\nonumber\\
&&\ \ \ \ \ +\left(\frac{1}{64}+\frac{5}{24 \pi ^2}\right)\frac{1}{\hat\nu^6}-
\left(\frac{7}{256}+\frac{535}{1152 \pi ^2}\right)\frac{1}{\hat\nu^8} \nonumber\\ 
&& \ \ \ \ \ + \left(\frac{49}{1024}+\frac{1123}{1440 \pi ^2}\right) 
\frac{1}{\hat\nu^{10}}-\left(\frac{173}{2048}+\frac{74677}{57600 \pi ^2}\right) 
\frac{1}{\hat\nu^{12}}+\ldots\nonumber\\
&&{\cal W}_2=\frac{7}{32 \pi ^2}+\left(-\frac{1}{12}+\frac{11}{16 \pi ^2}\right) \frac{1}{\hat\nu^2}+\left(-\frac{1}{16}+\frac{23}{32 \pi ^2}\right) \frac{1}{\hat\nu^4}-\frac{1}{12 \pi ^2}\frac{1}{\hat\nu^6}
+\frac{41}{576 \pi ^2}\frac{1}{\hat\nu^8}\nonumber\\
&&\ \ \ \ \ \ \ -\frac{1}{16\pi^2}\frac{1}{\hat\nu^{10}}+\frac{179}{3200 \pi ^2}
 \frac{1}{\hat\nu^{12}}+\ldots \\
&&{\cal W}_3=-\frac{9}{64 \pi ^2}+\left(\frac{1}{12}-\frac{13}{32 \pi ^2}\right) \frac{1}{\hat\nu^2}
-\frac{2}{3 \pi ^2}\frac{1}{\hat\nu^3}+\left(\frac{1}{16}-\frac{9}{64 \pi ^2}\right) \frac{1}{\hat\nu^4}
-\frac{1}{9 \pi ^2}\frac{1}{\hat\nu^5}
-\frac{1}{24 \pi ^2}\frac{1}{\hat\nu^6}\nonumber\\
&&\ \ -\frac{29}{2100 \pi ^2}\frac{1}{\hat\nu^7}+\frac{41}{1152 \pi ^2}\frac{1}{\hat\nu^8}
+\frac{2099}{88200 \pi ^2}\frac{1}{\hat\nu^9}-\frac{1}{32 \pi ^2}\frac{1}{\hat\nu^{10}}
-\frac{333463}{13970880 \pi ^2}\frac{1}{\hat\nu^{11}}
+\frac{179}{6400 \pi ^2} \frac{1}{\hat\nu^{12}}+\ldots\nonumber
\ee
\vskip 0.2cm
Similarly, one can derive the following small $\hat\nu$ expansions:
\be 
&&{\cal W}_1=-\frac{\KK}{8 \pi ^2}-\frac{(-1+2 \KK+2\ln 2) \hat\nu^2}{32 \pi ^2}+\frac{7 (3+6
\KK+8\ln 2) \hat\nu^4}{1536 \pi ^2}-\frac{(1147+630 \KK+1152 \ln 2) \hat\nu^6}{46080 \pi
 ^2}\nonumber\\
&&+\frac{(19837+5490 \KK+13824 \ln 2) \hat\nu^8}{737280 \pi ^2}-\frac{(449031+74550
 \KK+256000 \ln 2) \hat\nu^{10}}{17203200 \pi ^2}+\ldots\nonumber\\
&&{\cal W}_2=\frac{(7-12 \ln \hat\nu) \hat\nu^2}{48 \pi ^2}+\frac{(17+168 \ln \hat\nu) \hat\nu^4}{1152 \pi ^2}+\frac{(-37-80 \ln \hat\nu) \hat\nu^6}{800 \pi ^2}+\frac{(127+180 \ln \hat\nu) \hat\nu^8}{2400 \pi ^2}\nonumber\\
&&\ \ \ -\frac{5 (149+168 \ln \hat\nu) \hat\nu^{10}}{14112 \pi ^2}+\ldots\\
&&{\cal W}_3=-\frac{\ln 2}{8 \pi ^2}\hat\nu^2+\frac{(5+28 \ln 2) \hat\nu^4}{384 \pi ^2}+\frac{(-41-128 \ln 2) \hat\nu^6}{2560 \pi ^2}+\frac{3 (7+16 \ln 2) \hat\nu^8}{1280 \pi ^2}\nonumber\\
&&\ \ \ \ +\frac{(-8249-15360 \ln 2) \hat\nu^{10}}{516096 \pi ^2}+\ldots \no 
\ee

\section{Two-loop contributions to the string free energy on the cylinder}
\label{ExplicitList}

Here we  list all the terms entering the computation of the 2-loop free energy 
on $\mathbb{R}\times S^1$, discarding purely power-like divergences in the 
continuous momentum integral. The overall factor of $4$ appearing in the definition
of $A_{\text{sunset}}^{\text{BBB}}$,
$A_{\text{double-bubble}}^{\text{BB}}$,
$A_{\text{sunset}}^{\text{BFF}}$,
$A_{\text{double-bubble}}^{\text{BF}}$,
$A_{\text{double-bubble}}^{\text{FF}}$ and
$A_{\text{non-1PI}}$ is a consequence of the closed string 
normalization we are using.

$\bullet$ Bosonic sunset:
\be
&&A_{\text{sunset}}^{\text{BBB}}=
4\int\frac{dp_0dq_0dr_0}{(2\pi)^4}\sum_{p_1,q_1,r_1}
\delta^{(2)}(p+q+r)\left[
B_{0,0,4}+B_{2,2,4}+B_{4,4,4}\right]\\
&&
B_{0,0,4}=
-\frac{5}{4}\, \frac{1}{P[p,0]P[q,0]}
+\frac{5}{2}\,  \frac{1-2 p_0^2}{P(p,0)P(r,4)}
+{5}\, \frac{(1+ p_0\, q_0)^2}{P(p,0)P(q,0)P(r,4)}
\\
&&B_{2,2,4}=
-4\frac{1+p_0^2}{P(p,2)P(r,4)}
+2\, \frac{(1+p_0^2)(1+q_0^2)}{P(p,2)P(q,2)P(r,4)}
\\
&&B_{4,4,4}=
-\frac{\frac{7}{4}+p_0^2}{P(p,4)P(r,4)}
+\frac{{3}+p_0^2\left(3+\frac{1}{4}p_0^2+\frac{1}{2}q_0^2\right)}
{P(p,4)P(q,4)P(r,4)}
\ee
$\bullet$ Bosonic double-bubble:
\be &&
A_{\text{double-bubble}}^{\text{BB}}=
4\int\frac{dp_0dq_0dr_0}{(2\pi)^4}\sum_{p_1,q_1,r_1}
\delta^{(2)}(p+q+r)\left[B_{2,4}+B_{4,4}\right]\ , \\
&&
B_{2,4}=
2\frac{1+p_0^2}{P(p,2)P(r,4)}\ , 
\ \ \ \ \ \ \ \ \ 
B_{4,4}=\frac{1}{2}\;\frac{1}{P(p,4)P(r,4)}\  
\ee
$\bullet$ Fermionic sunset:
\be
&&A_{\text{sunset}}^{\text{BFF}}=
4\int\frac{dp_0dq_0dr_0}{(2\pi)^4}\sum_{p_1,q_1,r_1}
\delta^{(2)}(p+q+r)\left[
F_{0,1,1}+ F_{2,1,1}+F_{4,1,1}\right]
\\
&&F_{0,1,1}=
-\frac{5 p_0^2}{P(p,1)P(r,0)}
+\frac{10 p_0 q_0 r_0^2}{P(p,1)P(q,1)P(r,0)}
\\
&&F_{2,1,1}=
\frac{4 p_0 q_0 (1+ r_0^2)}{P(p,1)P(q,1)P(r,2)}\ , 
\ \ \ \  \ 
F_{4,1,1}=
\,\frac{12 p_0^2}{P(p,1)P(r,4)}
+\frac{2 p_0\left(p_0-q_0\right)^2 q_0}{P(p,1)P(q,1)P(r,2)}
\ee
$\bullet$ Fermionic double-bubble: 
\be
A_{\text{double-bubble}}^{\text{FF}}=0
\ee
$\bullet$ Bosonic-fermionic double-bubble:
\be
&&A_{\text{double-bubble}}^{\text{BF}}=
4\int\frac{dp_0dq_0dr_0}{(2\pi)^4}\sum_{p_1,q_1,r_1}
\delta^{(2)}  (p+q+r)\left[C_{0,1/4}+C_{4,1}\right]\\ 
&&C_{0,1}=
\frac{5 p_0^2}{P(p,1)P(r,0)}
\ , \ \ \ \ \ \ \ \ \ \ 
C_{4,1}=
-\frac{4 p_0^2}{P(p,1)P(r,4)}
\ee
$\bullet$ Non-1PI:\\
 this diagram arises due to the presence of a tadpole $T(p)$ 
and has $\tilde\phi$ as internal leg
\be
&&A_{\text{non-1PI}}=
4\int\frac{dp_0dq_0}{(2\pi)^4}\sum_{p_1,q_1}
{\cal A}^{\text{non-1PI}}
~~, ~~~~~~~~~~~
{\cal A}^{\text{non-1PI}}=\frac{1}{2}T(p)T(q)\ , \\
&& \label{tad}
T(p)=\frac{1+\frac{1}{2}p_0^2}{P(p, 4)}
         +\frac{5}{4}\,\frac{p_0^2-p_1^2}{P(p,0)}
         +\frac{1+p_0^2}{P(p,2)}-4\,\frac{p_0^2}{P(p,1)}
\ee
The first three contributions in $T(p)$ come respectively from the bosonic vertices  
$\tilde \phi^3$, $\tilde\phi\,y^2$ and $\tilde\phi\,\tilde{x}\tilde{x}^{*}$, while the last term is due to the cubic vertex of $\tilde\phi$ and the fermions.

We notice that  $C_{0,1}$ cancels against the first term of $F_{0,1,1}$.
We also notice the partial cancellation of $C_{4,1}$ against 
the first term of $F_{4,1,1}$, of $B_{2,4}$ against the first term in $B_{2,2,4}$ 
and of $B_{4,4}$ with the first term in $B_{4,4,4}$.

\section{Sums and integrals \label{appInt}}

\begin{figure}[h]
\begin{center}
\includegraphics[width=90mm]{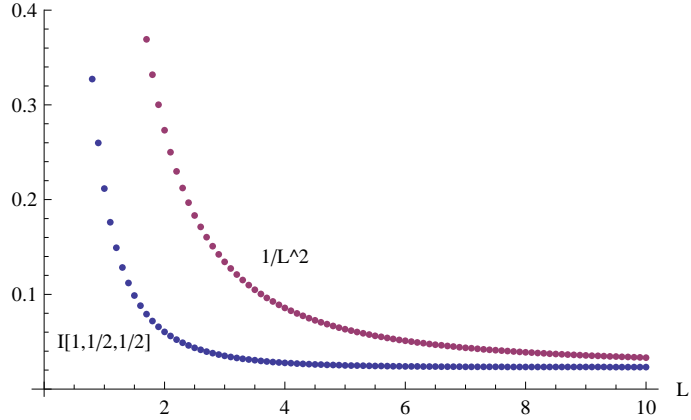}
\parbox{13cm}{\caption{The plots of the massive 2-loop integral ${\rm I}[1,1/2,1/2]$
 (blue) and $\frac{1}{L^2}+\frac{\KK}{4\pi^2}$ (red) as function of $L$. 
  ${\rm I}[1,1/2,1/2]$  quickly approaches the continuum limit $\frac{\KK}{4\pi^2} $.  }
\label{decay}}
\end{center}
\end{figure}

Some simple sums which occur in the calculation of  1-loop integrals are
\be
\frac{2\pi}{L}\sum_{n\in \IZ}\frac{1}{\left(\frac{2\pi n}{L}\right)^2+p_0^2}&=&
\frac{\pi}{p_0}
\coth\left(\ha Lp_0\right)\ , 
\\
\frac{2\pi}{L}\sum_{n\in \IZ}\frac{1}{\left(\frac{2\pi n}{L}\right)^2+p_0^2+m^2}&=&
\frac{\pi}{ \sqrt{p_0^2+m^2}}
\coth\big(\ha L\sqrt{p_0^2+m^2}\big)~~.
\label{1loopsum}
\ee
As already observed in the main text these expressions  imply
 that massive one loop integrals are exponentially suppressed.
Let us  now consider an example of a purely massive 2-loop integral 
 and show explicitly show that it  decays faster than $1/L^2$, {\it i.e.}  
  does not contribute to the leading finite size term.
   The prototype of a   2-loop integral in a theory on a 2d  cylinder is 
\be
\begin{aligned}
&~~~~~~~~~~~~~~~~~~~~~~~~~~~~~~~~~~~~~~~{\rm I}[m_1^2,m_2^2,m_3^2]=\int \frac{dp_0\, dq_0}{(2\pi)^4}~\Sigma[m_1^2,m_2^2,m_3^2]\,,\\
&\Sigma[m_1^2,m_2^2,m_3^2]\equiv
\left(\frac{2\pi}{L}\right)^2 \sum_{n,m=-\infty}^\infty\frac{1}{\Big[\big(
\frac{2\pi n}{L}\big)^2+p_0^2+m_1^2\Big]\Big[\big(\frac{2\pi m}{L}\big)^2
+q_0^2+m_2^2\Big]\Big[\left(\frac{2\pi (n+m)}{L}\right)^2+A^2\Big]} \\
& ~~~~~~~~~~~~~~~~~~~~~~~~~~~~~~~~~~~~A\equiv\sqrt{(p_0+q_0)^2+m_3^2}
\end{aligned}
\ee
When all  the masses are  non-vanishing the integral ${\rm I}[m_1^2,m_2^2,m_3^2]$ is convergent.  To study its behaviour as a function of $L$ we can first compute the sum over $n,m$, $\Sigma[m_1^2,m_2^2,m_3^2]$. This can be found analytically. If we define the function
\be
\begin{aligned}
\Theta[a,b,c]=&-\frac{\pi}{2 a} \coth[\pi a]\left(\frac{1}{(a+b)^2-c^2}
+\frac{1}{(a-b)^2-c^2}\right)+\frac{\pi}{2 c} \coth\big[\pi (b-c)\big]\frac{1}{(b-c)^2-a^2}\\
&-\frac{\pi}{2 c} \coth\big[\pi (b+c)\big]\frac{1}{(b+c)^2-a^2}
\end{aligned}
\ee
we then have the following explicit expression for the sum 
\be
\begin{aligned}
\Sigma[m_1^2,m_2^2,m_3^2]=&\Big(\frac{L}{2\pi}\Big)^4\Big(\frac{2\pi^2}{L\sqrt{q_0^2+m_2^2}}\coth\Big[\frac{L\sqrt{q_0^2+m_2^2}}{2}\Big]\Theta\Big[\frac{L\sqrt{p_0^2+m_1^2}}{2\pi},\frac{L\sqrt{q_0^2+m_2^2}}{2\pi},\frac{L A}{2\pi}\Big]\Big.\\
&+\Big.\frac{2\pi^2}{A L}\coth\Big[\frac{L A}{2}
\Big]\Theta\Big[\frac{L\sqrt{p_0^2+m_1^2}}{2\pi},
\frac{L A}{2\pi},\frac{L\sqrt{q_0^2+m_2^2}}{2\pi}\Big]\Big)\,.
\end{aligned}
\ee
The remaining continuous integrals over $p_0$ and $q_0$ can be computed 
numerically for various values of the size $L$. To exemplify 
the behaviour of a purely massive integral we have plotted 
${\rm I}[1,1/2,1/2]$ for the range $1<L<10$ in Fig.~\ref{decay}. Note that
 the function quickly approaches the continuum limit 
 ($L\rightarrow \infty$) value of $\frac{\KK}{4\pi^2} $ and in doing so its decay
  is faster than $1/L^2$. For this reason it cannot contribute to finite size effects.

\section{Some useful 1-dimensional 2-loop integrals \label{2Lints}}

\be
&&\int dp_0 d q_0 dr_0\frac{\delta(p_0+q_0+r_0)}{P(p,m_p)P(q,m_q)P(r,m_r)}
\cr
&& \ \ \ \ = \frac{1}{2\pi}
\frac{2\pi^3}{\sqrt{p_1^2+m_p^2}\sqrt{q_1^2+m_q^2}\sqrt{r_1^2+m_r^2}
(\sqrt{p_1^2+m_p^2}+\sqrt{q_1^2+m_q^2}+\sqrt{r_1^2+m_r^2})}
\\
&&
\int dp_0 d q_0 dr_0\frac{p_0 q_0\ \delta(p_0+q_0+r_0)}{P(p,m_p)P(q,m_q)P(r,m_r)}
\cr
&& \ \ \ = 
-\frac{1}{2\pi}\frac{2\pi^3}{\sqrt{r_1^2+m_r^2}
(\sqrt{p_1^2+m_p^2}+\sqrt{q_1^2+m_q^2}+\sqrt{r_1^2+m_r^2})}
\\
&&
\int dp_0 d q_0 dr_0\frac{p_0^2\ \delta(p_0+q_0+r_0)}{P(p,m_p)P(q,m_q)P(r,m_r)}
\cr
&&\ \ \  =  
-\frac{1}{2\pi}\frac{2\pi^3\,\sqrt{p_1^2+m_p^2}}{\sqrt{q_1^2+m_q^2}\sqrt{r_1^2+m_r^2}
(\sqrt{p_1^2+m_p^2}+\sqrt{q_1^2+m_q^2}+\sqrt{r_1^2+m_r^2})}
\ee


\baselineskip13pt
\begin{thebibliography}{99}


\bibitem{gkp}
  S.~S.~Gubser, I.~R.~Klebanov and A.~M.~Polyakov,
  ``A semi-classical limit of the gauge/string correspondence,''
  Nucl.\ Phys.\ B {\bf 636}, 99 (2002)
  [hep-th/0204051].
  
  
  \bibitem{ft1}
  S.~Frolov and A.A.~Tseytlin,
  ``Semiclassical quantization of rotating superstring in AdS(5) x S(5),''
  JHEP {\bf 0206}, 007 (2002)
  [hep-th/0204226].
  
    \bibitem{cusp}
  S.~Giombi, R.~Ricci, R.~Roiban, A.A.~Tseytlin and C.~Vergu,
  ``Quantum $AdS_5 \times S^5$  superstring in the AdS light-cone gauge,''
  JHEP {\bf 1003}, 1 (2010)
  [arXiv:0912.5105 [hep-th]].

   \bibitem{cuspJ}
  S.~Giombi, R.~Ricci, R.~Roiban, A.A.~Tseytlin and C.~Vergu,
  ``Generalized scaling function from light-cone gauge $AdS_5 \times S^5$
  superstring,''
  JHEP {\bf 1006} (2010) 060
  [arXiv:1002.0018].
  
  \bi{serban}
D.~Serban,
  ``Integrability and the AdS/CFT correspondence,''
  arXiv:1003.4214.\\
A.~Rej,
  ``Integrability and the AdS/CFT correspondence,''
  J.\ Phys.\ A  {\bf 42}, 254002 (2009)
  [arXiv:0907.3468].


\bi{bmsz}  
 N.~Beisert, J.~A.~Minahan, M.~Staudacher and K.~Zarembo,
  ``Stringing spins and spinning strings,''
  JHEP {\bf 0309}, 010 (2003)
  [arXiv:hep-th/0306139].
  
 
\bi{btz}
N.~Beisert, A.A.~Tseytlin and K.~Zarembo,
  ``Matching quantum strings to quantum spins: One-loop vs. 
finite-size
  corrections,''
  Nucl.\ Phys.\  B {\bf 715}, 190 (2005)
  [arXiv:hep-th/0502173].
  
  
  \bi{ft3}   S.~Frolov and A.A.~Tseytlin,
  ``Rotating string solutions: AdS/CFT duality in non-supersymmetric
  sectors,''
  Phys.\ Lett.\  B {\bf 570}, 96 (2003)
  [arXiv:hep-th/0306143].

  \bi{bt} N.~Beisert and A.A.~Tseytlin,
  ``On quantum corrections to spinning strings and Bethe equations,''
  Phys.\ Lett.\  B {\bf 629}, 102 (2005)
  [arXiv:hep-th/0509084].


\bi{bes}
N.~Beisert, B.~Eden and M.~Staudacher,
  ``Transcendentality and crossing,''
  J.\ Stat.\ Mech.\  {\bf 0701}, P021 (2007)
  [arXiv:hep-th/0610251].
  
  \bi{call}
C.~G.~Callan, T.~McLoughlin and I.~Swanson,
  ``Holography beyond the Penrose limit,''
  Nucl.\ Phys.\  B {\bf 694}, 115 (2004)
  [arXiv:hep-th/0404007].
C.~G.~Callan, H.~K.~Lee, T.~McLoughlin, J.~H.~Schwarz, I.~Swanson and X.~Wu,
  ``Quantizing string theory in AdS(5) x S5: Beyond the pp-wave,''
  Nucl.\ Phys.\  B {\bf 673}, 3 (2003)
  [arXiv:hep-th/0307032].
  
  \bi{ll}
J.~A.~Minahan, A.~Tirziu and A.A.~Tseytlin,
  ``$1/J^2$ corrections to BMN energies from the quantum long range
  Landau-Lifshitz model,''
  JHEP {\bf 0511}, 031 (2005)
  [arXiv:hep-th/0510080].
  



\bibitem{bgk}
  A.~V.~Belitsky, A.~S.~Gorsky and G.~P.~Korchemsky,
  ``Logarithmic scaling in gauge / string correspondence,''
  Nucl.\ Phys.\ B {\bf 748}, 24 (2006)
  [hep-th/0601112].


 \bibitem{ftt}
  S.~Frolov, A.~Tirziu and A.A.~Tseytlin,
  ``Logarithmic corrections to higher twist scaling at strong coupling from
  AdS/CFT,''
  Nucl.\ Phys.\  B {\bf 766} (2007) 232
  [arXiv:hep-th/0611269].

\bi{am2}
 L.~F.~Alday and J.~M.~Maldacena,
  ``Comments on operators with large spin,''
  JHEP {\bf 0711}, 019 (2007)
  [arXiv:0708.0672].



\bi{frs}
L.~Freyhult, A.~Rej and M.~Staudacher,
  ``A Generalized Scaling Function for AdS/CFT,''
  J.\ Stat.\ Mech.\  {\bf 0807}, P07015 (2008)
  [arXiv:0712.2743].
  

 \bi{bfst}
N.~Beisert, S.~Frolov, M.~Staudacher and A.A.~Tseytlin,
  ``Precision spectroscopy of AdS/CFT,''
  JHEP {\bf 0310}, 037 (2003)
  [arXiv:hep-th/0308117].





  
 \bibitem{rt1}
	R.~Roiban and A.A.~Tseytlin,
``Strong-coupling expansion of cusp anomaly from quantum superstring,''
  JHEP {\bf 0711}, 016 (2007)
  [arXiv:0709.0681].
  R.~Roiban, A.~Tirziu and A.A.~Tseytlin,
  ``Two-loop world-sheet corrections in $AdS_5 \times  S^5$ superstring,''
  JHEP {\bf 0707}, 056 (2007)
  [arXiv:0704.3638].

  
  
  \bibitem{rt2}
  R.~Roiban and A.A.~Tseytlin,
  ``Spinning superstrings at two loops: strong-coupling corrections to
  dimensions of large-twist SYM operators,''
  Phys.\ Rev.\  D {\bf 77} (2008) 066006
  [arXiv:0712.2479].


  
    \bibitem{gromov}
  N.~Gromov,
  ``Generalized Scaling Function at Strong Coupling,''
  JHEP {\bf 0811} (2008) 085
  [arXiv:0805.4615].
  
  
 
  \bi{basso}
Z.~Bajnok, J.~Balog, B.~Basso, G.~P.~Korchemsky and 
L.~Palla,
  ``Scaling function in AdS/CFT from the O(6) sigma model,''
  Nucl.\ Phys.\  B {\bf 811}, 438 (2009)
  [arXiv:0809.4952].

 
   \bi{volin08}
 D.~Volin,
  ``The 2-loop generalized scaling function from the BES/FRS equation,''
  arXiv:0812.4407.

  \bibitem{beccaria}
  M.~Beccaria,
  ``The generalized scaling function of AdS/CFT and semiclassical string
  theory,''
  JHEP {\bf 0807} (2008) 082
  [arXiv:0806.3704].

    \bibitem{volin}
  D.~Volin,
  ``Quantum integrability and functional equations,''
  arXiv:1003.4725.
  
  \bi{es}
B.~Eden and M.~Staudacher,
  ``Integrability and transcendentality,''
  J.\ Stat.\ Mech.\  {\bf 0611}, P014 (2006)
  [arXiv:hep-th/0603157].

\bi{frz}
 L.~Freyhult and S.~Zieme,
  ``The virtual scaling function of AdS/CFT,''
  Phys.\ Rev.\  D {\bf 79}, 105009 (2009)
  [arXiv:0901.2749].

\bi{fio_large_spin}
  D.~Bombardelli, D.~Fioravanti and M.~Rossi,
  ``Large spin corrections in ${\cal N}=4$ SYM sl(2): still a linear integral
  equation,''
  Nucl.\ Phys.\  B {\bf 810}, 460 (2009)
  [arXiv:0802.0027 [hep-th]].

  \bi{ban}
Z.~Bajnok, R.~A.~Janik and T.~Lukowski,
  ``Four loop twist two, BFKL, wrapping and strings,''
  Nucl.\ Phys.\  B {\bf 816}, 376 (2009)
  [arXiv:0811.4448].
T.~Lukowski, A.~Rej and V.~N.~Velizhanin,
  ``Five-Loop Anomalous Dimension of Twist-Two Operators,''
  Nucl.\ Phys.\  B {\bf 831}, 105 (2010)
  [arXiv:0912.1624].

  \bi{fio}
    D.~Fioravanti, G.~Infusino and M.~Rossi,
  ``On the high spin expansion in the $sl(2)$ ${\cal N}=4$ SYM theory,''
  Nucl.\ Phys.\  B {\bf 822}, 467 (2009)
  [arXiv:0901.3147 [hep-th]].
 D.~Fioravanti, P.~Grinza and M.~Rossi,
  ``On the logarithmic powers of $sl(2)$ SYM$_4$,''
  Phys.\ Lett.\  B {\bf 684}, 52 (2010)
  [arXiv:0911.2425].
D.~Fioravanti, P.~Grinza and M.~Rossi,
  ``Beyond cusp anomalous dimension from integrability,''
  Phys.\ Lett.\  B {\bf 675}, 137 (2009)
  [arXiv:0901.3161].


\bi{zama}
S.~Schafer-Nameki and M.~Zamaklar,
  ``Stringy sums and corrections to the quantum string Bethe 
ansatz,''
  JHEP {\bf 0510}, 044 (2005)
  [arXiv:hep-th/0509096].
S.~Schafer-Nameki, M.~Zamaklar and K.~Zarembo,
  ``How Accurate is the Quantum String Bethe Ansatz?,''
  JHEP {\bf 0612}, 020 (2006)
  [arXiv:hep-th/0610250].
  
  
\bi{ft2}
S.~Frolov and A.A.~Tseytlin,
  ``Multi-spin string solutions in AdS(5) x S5,''
  Nucl.\ Phys.\  B {\bf 668}, 77 (2003)
  [arXiv:hep-th/0304255].
 \bibitem{bed}
  M.~Beccaria, G.~V.~Dunne, V.~Forini, M.~Pawellek and A.A.~Tseytlin,
  ``Exact computation of one-loop correction to energy of spinning folded
  string in \adss,''
  J.\ Phys.\ A {\bf 43}, 165402 (2010)
  [arXiv:1001.4018].




\bi{grom}
N.~Gromov,
  ``Y-system and Quasi-Classical Strings,''
  JHEP {\bf 1001}, 112 (2010)
  [arXiv:0910.3608].
  
  \bi{krt}
M.~Kruczenski,
  ``Spin chains and string theory,''
  Phys.\ Rev.\ Lett.\  {\bf 93}, 161602 (2004)
  [arXiv:hep-th/0311203].
M.~Kruczenski, A.~V.~Ryzhov and A.A.~Tseytlin,
  ``Large spin limit of AdS(5) x S5 string theory and low 
energy  expansion
  of ferromagnetic spin chains,''
  Nucl.\ Phys.\  B {\bf 692}, 3 (2004)
  [arXiv:hep-th/0403120].

  
  
  \bi{ste}
B.~J.~Stefanski and A.A.~Tseytlin,
  ``Large spin limits of AdS/CFT and generalized Landau-
Lifshitz equations,''
  JHEP {\bf 0405}, 042 (2004)
  [arXiv:hep-th/0404133].
  
  \bi{ptt}
I.~Y.~Park, A.~Tirziu and A.A.~Tseytlin,
  ``Spinning strings in AdS(5) x S5: One-loop correction to 
energy in  SL(2)
  sector,''
  JHEP {\bf 0503}, 013 (2005)
  [arXiv:hep-th/0501203].
  
 \bibitem{mtt}
  J.~A.~Minahan, A.~Tirziu and A.A.~Tseytlin,
  ``1/J corrections to semiclassical AdS/CFT states from quantum
  Landau-Lifshitz model,''
  Nucl.\ Phys.\  B {\bf 735}, 127 (2006)
  [arXiv:hep-th/0509071].
  
\bibitem{mt2}
R.R. ~Metsaev and A.A.~Tseytlin,
``Superstring action in $AdS_5\times S^5$: $\k$-symmetry light cone gauge,''
Phys.\ Rev.\ D {\bf 63}, 046002 (2001)
[arXiv:hep-th/0007036].
R.R. ~Metsaev, C.B.~Thorn and A.A.~Tseytlin,
``Light-cone Superstring in AdS Space-time,''
Nucl.\ Phys.\ B {\bf 596}, 151 (2001)
[arXiv:hep-th/0009171].

\bi{amm}
 L.~F.~Alday and J.~M.~Maldacena,
  ``Gluon scattering amplitudes at strong coupling,''
  JHEP {\bf 0706}, 064 (2007)
  [arXiv:0705.0303].

\bi{krtt}
M.~Kruczenski, R.~Roiban, A.~Tirziu and A.A.~Tseytlin,
  ``Strong-coupling expansion of cusp anomaly and gluon amplitudes from quantum
  open strings in $AdS_5 \times S^5$,''
  Nucl.\ Phys.\  B {\bf 791}, 93 (2008)
  [arXiv:0707.4254].


   \bi{gro}
N. Gromov, private communication. 


\bi{krc}
P.~Y.~Casteill and C.~Kristjansen,
  ``The Strong Coupling Limit of the Scaling Function from 
the Quantum   String
  Bethe Ansatz,''
  Nucl.\ Phys.\  B {\bf 785}, 1 (2007)
  [arXiv:0705.0890].




\end{thebibliography}
\end{document}